\documentclass[aps,prd,superscriptaddress,nofootinbib,amsmath,amsfonts,preprintnumbers,groupedaddress,10pt,english]{revtex4}
\usepackage{amsmath}
\usepackage{amssymb}
\usepackage{babel}
\usepackage{wrapfig}
\usepackage{cancel}

\usepackage{relsize,exscale}
\makeatletter

\usepackage{array,multirow,graphicx}
\usepackage{dcolumn}
\usepackage{newlfont}
\usepackage{bm}
\usepackage[colorlinks,citecolor=blue,urlcolor=blue,linkcolor=blue]{hyperref}
\usepackage[figtopcap]{subfigure}
\usepackage{color}
\usepackage{verbatim}

\usepackage{scalerel}
\usepackage{tikz}
\usetikzlibrary{svg.path}
\definecolor{orcidlogocol}{HTML}{A6CE39}
\tikzset{
  orcidlogo/.pic={
    \fill[orcidlogocol] svg{M256,128c0,70.7-57.3,128-128,128C57.3,256,0,198.7,0,128C0,57.3,57.3,0,128,0C198.7,0,256,57.3,256,128z};
    \fill[white] svg{M86.3,186.2H70.9V79.1h15.4v48.4V186.2z}
                 svg{M108.9,79.1h41.6c39.6,0,57,28.3,57,53.6c0,27.5-21.5,53.6-56.8,53.6h-41.8V79.1z M124.3,172.4h24.5c34.9,0,42.9-26.5,42.9-39.7c0-21.5-13.7-39.7-43.7-39.7h-23.7V172.4z}
                 svg{M88.7,56.8c0,5.5-4.5,10.1-10.1,10.1c-5.6,0-10.1-4.6-10.1-10.1c0-5.6,4.5-10.1,10.1-10.1C84.2,46.7,88.7,51.3,88.7,56.8z};}}
\newcommand\orcid[1]{\href{https://orcid.org/#1}{\mbox{\scalerel*{
\begin{tikzpicture}[yscale=-1,transform shape]
\pic{orcidlogo};
\end{tikzpicture}
}{|}}}}
\graphicspath{{./}{Figs/}}
\begin{document}
\date{\today}
\title{Constraining linear form of $f(\mathcal{R,G,T})$  gravity from astrophysical observations of the Pulsar U1724}

\author{G.~G.~L.~Nashed~\orcid{0000-0001-5544-1119}}
\email{nashed@bue.edu.eg}
\affiliation {Centre for Theoretical Physics, The British University, P.O. Box
43, El Sherouk City, Cairo 11837, Egypt\\Center for Space Research, North-West University, Potchefstroom 2520, South Africa}
\begin{abstract}
In this work we examine the internal structure of compact stars within an extended gravitational framework described by the function $f(\mathcal{R},\mathcal{G},\mathcal{T})$. Throughout this work, the quantity $\mathcal{R}$ refers to the curvature scalar formed from the Ricci tensor.
The term $\mathcal{G}$ denotes the Gauss--Bonnet curvature invariant, while $\mathcal{T}$ corresponds to the trace obtained by contracting the matter energy-momentum tensor. Our analysis is directed toward massive radio pulsars with masses above $1.8\,M_{\odot}$, which provide an exceptional testing ground for gravity under conditions inaccessible to laboratory experiments. Adopting the linear form $f(\mathcal{R},\mathcal{G},\mathcal{T})=\mathcal{R}+\alpha\,\mathcal{G}+\beta\,\mathcal{T}$
where $\alpha$ and $\beta$ are parameters of suitable dimensionality,\footnote{$\alpha$ has dimensions of $[L^{2}]$ and $\beta$ carries units of $[N^{-1}]$.}
we obtain an exact analytic solution for static anisotropic stellar matter in hydrostatic equilibrium. This solution allows all physical quantities to be expressed in terms of the dimensionless parameters $
\alpha_{1}=\alpha/R^{2},\qquad
\beta_{1}=\beta/\kappa^{2}$
together with the compactness $C=2GM/(Rc^{2})$. We constraint the two parameters $\alpha$ and $\beta$ by matching the model with the mass and radius of pulsar \textit{U1724} requires restricting these parameters to
$\alpha_{1}=\pm0.023$ and $\beta_{1}=\pm0.001$, where $\kappa^{2}=8\pi G/c^{4}$ is the standard Einstein coupling. The resulting stellar configuration satisfies the causal bound on the radial sound speed, $c_{s}^{2}<c^{2}/3$, distinguishing it from the corresponding behaviour in general relativity. No specific equation of state is introduced; nevertheless, the solution exhibits an effectively linear relation between pressure and density. The model predicts central densities several times higher than the nuclear saturation value $\rho_{\mathrm{nuc}}=2.6\times10^{14}\,\mathrm{g\,cm^{-3}}$, while the surface density $\rho_{s}$ also remains above this threshold. A mass--radius curve derived from the boundary density is presented, showing consistency with existing astrophysical constraints.
\end{abstract}

\maketitle
\section{Introduction}\label{Sec:Introduction}

A wide range of astronomical observations---including type Ia supernova measurements, the cosmic microwave background, and the large-scale distribution of matter---have firmly established that the Universe is undergoing accelerated expansion \cite{SupernovaCosmologyProject:1997czu,SupernovaSearchTeam:1998fmf,SDSS:2003eyi,WMAP:2006bqn}. This difficulty has motivated the exploration of modified theories of gravity (MTGs), which attempt to address cosmic acceleration through adjustments to the gravitational action. One of these extensions is $f(\mathcal{R})$  \cite{DeFelice:2010aj}, where the  Hilbert action $\mathcal{R}$ is extended  to an arbitrary function $f(\mathcal{R})$. Models of this type have been shown to reproduce the late-time accelerated expansion \cite{Carroll:2003wy}. A consistent formulation that remains compatible with Newtonian gravity was introduced in \cite{Nojiri:2003ft}, and subsequent work has established criteria for constructing cosmologically viable models within this framework \cite{Capozziello:2006dj,Amendola:2006kh,Nojiri:2006gh,Santos:2007bs,Nojiri:2007jr,Tsujikawa:2007xu,Nojiri:2004bi,Allemandi:2005qs}.

An alternative extension of $f(\mathcal{R})$ gravity was put forward in Ref.~\cite{Nojiri:2004bi},
where the usual separation between matter and curvature is relaxed.
In that formulation, the action contains a term that ties the matter Lagrangian to the geometric part of the theory, producing a nonminimal interaction.
This additional coupling modifies the effective gravitational dynamics and has been discussed as a possible route to account for the observed late-time cosmic acceleration c.f. \cite{Bertolami:2007gv,Nojiri:2017ncd,Nojiri:2010wj,Nashed:2018oaf,Nashed:2018efg,Nashed:2018piz}. The presence of a non-minimal coupling between matter and geometry is known to generate an additional force, which becomes apparent once different matter Lagrangians are considered \cite{Bertolami:2008ab}. In fact, as shown in \cite{Bertolami:2007vu,Bertolami:2009cd}, even the most natural choices of the matter action do not eliminate this extra interaction. The consequences of such couplings for the equilibrium and structure of relativistic stars have therefore attracted significant attention.

Among the many extensions of general relativity (GR), Gauss--Bonnet (GB) gravity occupies a distinguished position. The GB term is constructed from curvature invariants and takes the form\footnote{The Riemann tensor $\mathcal{R}^{\alpha}{}_{\beta\xi\eta}$, the Ricci tensor $\mathcal{R}_{\beta\xi}$, and the Ricci scalar $\mathcal{R}$ are defined in the usual manner via the Levi--Civita connection $\overset{\mathbf{L}}{\Gamma}{}^{\mu}{}_{\nu\sigma}$. Explicitly,
\[
R^{\mu}{}_{\nu\rho\sigma}
=\partial_{\rho}\overset{\mathbf{L}}{\Gamma}{}^{\mu}{}_{\nu\sigma}
-\partial_{\sigma}\overset{\mathbf{L}}{\Gamma}{}^{\mu}{}_{\nu\rho}
+\overset{\mathbf{L}}{\Gamma}{}^{\mu}{}_{\tau\rho}\overset{\mathbf{L}}{\Gamma}{}^{\tau}{}_{\nu\sigma}
-\overset{\mathbf{L}}{\Gamma}{}^{\mu}{}_{\tau\sigma}\overset{\mathbf{L}}{\Gamma}{}^{\tau}{}_{\nu\rho},
\]
with
\[
\overset{\mathbf{L}}{\Gamma}{}^{\mu}{}_{\nu\sigma}
=g^{\mu\alpha}\left(g_{\alpha\nu,\sigma}+g_{\alpha\sigma,\nu}-g_{\sigma\nu,\alpha}\right).
\]
The Ricci tensor is $R_{\mu\nu}=R^{\rho}{}_{\mu\rho\nu}$ and $R=g^{\mu\nu}R_{\mu\nu}$.}
\begin{equation}
\mathcal{G}
=\mathcal{R}_{\alpha\beta\xi\eta}\mathcal{R}^{\alpha\beta\xi\eta}
-4\mathcal{R}_{\beta\xi}\mathcal{R}^{\beta\xi}
+\mathcal{R}^{2},
\label{GB}
\end{equation}
where $\mathcal{G}$ denotes the Gauss--Bonnet invariant. This combination of curvature terms is free from spin-2 ghost instabilities \cite{Calcagni:2005im,DeFelice:2006pg,DeFelice:2008wz} and arises naturally as the second-order Lovelock scalar. In four dimensions,the GB expression acts solely as a topological term and leaves the equation  of motions unchanged unless it is multiplied by a scalar field or promoted to a general function $f(\mathcal{G})$ \cite{Metsaev:1987zx,Nojiri:2006je,Amendola:2007ni}.

The idea of treating the GB term through an arbitrary function was introduced by Nojiri and Odintsov \cite{Nojiri:2005jg}, leading to the class of modified $f(\mathcal{G})$ gravity theories. This extension has been shown to provide viable models of dark energy while remaining compatible with solar-system bounds \cite{DeFelice:2009aj}. Within this approach, one can account for the transition from deceleration to acceleration, accommodate both non-phantom and phantom phases, and even unify early- and late-time cosmological dynamics \cite{Cognola:2006eg,Nojiri:2006ri}. As a result, MTGs continue to offer a promising theoretical laborato

In recent investigations of highly dense astrophysical systems, considerable attention has been directed toward objects such as neutron stars, pulsars, and black holes. A broad spectrum of modified theories of gravity (MTGs) has been employed to explore their internal composition and gravitational behavior \cite{Yousaf:2017lto,Bhatti:2017fov}.

A convenient way to formulate stellar models is to work within a spherically symmetric spacetime, since this geometric setting allows a wide range of possibilities for specifying the matter configuration. While many classical analyses treat the stellar interior as an idealized perfect fluid, later studies have emphasized that pressures within compact bodies need not be isotropic. Introducing anisotropy alters the conditions for equilibrium and affects the overall stability of the configuration, and such effects can be tracked through an appropriately chosen equation of state \cite{Bowers:1974tgi}.

For these reasons, MTG frameworks often incorporate anisotropic matter profiles as a more versatile and physically meaningful option. Various investigations have already demonstrated that several qualitative and quantitative features of compact objects are influenced when anisotropic stresses are considered together with electromagnetic contributions \cite{Bhar:2014jta}.\\

 The remarkable advancements in astrophysical observations of pulsar masses and radii have been achieved through the amalgamation of Shapiro time delay (radio signals),  gravitational wave signals, and X-rays. This convergence imparts a robust constraint on the proposed equations of state (EoSs). Noteworthy among these mass-radius measurements are as follows:  The pulsar \textbf{PSR J0348+0432} possesses a mass of $M = 2.01 \pm 0.04~M_\odot$ and an estimated radius of $R = 13 \pm 2$km\citep{Antoniadis:2013pzd}.
Similarly, \textbf{PSR J0740+6620} has been measured to have a mass of $M = 2.08 \pm 0.07~M_\odot$\citep{NANOGrav:2019jur,Fonseca:2021wxt} and a radius of $R = 13.7^{+2.6}_{-1.5}$~km\cite{Miller:2021qha}.
An independent NICER observation further constrains its radius to $R = 12.39^{+1.30}_{-0.98}$km\cite{Riley:2021pdl}.
The stellar \textbf{PSR J1614--2230} has $M = 1.908 \pm 0.016~M_\odot$ and a radius of $R = 13 \pm 2$km\citep{Demorest:2010bx,Fonseca:2016tux,NANOGRAV:2018hou}.

In addition, \textbf{PSR J0437--4715} is reported to have a mass of $M = 1.44 \pm 0.07~M_\odot$\citep{Reardon:2015kba}, with a radius of $R = 13.6 \pm 0.9$~km, derived from analyses of surface X-ray thermal emissions\citep{Gonzalez-Caniulef:2019wzi}.
The pulsar \textbf{PSR J0030+0451} has a mass of $M = 1.44^{+0.15}{-0.14}~M\odot$ and a radius of $R = 13.02^{+1.24}_{-1.06}$km, as determined by NICER\citep{Miller:2019cac}.

We also consider three additional \textbf{mass--radius} constraints inferred from gravitational wave detections by the LIGO/Virgo collaboration.
The first observed neutron star merger, \textbf{GW170817}, provides two component measurements:
GW170817-1 with $M = 1.45 \pm 0.09~M_\odot$ and $R = 11.9 \pm 1.4$km, and GW170817-2 with $M = 1.27 \pm 0.09~M_\odot$ and the same radius estimate of $R = 11.9 \pm 1.4$~km\citep{LIGOScientific:2018cki}.
Combined analyses of \textbf{GW170817} and \textbf{GW190814} suggest that a canonical neutron star of $M = 1.4~M_\odot$ has a radius of $R = 12.9 \pm 0.8$km\citep{LIGOScientific:2020zkf}.

Moreover, we include data for what is potentially the \textbf{lightest known neutron star}, situated in the supernova remnant \textbf{HESS J1731--347}, with an estimated mass of $M = 0.77^{+0.20}{-0.17}~M\odot$ and a radius of $R = 10.4^{+0.86}{-0.78}$~km.
These measurements are derived from X-ray spectral analyses combined with distance estimates obtained via Gaia~\citep{2022NatAs...6.1444D}.
At the opposite end of the spectrum lies \textbf{PSR J0952--0607}, potentially the most massive neutron star observed to date, with a mass of $M = 2.35 \pm 0.17~M_\odot$\cite{Romani:2022jhd} and an estimated radius of $R = 14.087 \pm 1.0186$~km\cite{2023arXiv230514953E}.

Recent multi-messenger observations have yielded increasingly precise measurements of neutron star masses and radii, offering valuable information about the behavior of matter at supranuclear densities. At the same time, these data expose several theoretical tensions. One example is the contrast between \textbf{PSR~J0740+6620} and \textbf{PSR~J0030+0451}: although both stars exhibit similar radii in the range $R\approx12$--$13\,$km, their masses differ substantially. In order for PSR~J0740+6620 to sustain a mass near $2\,M_{\odot}$, its core must support a very large sound speed, with estimates suggesting $c_{s}^{2}\sim0.75\,c^{2}$ \cite{Legred:2021hdx}. On the other hand, tidal deformability bounds inferred from gravitational-wave events favour comparatively soft equations of state, implying lower sound speeds \cite{LIGOScientific:2018jsj,LIGOScientific:2020aai}.

Another difficulty appears when one considers isotropic stellar models. Their evolutionary sequences leave an almost empty interval between roughly $2.2\,M_\odot$ and $5\,M_\odot$, a range separating the heaviest neutron stars from the lightest black holes \cite{Yang:2020xyi}. While assuming $p_r = p_t$ can be convenient, it becomes inadequate in the inner regions of neutron stars, where a number of microphysical mechanisms---superfluidity, crystalline phases, strong magnetic fields, hyperons, deconfined quarks, and meson condensates---naturally produce anisotropy. A sufficiently large anisotropic stress acts as an additional repulsive contribution, allowing the star to approach the compactness limit $C = 2GM/(c^{2}R) \to 1$ within GR \cite{Alho:2022bki}. Such situations require the imposition of further physical constraints to keep the configuration within admissible bounds \cite{Alho:2021sli,Roupas:2020mvs,Raposo:2018rjn,Cardoso:2019rvt}. Similar considerations arise in modified gravity theories, particularly when matter couples non-minimally to curvature \cite{Nashed:2022zyi,ElHanafy:2022kjl,2023arXiv230514953E}.

Although GR has passed stringent tests in both weak and strong gravity regimes, it remains important to examine controlled departures from the theory. High-precision astrophysical observations now play a central role in restricting modified gravity models. A natural and widely studied extension of GR is obtained by replacing the Einstein-Hilbert term $\mathcal{R}$ with a general function $f(\mathcal{R})$. A large body of work has explored neutron star solutions within this context \cite{Kobayashi:2008tq,Upadhye:2009kt,Feng:2017hje,TeppaPannia:2016vsb,Wojnar:2016bzk,Arapoglu:2016ozr,Katsuragawa:2015lbl,
Fiziev:2015xpa,Hendi:2015pua,Momeni:2015vwa,Zubair:2016kov,Bakirova:2016ffk,AparicioResco:2016xcm,Moraes:2015uxq,
Sharif:2015jaa,Sotani:2017pfj,Capozziello:2011nr,Arapoglu:2010rz,Astashenok:2013vza,Astashenok:2014pua,
Astashenok:2014gda,Astashenok:2014nua}. The linear extension $f(\mathcal{R},\mathcal{T})$, introduced by Harko \emph{et al.}~\cite{Harko:2011kv}, has likewise been applied to isotropic \cite{Moraes:2015uxq,Das:2016mxq,Deb:2017rhd,Lobato:2020fxt} and anisotropic \cite{Deb:2018sgt,Maurya:2019sfm,Maurya:2019iup,Nashed:2023uvk,Chandrasekhar:1964zz} matter configurations. The addition of the Gauss-Bonnet term in $f(\mathcal{R},\mathcal{G})$ models introduces further freedom and can address some of the shortcomings found in $f(\mathcal{R})$ gravity when modelling compact stars \cite{Das:2023bff,Shamir:2017ndy}.

In this work we focus on the linear model
\[
f(\mathcal{R},\mathcal{G},\mathcal{T})
= \mathcal{R} + \alpha\,\mathcal{G} + \beta\,\mathcal{T},
\]
and constrain the parameters $\alpha$ and $\beta$ using the observationally inferred mass and radius of the pulsar \textit{U1724}. Within this framework, we get $v_{r}^{2}/c^{2} < 1/3$, unlike what is typically found in $f(\mathcal{R},\mathcal{T})$ or $f(\mathcal{R},\mathcal{G})$ models. A further objective is to analyse the stability of the resulting configuration. Instead of introducing an explicit equation of state---or multiple equations of state in anisotropic cases---we adopt the standard interior form for the metric potential $g_{rr}$ and suppress the contribution of $g_{tt}$ to the anisotropy, following a commonly used simplification in the literature.

The outline  is as follows.
In Section~\ref{Sec:fR_gravity}, we outline the ingredient tools of the extended gravity model considered here, namely $f(\mathcal{R},\mathcal{G},\mathcal{T})$, and summarizes the  equation of motions of spherically symmetric geometry with anisotropic fluid.

In Section~\ref{Sec:Model}, we specialize to the linear form
\[
f(\mathcal{R},\mathcal{G},\mathcal{T})=\mathcal{R}+\alpha\,\mathcal{G}+\beta\,\mathcal{T},
\]
and apply it to an anisotropic stellar configuration. The resulting system consists of three coupled non-linear differential equations for five unknown functions: the two metric potentials together with the density, radial pressure, and tangential pressure.

To render the system tractable without introducing an explicit equation of state, two simplifying conditions are imposed. One metric potential, $g_{rr}$, is taken in the standard form frequently used for interior solutions, while the dependence of the anisotropy on $g_{tt}$ is removed. With these assumptions, closed-form expressions for the density and pressures can be obtained, and they satisfy the linear $f(\mathcal{R},\mathcal{G},\mathcal{T})$ field equations.

Section~\ref{Sec:Stability} employs observational estimates of $\mathcal{R}$and $M$ of the pulsar \textit{U1724} to restrict the parameters $\alpha$ and $\beta$. The resulting stellar model is then examined against a range of stability criteria, addressing both geometric and material conditions.

In Section~\ref{Sec:EoS_MR}, we extract the EoS implied by the solutions and discuss the corresponding mass--radius relation, commenting on the physical implications.

A summary of the main results and their significance is presented in Section~\ref{Sec:Conclusion}.
\section{$f(\mathcal{R,G,T})$  gravity}\label{Sec:fR_gravity}

To formulate the theory of gravity $f(\mathcal{R,G,T})$ and derive its field equations, we consider the subsequent action functional \cite{Shamir:2017rjz}:
\begin{equation}\label{1}
\mathcal{S}_{f(\mathcal{R,G,T})}=\frac{1}{2\kappa^{2}}\int d^{4}x\sqrt{-g}[f(\mathcal{R,G,T})+\mathcal{L}_{m}]\,.
\end{equation}
In the formulation adopted here, $\kappa$ plays the role of the gravitational interaction parameter, and $g$ denotes the determinant associated with the metric tensor $g_{\alpha\beta}$.
The energy--momentum tensor, indicated by $\mathcal{T}_{\gamma\delta}$, is obtained from the matter sector through
\[
\mathcal{T}_{\gamma\delta}
   = -\,\frac{2}{\sqrt{-g}}
      \frac{\delta\!\left( \sqrt{-g}\,\mathcal{L}_{m} \right)}
           {\delta g^{\gamma\delta}},
\]
with $\mathcal{L}_{m}$ representing the matter Lagrangian density that characterizes the non-gravitational fields in the model \cite{landau2013classical} 
which further equals to
\begin{equation}\label{3}
\mathcal{T}_{\gamma\delta}=g_{\gamma\delta
}\mathcal{L}_{m}-2\frac{\partial\mathcal{L}_{m}}{\partial
g^{\gamma\delta }}.
\end{equation}
By performing the variation of the action introduced in Eq.~(\ref{1}), we obtain 
\begin{eqnarray}\label{4}\nonumber
\delta {\cal S} &= \frac{{{1}}}{2\kappa ^{2}}\int {{d^4}} x[(f({\cal R},{\cal G},{\cal T})\delta \sqrt { - g}  + \sqrt { - g} ({f_{\cal R}}({\cal R},{\cal G},{\cal T})\delta {\cal R} + {f_{\cal G}}({\cal R},{\cal G},{\cal T})\delta {\cal G} + {f_{\cal T}}({\cal R},{\cal G},{\cal T})\delta {\cal T}) + \sqrt { - g} \delta {{\cal L}_m}] = 0,
\end{eqnarray}
with  \[f_{\mathcal{R}}(\mathcal{R,G,T})=\frac{\partial
f(\mathcal{R,G,T})}{\partial\mathcal{R}}, \quad 
f_{\mathcal{G}}(\mathcal{R,G,T})=\frac{\partial
f(\mathcal{R,G,T})}{\partial\mathcal{G}}, \quad \mbox{and} \quad 
f_{\mathcal{T}}(\mathcal{R,G,T})=\frac{\partial f(\mathcal{R,G,T})}{\partial \mathcal{T}}.\] 
Upon varying the quantities $\sqrt{-g}$, $\mathcal{R}^{\xi}{}_{\alpha\beta\eta}$, $\mathcal{R}_{\alpha\eta}$, and $\mathcal{R}$, the resulting identities can be written as follows:
\begin{eqnarray}\nonumber
\delta\sqrt{-g}&=&-\frac{1}{2}\sqrt{-g}g_{\rho\sigma}\delta
g^{\rho\sigma },\\\nonumber\delta
\mathcal{R}^{\xi}_{\rho\sigma\eta}&=&\nabla_{\sigma}(\delta\Gamma^{\xi}_{\eta\rho})
-\nabla_{\eta}(\delta\Gamma^{\xi}_{\sigma\rho})\equiv (g_{\rho\lambda}
\nabla_{[\eta}\nabla_{\sigma
]}+g_{\lambda[\sigma}\nabla_{\eta]}\nabla_{\rho}) \delta
g^{\xi\lambda}+\nabla_{[\eta}\nabla^{\xi}\delta
g_{\sigma]\rho},\\\label{4a}\delta R_{\rho\eta}&=&\delta
\mathcal{R}^{\xi}_{\rho\xi\eta}\,, \qquad \qquad\delta
\mathcal{R}=(\mathcal{R}_{\rho\sigma}+g_{\rho\sigma}\nabla^2-\nabla_{\rho}\nabla_{\sigma})
\delta g^{\rho\sigma}.
\end{eqnarray}
The operator $\nabla_{\rho}$ is understood to be the covariant derivative, and $\Gamma^{\xi}_{\rho\sigma}$ refers to the associated Christoffel symbols.
When the scalar quantities $\mathcal{R}$, $\mathcal{G}$, and $\mathcal{T}$ are varied, one obtains
\begin{eqnarray}\nonumber
\delta \mathcal{R} &=& {\mathcal{R}_{\rho \sigma }}\delta {g^{\rho \sigma }} - {\nabla _\rho }{\nabla _\sigma }\delta {g^{\rho \sigma }} - {g_{\rho \sigma }}\Box\delta {g^{\rho \sigma }}\,,\\\nonumber
\delta\mathcal{G}&=&2\mathcal{R}\delta
\mathcal{R}-4\delta(R_{\rho\sigma}\mathcal{R}^{\rho\sigma})+\delta(\mathcal{R}_{\rho
\sigma\xi\eta}R^{\rho\sigma\xi\eta}),
\\\label{4b}\delta \mathcal{T}&=&(\mathcal{T}_{\rho\sigma}+\Theta_{\rho\sigma})\delta
g^{\rho\sigma},\qquad \qquad\Theta_{\rho\sigma}=g^{\xi\eta}\frac{\delta
\mathcal{T}_{\xi\eta}}{\delta g_{\rho\sigma}}.
\end{eqnarray}

Employing Eq.~(\ref{4}) together with the previously derived variation terms enables the construction of the field equations associated with the $f(\mathcal{R},\mathcal{G},\mathcal{T})$ framework. They can be written in the form \cite{Shamir:2017rjz}:
\begin{align}\label{5}
&{\mathcal{G}_{\rho \sigma }} = \frac{1}{{{f_{\cal R}}\left( {{\cal R},{\cal G},{\cal T}} \right)}}\left[ {{\kappa ^2}{{\cal T}_{\rho \sigma }}} \right. - \left( {{{\cal T}_{\rho \sigma }} + {\Theta _{\rho \sigma }}} \right){f_T}\left( {{\cal R},{\cal G},{\cal T}} \right)
 + \frac{1}{2}{g_{\rho \sigma }}(f\left( {{\cal R},{\cal G},{\cal T}} \right) + \mathcal{R}{f_{\cal R}}\left( {{\cal R},{\cal G},{\cal T}} \right))+ {\nabla _\rho }{\nabla _\sigma }{f_{\cal R}}\left( {{\cal R},{\cal G},{\cal T}} \right)\nonumber\\
& - {g_{\rho \sigma }}\Box{f_{\cal R}}\left( {{\cal R},{\cal G},{\cal T}} \right) - (2\mathcal{R}{\mathcal{R}_{\rho \sigma }} - 4\mathcal{R}_\rho ^\xi {\mathcal{R}_{\xi \sigma }} - 4{\mathcal{R}_{\rho \xi \sigma \eta }}{\mathcal{R}^{\xi \eta }} + 2\mathcal{R}_\rho ^{\xi \eta \delta }{R_{\sigma \xi \eta \delta }}){f_{\cal G}}\left( {{\cal R},{\cal G},{\cal T}} \right) - (2\mathcal{R}{g_{\rho \sigma }}{\nabla ^2} - 2\mathcal{R}{\nabla _\rho }{\nabla _\sigma } \\
& - 4{g_{\rho \sigma }}{\mathcal{R}^{\xi \eta }}{\nabla _\xi }{\nabla _\eta }- 4{\mathcal{R}_{\rho \sigma }}{\nabla ^2} + 4\mathcal{R}_\rho ^\xi {\nabla _\sigma }{\nabla _\xi } + 4\mathcal{R}_\sigma ^\xi {\nabla _\rho }{\nabla _\xi } + 4{\mathcal{R}_{\rho \xi \sigma \eta }}{\nabla ^\xi }{\nabla ^\eta })\left. {{f_{\cal G}}\left( {{\cal R},{\cal G},{\cal T}} \right)} \right]\,,\nonumber
\end{align}
The quantity $\mathcal{G}_{\rho\sigma}$ is constructed from the Ricci tensor and scalar via
\[
\mathcal{G}_{\rho\sigma} = \mathcal{R}_{\rho\sigma} - \frac{1}{2} g_{\rho\sigma}\,\mathcal{R},
\]
which is the standard definition of the Einstein tensor. \\[2mm]
It should also be observed that, by choosing particular functional forms of $f(\mathcal{R},\mathcal{G},\mathcal{T})$, the theory naturally collapses into several familiar modified-gravity frameworks:
$$f(\mathcal{R,G,T}) \equiv \left. {\left| {\begin{array}{*{40}{cc}}
GR & \cite{Einstein:1916vd}\\
{f({\mathcal R})} & \cite{Starobinsky:2007hu,Sotiriou:2008rp,Nojiri:2010wj,DeFelice:2010aj,Nashed:2019tuk,Nashed:2022xmv} \\
{f({\mathcal R},{\cal T})} & \cite{Harko:2011kv,Nashed:2023uvk}\\
{f({\mathcal G})} & \cite{DeFelice:2009aj,Nashed:2022mij,Nashed:2021cfs}\\
{f({\mathcal R},{\mathcal G})} & \cite{Bamba:2010wfw}\\
{f({\mathcal G},{\mathcal T})} & \cite{Shamir:2017rjz}
\end{array}} \right.} \right\}\,.$$
The dynamical relations associated with each of the previously mentioned subclasses follow as specific realizations of Eq.~(\ref{5}).
When the function is limited to $f(\mathcal{R},\mathcal{G},\mathcal{T}) = \mathcal{R}$, the formulation collapses to the usual general-relativistic case.
Under this restriction, the trace of Eq.~(\ref{5}) assumes the form:
\begin{align}\nonumber
&{\kappa ^2}{\cal T} - ({\cal T} + \Theta ){f_{\cal T}}({\cal R},{\cal G},{\cal T})  - {f_{\cal R}}({\cal R},{\cal G},{\cal T})R - 3\Box {f_{\cal R}}({\cal R},{\cal G},{\cal T})+ 2{\cal G}{f_{\cal G}}({\cal R},{\cal G},{\cal T})\\
&- 2\mathcal{R}{\nabla ^2}{f_{\cal G}}({\cal R},{\cal G},{\cal T}) + 4{R^{\rho \sigma }}{\nabla _\rho }{\nabla _\sigma }{f_{\cal G}}({\cal R},{\cal G},{\cal T})+ 2f({\cal R},{\cal G},{\cal T}) = 0.
\end{align}

It is also noteworthy that the covariant divergence of Eq.~(\ref{5}) does not vanish.
Carrying out this operation leads to the following expression:
\begin{align}\label{5a}\nonumber
{\nabla ^\rho }{{\cal T}_{\rho \sigma }} &= \frac{{{f_{\cal T}}({\cal R},{\cal G},{\cal T})}}{{{\kappa ^2} - {f_{\cal T}}({\cal R},{\cal G},{\cal T})}}\left[ {\left\{ {{{\cal T}_{\rho \sigma }} + {\Theta _{\rho \sigma }}} \right\}{\nabla ^\rho }\left\{ {\ln {f_{\cal T}}({\cal R},{\cal G},{\cal T})} \right\}}- {\frac{1}{2}{g_{\rho \sigma }}{\nabla ^\rho }{\cal T} + {\nabla ^\rho }{\Theta _{\rho \sigma }}} \right].
\end{align}
In obtaining the above expression, the relations given below have been employed:
\begin{eqnarray}
&& \left( \nabla_\sigma \Box - \Box \nabla_\sigma \right) \psi = g^{\alpha\beta}
\left( \nabla_\sigma \nabla_\alpha \nabla_\beta - \nabla_\alpha \nabla_\beta
\nabla_\sigma \right) \psi  \notag \\
&=& g^{\alpha\beta} \left( \nabla_\sigma \nabla_\alpha - \nabla_\alpha
\nabla_\sigma \right) \nabla_\beta \psi = g^{\alpha \beta} \mathcal{R}_{\thickspace \beta
\alpha \sigma}^\rho \nabla_\rho \psi  =- \mathcal{R}_{\rho \sigma} \nabla^\rho \psi,
\end{eqnarray}
and
\begin{equation}
\nabla_\nu f \left( \mathcal{R,G,T} \right) = f_{\mathcal{R}} \nabla_\nu \mathcal{R} +f_{\mathcal{G}} \nabla_\nu \mathcal{G} + f_{\mathcal{T}} \nabla_\nu \mathcal{T},
\label{eq:nablaf}
\end{equation}
respectively.

To express $\Theta_{\sigma\rho}$ in a more convenient form, we apply a differentiation to Eq.~(\ref{3}) in the manner indicated below:
\begin{equation}\label{6}
\frac{\delta \mathcal{T}_{\sigma\rho}}{\delta g^{\xi\eta}}=\frac{\delta
g_{\sigma\rho}}{\delta g^{\xi\eta}}\mathcal{L}_{m}+g_{\sigma\rho}
\frac{\partial\mathcal{L}_{m}}{\partial
g^{\xi\eta}}-2\frac{\partial^2\mathcal{L}_{m}}{\partial
g^{\xi\eta}\partial g^{\sigma\rho}}.
\end{equation}
Using the relations
\begin{equation}\nonumber
\frac{\delta g_{\sigma\rho}}{\delta
g^{\xi\eta}}=-g_{\sigma\mu}g_{\rho\nu}\delta_{\xi\eta}^{\mu\nu},\qquad \mbox {and} \qquad
\delta_{\xi\eta}^{\mu\nu}=\frac{\delta g^{\mu\nu}}{\delta
g^{\xi\eta}}.
\end{equation}
The symbol $\delta_{\xi\eta}^{\mu\nu}$ corresponds to the generalized Kronecker delta.
By replacing Eq.~(\ref{4b}) with the expression given in Eq.~(\ref{6}), we arrive at
\begin{equation}\label{7}
\Theta_{\sigma\rho}=-2\mathcal{T}_{\sigma\rho}+g_{\sigma\rho}\mathcal{L}_{m}-2g^{\xi\eta}
\frac{\partial^{2}\mathcal{L}_{m}}{\partial g^{\sigma\rho}\partial
g^{\xi\eta}}.
\end{equation}
This facilitates the determination of the tensor $\Theta_{\sigma\rho}$.
Meanwhile, the energy--momentum tensor $\mathcal{T}_{\mu\nu}$ from
Eq.~\eqref{5}, which corresponds to the matter contribution, can be
modeled as an anisotropic fluid, that is,
\begin{equation}\label{Tmn-anisotropy}
   \mathcal{T}^\mu{}_\nu=  {(p_t+\rho c^2)v{^\mu} v{_\nu}+p_t \delta ^\mu _\nu + (p_{r}-p_t) w{^\mu} w{_\nu}}\,.
\end{equation}
Accordingly, the stress--energy tensor takes the diagonal form:  \[{ \mathcal{T}{^\mu}{_\nu}}={diag(-\rho c^2\, ,p_{r}\, ,p_t\, ,p_t)}.\]
Assuming   $\mathcal{L}_{m}$ in the case of anisotropic to have the form:
 \[\mathcal{L}_{m}=-\frac{1}3(p_r+2p_t)\,,\] which in the case of isotropic reproduce the one presented in \cite{Harko:2011kv}. Now using
Eq.(\ref{7}) we get:
\begin{equation}\label{9}
\Theta_{\sigma\lambda}=-2\mathcal{T}_{\sigma\lambda}-\frac{1}3(p_r+2p_t)g_{\sigma\lambda}\,.
\end{equation}

\section{The matter-coupled anisotropic Gauss-Bonnet model}\label{Sec:Model}
In this section, we are going to derive anisotropic model in the linear frame of $f \left( \mathcal{R,G,T} \right)$ theory.
   In what follows, our analysis will be based on choosing a particular functional form of the model, given by, \begin{align}\label{RGT} \left( \mathcal{R,G,T} \right) =\mathcal{R}+ \alpha \mathcal{G} + \beta \mathcal{T}\,,\end{align}  which includes Ricci scalar,  Gauss-Bonnet scalar as well as the trace of the matter where $\alpha$ and $\beta$ are  dimensionful parameters with unit of [${\textit L}^2$, $\frac{1}{N}$, where N is the unit of Newton], respectively.
   In this study we are going to impose the following line-element:
   \begin{align}\label{RG}
   {    ds^2 = g_{\mu \nu} dx^\mu dx^\nu=-e^{\mu}c^2dt^2 + e^{\nu}dr^2 + r^2(d\theta^2 + \sin^2\theta d\phi^2)}\,,
   \end{align}
   The metric functions, denoted as $\mu$ and $\nu$, are two functions depend on the radial coordinate $r$ only. Consequently, we determine $\sqrt{-g}=r^2 \sin \theta\,e^{(\mu + \nu)/2} $, and the four-velocity vector $v^\mu=(ce^{-\mu/2}, 0, 0, 0)$.

   By employing Eq. (\ref{RG}) to calculate, the $R$ and GB scalars we get:
\begin{align}\label{RaG}
&R=\frac {\left(\nu'{r}^{2} -4\,r \right) \mu'-2\,\mu''{r}^{2}- \mu'^{2}{r}^{2}  +4\, \nu' r+4\,{e^{\nu  }}-4}{2{r}^{2}{e^{\nu  }}} \,,\nonumber\\
&G=\frac {{e^{-2\,\nu  }} \left\{  \left[  \nu'\left(e^{\nu}-3 \right) - \left( {e^{\nu  } }-1 \right) \mu' \right] \mu'  -\left( 2\,{ e^{\nu  }}-2 \right)  \mu'' \right\} }{2{r}^{2}}\,,
\end{align}
where $'\equiv d/dr$ and $''\equiv d^2/dr^2$ etc. Using the value of $\mathcal{R}$ and GB given by (\ref{RG}) in  (\ref{5}) we get:
\begin{align}
    &{  \rho} =\frac{5{e^{-2\nu  }}}{6c^2 \left( {\kappa}^{2}+2 \beta \right)\left( 8\beta+{\kappa}^{2} \right){r}^{2}} \bigg\{  \left[  \left\{  \left( -{\frac {24\alpha}{5}}+2{r}^{ 2} \right) \beta-{\frac {12{\kappa}^{2}\alpha }{5}}\right\} e^ {\nu}+{\frac {24\alpha }{5}}\left( \frac{{\kappa }^{2}}2+\beta \right)  \right] \mu''  + \left( {\frac {32\beta}{5}}+\frac{6{\kappa}^{2}}5 \right) { e^{2\nu }}+ \bigg\{  \bigg[\bigg({r}^{2}  \nonumber\\
    & -{\frac { 12\alpha}{5}} \bigg) \beta-\frac{6{\kappa}^{2}\alpha}5 \bigg] {e^{\nu  }}+{\frac {12\alpha }{5}}\left( \frac{{\kappa}^{2}}2 +\beta \right)  \bigg\}  \mu'^{2}+ \left[  \left\{  \left[  \left( { \frac {12\alpha}{5}}-{r}^{2} \right) \beta+\frac{6{\kappa}^{2}\alpha }5\right]e^{\nu}-{\frac {36\alpha }{5}}\left( \frac{{\kappa}^{2}}2+\beta \right)  \right\} \nu'+4re^{\nu}\beta \right] \mu' \nonumber\\
    &   +{\frac {32}{5}} \left( \beta+3/ 16{\kappa}^{2} \right) {e^{\nu  }} \left( \nu' r-1 \right) \bigg\}\,,  \label{9}  \\
&{ p_r}=\frac{-5{e^{-2\nu  }}}{\left( 96{r}^{2}{\beta}^{2}+60{\kappa}^{2}{r}^{ 2}\beta+6{\kappa}^{4}{r}^{2} \right)} \bigg\{  \left[  \left( (2{r}^{2} -{\frac {24}{5}}\alpha ) \beta-{\frac {12}{ 5}}{\kappa}^{2}\alpha \right) {e^{\nu  }}+{ \frac {24}{5}}\alpha \left( \frac{{\kappa}^{2}}2+\beta \right) \right] \mu'+ \left( { \frac {32}{5}}\beta+\frac{6{\kappa}^{2}}5 \right) {e^{2\nu  }}\nonumber\\
    &   + \left\{  \left[  \left(  {r}^{2}-{\frac {12}{5}}\alpha \right) \beta-\frac{6{\kappa}^{2}\alpha}5 \right] {e^{\nu  }}+{\frac {12}{5}}\alpha \left( \frac{{\kappa}^{2}}2+\beta \right)  \right\} \mu'^{2}+ \bigg\{  \left[  \left\{  \left( {\frac {12}{5}}\alpha- {r}^{2} \right) \beta+\frac{6{\kappa}^{2}\alpha}5 \right\} {e^{\nu  }}-{\frac {36}{5}}\alpha \left( 1/2{\kappa}^{2 }+\beta \right)  \right] \nu'\nonumber\\
    &  -{\frac { 28}{5}}r{e^{\nu  }} \left( \beta+3/14{\kappa }^{2} \right)  \bigg\} \mu'  -{\frac {16 }{5}}{e^{\nu  }} \left( \beta \nu' r+3/8{\kappa}^{2}+2\beta \right)  \bigg\}\, ,   \label{10}  \\
&{ p_t} =\frac{7{e^{-2\nu  }}}{6\left( 16{\beta}^{2}+ {\kappa}^{4}+10{\kappa}^{2}\beta \right) {r}^{2}} \bigg\{  \left[  \left(  ( {\frac {24}{7}}\alpha+2{r}^{2 }) \beta+\frac{3{\kappa}^{2}}7 ( 4\alpha+{r}^{2} ) \right) {e^{\nu }}-{\frac {24\alpha}{7}} \left( \frac{{\kappa}^{2}}{2}+\beta \right)  \right] \mu''  +{\frac {16\beta}{7}}{e^{2\nu  }}+ \bigg\{  \bigg[  \left( {\frac {12\alpha}{7}}+{r}^{ 2} \right) \nonumber\\
    & \beta+\frac{3{\kappa}^{2}}{14} \left( 4\alpha+{r}^{2} \right) \bigg] {e^{\nu  }}-{\frac {12}{7}}\alpha \left( \frac{{\kappa}^{2}}2+\beta \right)  \bigg\}  \mu'^{2}+ \bigg\{  \left[ {\frac {36}{7}}\alpha \left(\frac{{\kappa}^{2}}2+\beta \right) - \left\{  \left(  {r}^{2}+{\frac {12}{7}}\alpha \right) \beta+\frac{3{\kappa}^{2}}{14} \left( 4\alpha+{r}^{2} \right)  \right\} {e^{\nu  }} \right]\nu'\nonumber\\
    &   +\frac{4r}7 \left( \beta+3/4{\kappa}^{2} \right) {e^{\nu}} \bigg\} \mu'  -{\frac {8}{7}}{e ^{\nu  }} \left( r \left( \beta+\frac{3{\kappa}^{2}}8 \right)\nu'  +2\beta \right)  \bigg\}\,.  \label{11}
\end{align}
{Clearly, the matter density and pressures are altered and recover their general relativistic forms in the suitable limit~\cite{Nashed:2020kjh,Roupas:2020mvs}
when the coupling parameters are set to $\alpha=\beta=0$.
It is evident that the above system contains five unknowns functions, namely
$\rho$, $p_{r}$, $p_{t}$, $\mu$, and $\nu$; therefore, two  extra
conditions are assumed to close the system.
These assumptions  may be introduced by specifying equations of state (EoS)
relating   $p_{r}=p_{r}(\rho)$
and $p_{t}=p_{t}(\rho)$.
Alternatively, one may prescribe physically motivated forms for  $\mu(r)$ and $\nu(r)$.
In the present analysis, we adopt the latter approach by assuming a specific
functional form for the radial metric potential $\nu(r)$, given by~\cite{Nashed:2022zyi}:}
\begin{align}\label{nu}
\nu(r)=-ln\left(1-\frac{b_0{}^2 r^2}{R^2}\right)^4\,,\end{align}
where $b_0$ is a dimensionless constant and $R$ is the  star's radius.
Using Eq. (\ref{nu}) in Eq. (\ref{11}) we calculate the anisotropy $\Delta=p_t-p_r$ and get:
\begin{align}\label{ani}
&\Delta=\frac {6{R}^{4}{b_0}^{4}{r}^{3}-8 {R}^{2}{b_0}^{6}{r}^{5}+3{b_0}^{8}{r}^{7}}{{R}^{8}r \left( {\kappa}^{2}+2\beta \right)}+\frac { \left( r{R}^{8}-4{r}^{7}{R}^{2}{b_0}^{6}+6{r}^{ 5}{R}^{4}{b_0}^{4}-4{r}^{3}{R}^{6}{b_0}^{2}+{r}^{9}{b_0}^{8} \right)  \mu'^{ 2}}{4{R}^{8}r \left( {\kappa}^{2}+2\beta \right) }\nonumber\\
&+\frac{ \left(6{r}^{8}{b_0}^{8} -2{R}^{8}+12{R}^{4}{r}^{4}{b_0}^{4}-16{R}^{2}{b_0}^{6}{r}^{6}\right) \mu' }{4{R}^{8}r \left( {\kappa}^{2}+2\beta \right) }+\frac{2 \mu'' r{R}^{8}-8 \mu'' {r}^{7}{R}^{2}{b_0}^{6}+12 \mu''  {r}^{5}{R}^{4}{b_0}^{4}-8 \mu''  {r}^{3}{R }^{6}{b_0}^{2}+2\mu'' {r}^{9}{b_0}^{8}}{4{R}^{8}r \left( {\kappa}^{2}+ 2\beta \right)}\,.
\end{align}
If we assume the vanishing of the expressions that contains the component of $g_{tt}$, i.e., if we assume the anisotropy to have the following value:
\begin{align}\label{ani1}
\Delta=\frac {6{R}^{4}{b_0}^{4}{r}^{3}-8 {R}^{2}{b_0}^{6}{r}^{5}+3{b_0}^{8}{r}^{7}}{{R}^{8}r \left( {\kappa}^{2}+2\beta \right)}\,,
\end{align}
then we get
\begin{align}\label{diff}
&0=\frac { \left( r{R}^{8}-4{r}^{7}{R}^{2}{b_0}^{6}+6{r}^{ 5}{R}^{4}{b_0}^{4}-4{r}^{3}{R}^{6}{b_0}^{2}+{r}^{9}{b_0}^{8} \right)  \mu'^{ 2}}{4{R}^{8}r \left( {\kappa}^{2}+2\beta \right) }+\frac{ \left(6{r}^{8}{b_0}^{8} -2{R}^{8}+12{R}^{4}{r}^{4}{b_0}^{4}-16{R}^{2}{b_0}^{6}{r}^{6}\right) \mu' }{4{R}^{8}r \left( {\kappa}^{2}+2\beta \right) }\nonumber\\
&+\frac{2 \mu'' r{R}^{8}-8 \mu'' {r}^{7}{R}^{2}{b_0}^{6}+12 \mu''  {r}^{5}{R}^{4}{b_0}^{4}-8 \mu''  {r}^{3}{R }^{6}{b_0}^{2}+2\mu'' {r}^{9}{b_0}^{8}}{4{R}^{8}r \left( {\kappa}^{2}+ 2\beta \right)}\,.
\end{align}
The solution of the above differential equation, i.e. Eq. \eqref{diff},  has the following form:
\begin{align}\label{mu}
\mu=\ln  \left( \frac{b_1+2b_2{b_0}^{2}{R}^{2
}-2b_2{b_0}^{4}{r}^{2}}{4{b_0}^{2} \left( {R}^{2}-{b_0}^{2}{r}^{2} \right)} \right)^2\,,
\end{align}
where $b_1$ is a dimensional constant that has a unit of ${\textit length ^2}$, and $b_2$ is a dimensionless constant.
Using Eq.~(\ref{nu}) and Eq.~(\ref{mu}) in Eqs.~(\ref{9}), ~(\ref{10}), and (\ref{11}) we get the form of $\rho$, $p_r$ and $p_t$ which we listed in Appendix { A}\footnote{{ We stress on the fact that the metric anstaz $\mu$ and $\nu$ are physically acceptable because they are non-singular as $r\to 0$ which is a necessary condition for any stellar model.}}. In the next subsections we are going to discuss the viability the present model.

\subsection{Matching conditions}\label{Sec:Match}
Since the vacuum solutions in general relativity and in the
$f(\mathcal{R},\mathcal{G},\mathcal{T}) = \mathcal{R} + \alpha\,\mathcal{G} + \beta\,\mathcal{T}$
theory are equivalent~\citep{Ganguly:2013taa},
the exterior spacetime coincides with the vacuum solution.
Accordingly, we consider the exterior metric to  take  the  form:
\begin{equation}
 {   ds^2=-\left(1-\frac{2GM}{c^2r}\right) c^2 dt^2+\frac{dr^2}{\left(1-\frac{2GM}{c^2 r}\right)}+r^2 (d\theta^2+\sin^2 \theta d\phi^2)}\,.
\end{equation}
Using the interior spacetime described by Eqs.\eqref{nu} and\eqref{mu},
the junction conditions are imposed at the surface between the interior and exterior regions, ensuring the continuity of the metric potentials and their first derivatives across the stellar surface,
\begin{equation}\label{eq:bo}
 \mu(r={R_s})=\ln(1-C), \quad  \nu(r={R_s})=-\ln(1-C),\quad  \text{and} \quad {{p}_r}(r={R_s})=0.
\end{equation}
Here $C$ represents the compactness defined as,
\begin{align}\label{comp11}
 {  C=\frac{2GM}{c^2 {R_s}}}\,.
\end{align}
By substituting the metric potentials from Eqs.\eqref{nu} and\eqref{mu}, together with the expression for the radial pressure given in Eq.~\eqref{sol},
and applying the previously stated boundary conditions,
the constants $\{b_{0}, b_{1}, b_{2}\}$ can be determined as functions of the parameters $\alpha_{1}$, $\beta_{1}$, and the compactness factor of the stellar configuration.
Consequently, the parameter space of the linear model
$f(\mathcal{R},\mathcal{G},\mathcal{T}) = \mathcal{R} + \alpha\,\mathcal{G} + \beta\,\mathcal{T}$
within the present framework is essentially characterized by the set
$\{\alpha_{1},\,\beta_{1},\,C\}$.
Since both $M$ and $R$ of compact stars are limited by astrophysical measurements,
and thereby $C$ is known,
the remaining task is to identify the corresponding bounds on the theoretical parameters
$\alpha_{1}$ and $\beta_{1}$.
\section{Observational Constraints on Stability and Astrophysical Parameters of the Pulsar J0740+6620}\label{Sec:Stability}

In this section, we use the observational measurements of  $R$ and $M$ of the pulsar \textit{U1724}
to limit  the parameters $\alpha_{1}$ and $\beta_{1}$ associated with the linear correction terms in the modified
gravity model $f(\mathcal{R},\mathcal{G},\mathcal{T}) = \mathcal{R} + \alpha\,\mathcal{G} + \beta\,\mathcal{T}$.
Furthermore, we analyze the stability of the resulting stellar configuration by applying several physical viability conditions.
As emphasized in the introduction, precise astrophysical data are crucial for narrowing down the admissible parameter space of such theoretical frameworks.
We therefore select the pulsar \textit{U1724} as a representative object for testing the viability of the present model.
Its measured physical properties are reported as a mass of $M = (1.81 \pm 0.27)\,M_{\odot}$
and a radius of $R = (12.2 \pm 1.4)\,\mathrm{km}$~\citep{Roupas:2020mvs}.
It is worth noting that this mass value lies close to the upper limit expected for neutron stars,
where deviations from general relativity and the influence of modified gravity effects may become significant.

\subsection{Matter sector}\label{Sec:matt}
In this part of the study, we utilize the observational data on the pulsar \textit{U1724},
specifically its measured mass and radius, to place bounds on the model parameters
$\alpha_{1}$ and $\beta_{1}$ that quantify the linear modifications in the extended
gravitational theory $f(\mathcal{R},\mathcal{G},\mathcal{T}) = \mathcal{R} + \alpha\,\mathcal{G} + \beta\,\mathcal{T}$.
We also investigate the stability characteristics of the obtained stellar configuration
by examining a set of well-established physical and dynamical criteria.
As discussed earlier, high-precision astrophysical observations are essential for constraining
the viable region of the parameter space in any modified gravity scenario.
Accordingly, the pulsar \textit{U1724} has been chosen as a suitable test candidate for our analysis.
Observational studies indicate that its mass and radius are
$M = (1.81 \pm 0.27)\,M_{\odot}$ and $R = (12.2 \pm 1.4)\,\mathrm{km}$, respectively~\citep{Roupas:2020mvs}.
It is remarkable that this source possesses a mass close to the upper bound typically associated
with neutron stars, where departures from general relativity and the influence of extended gravity terms
are expected to become increasingly relevant.
\begin{figure*}
\centering
\subfigure[~$\rho$]{\label{fig:density}\includegraphics[scale=0.3]{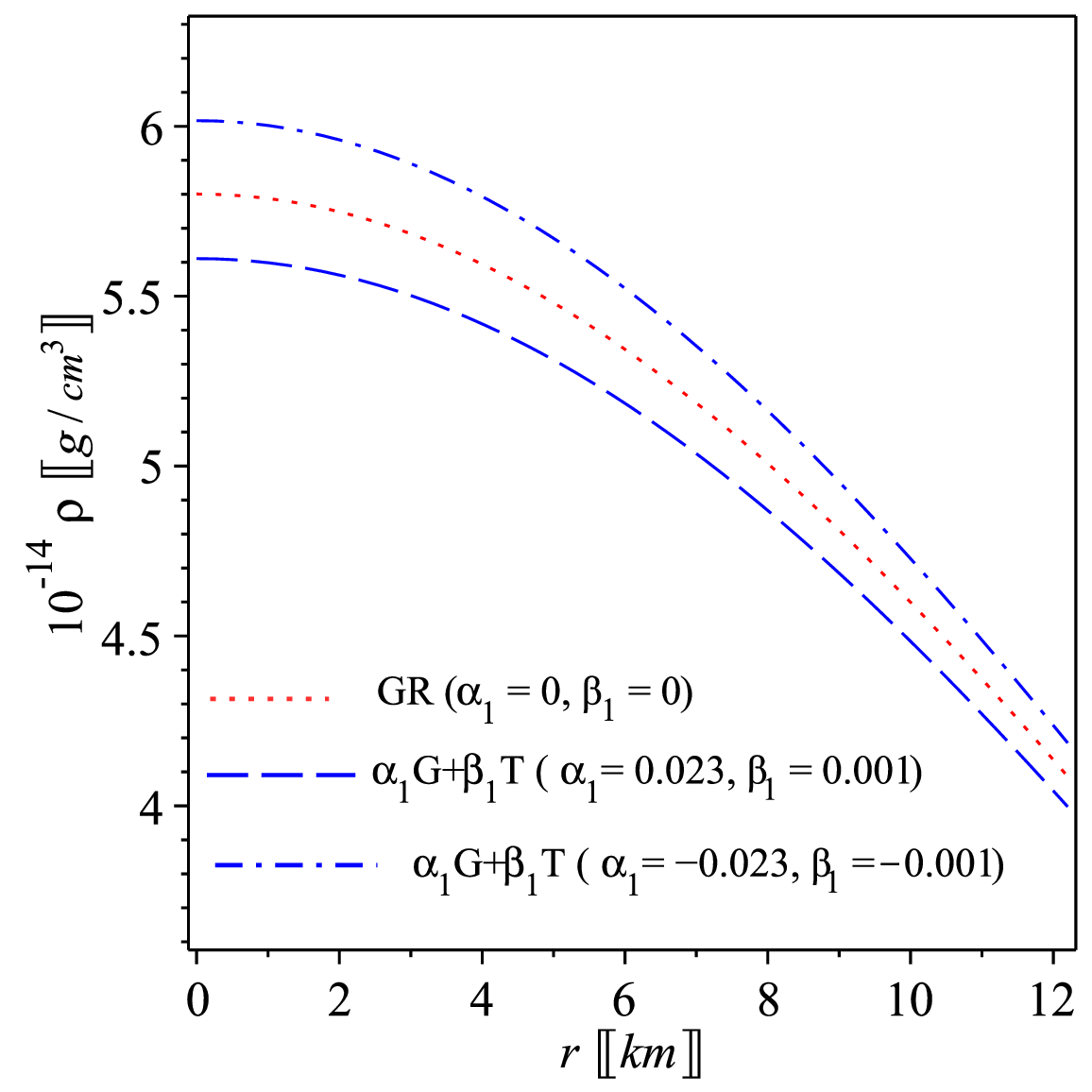}}\hspace{0.5cm}
\subfigure[~$p_r$]{\label{fig:radpressure}\includegraphics[scale=0.3]{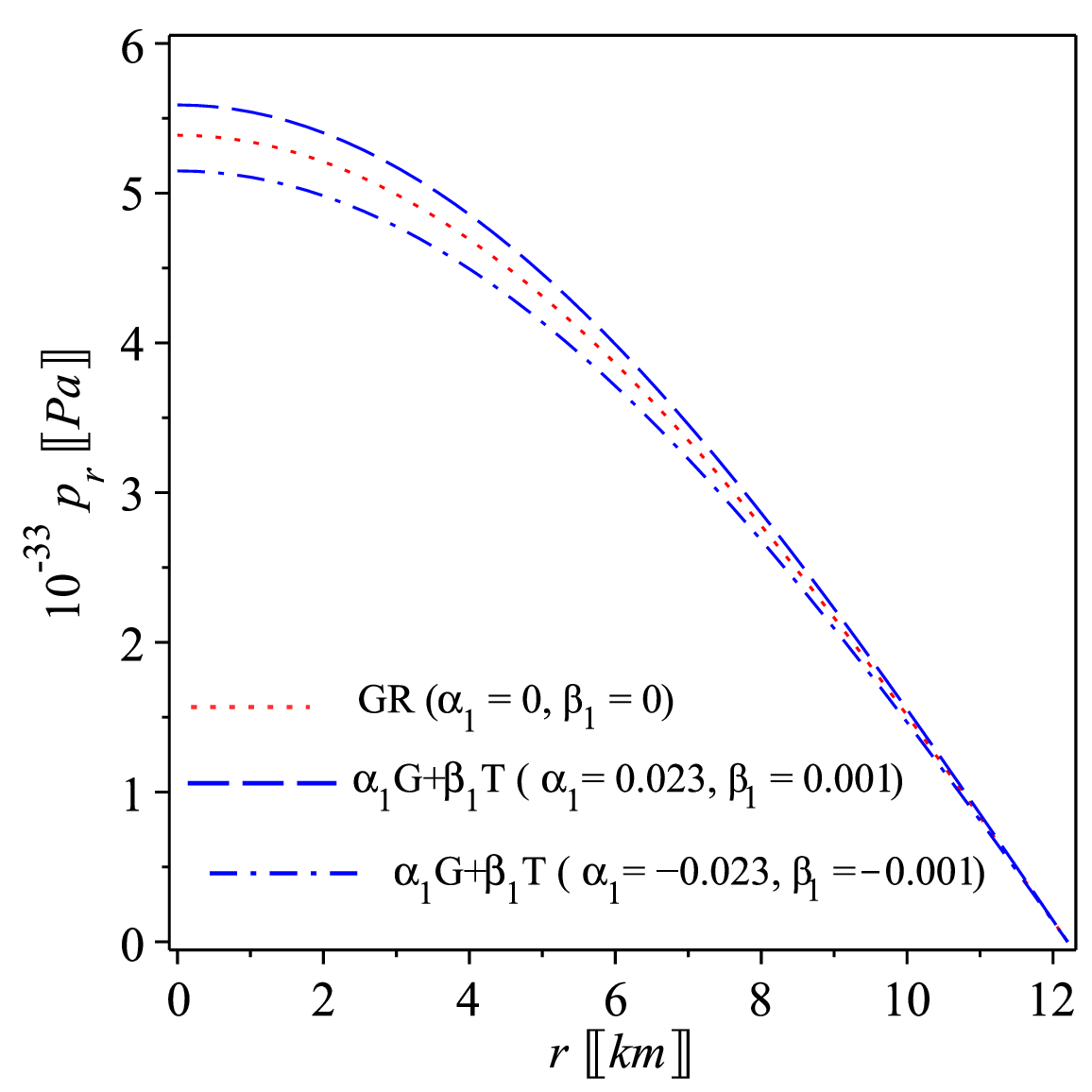}}\\
\subfigure[~$p_t$]{\label{fig:tangpressure}\includegraphics[scale=0.3]{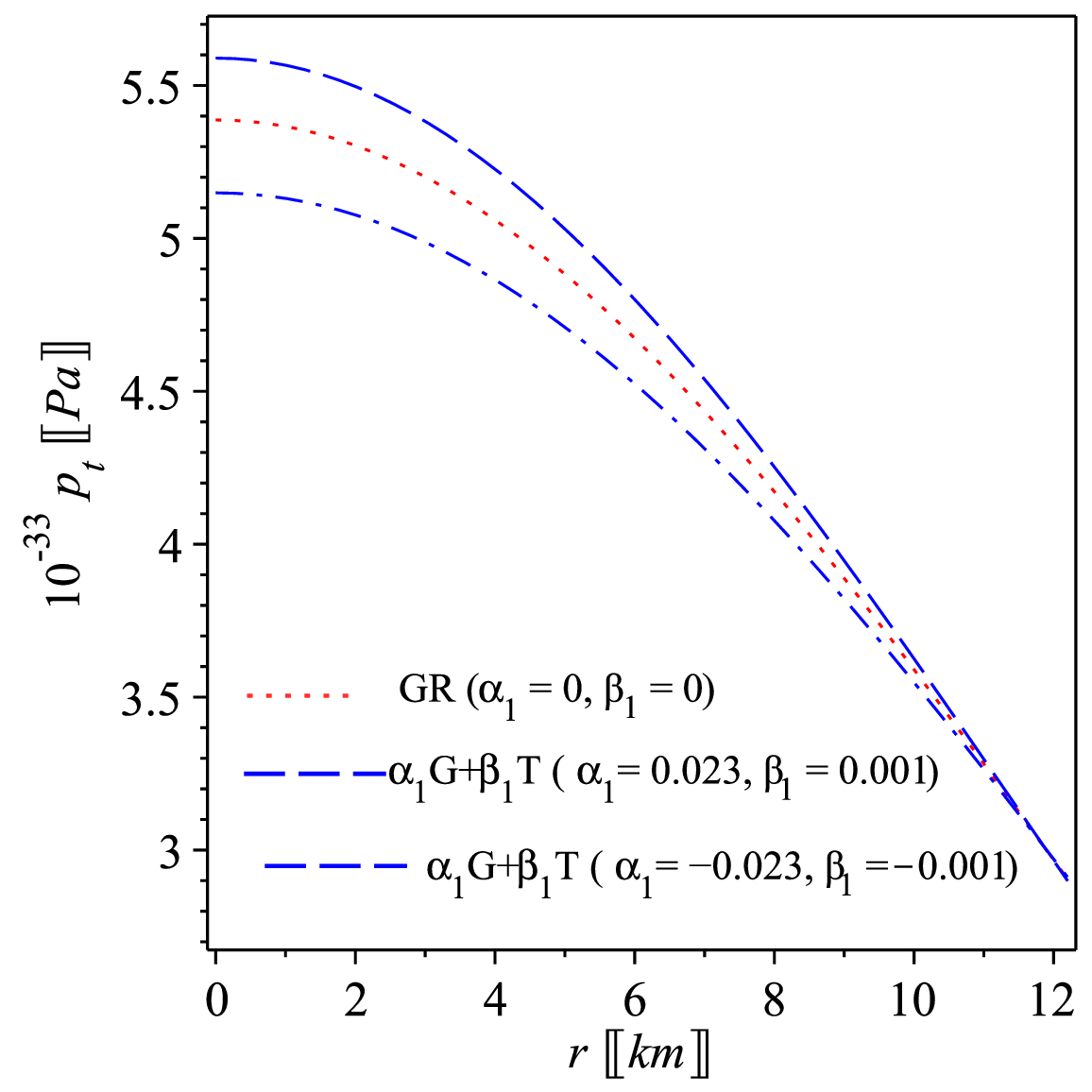}}\hspace{0.5cm}
\subfigure[~$\Delta$]{\label{fig:anisotf}\includegraphics[scale=0.3]{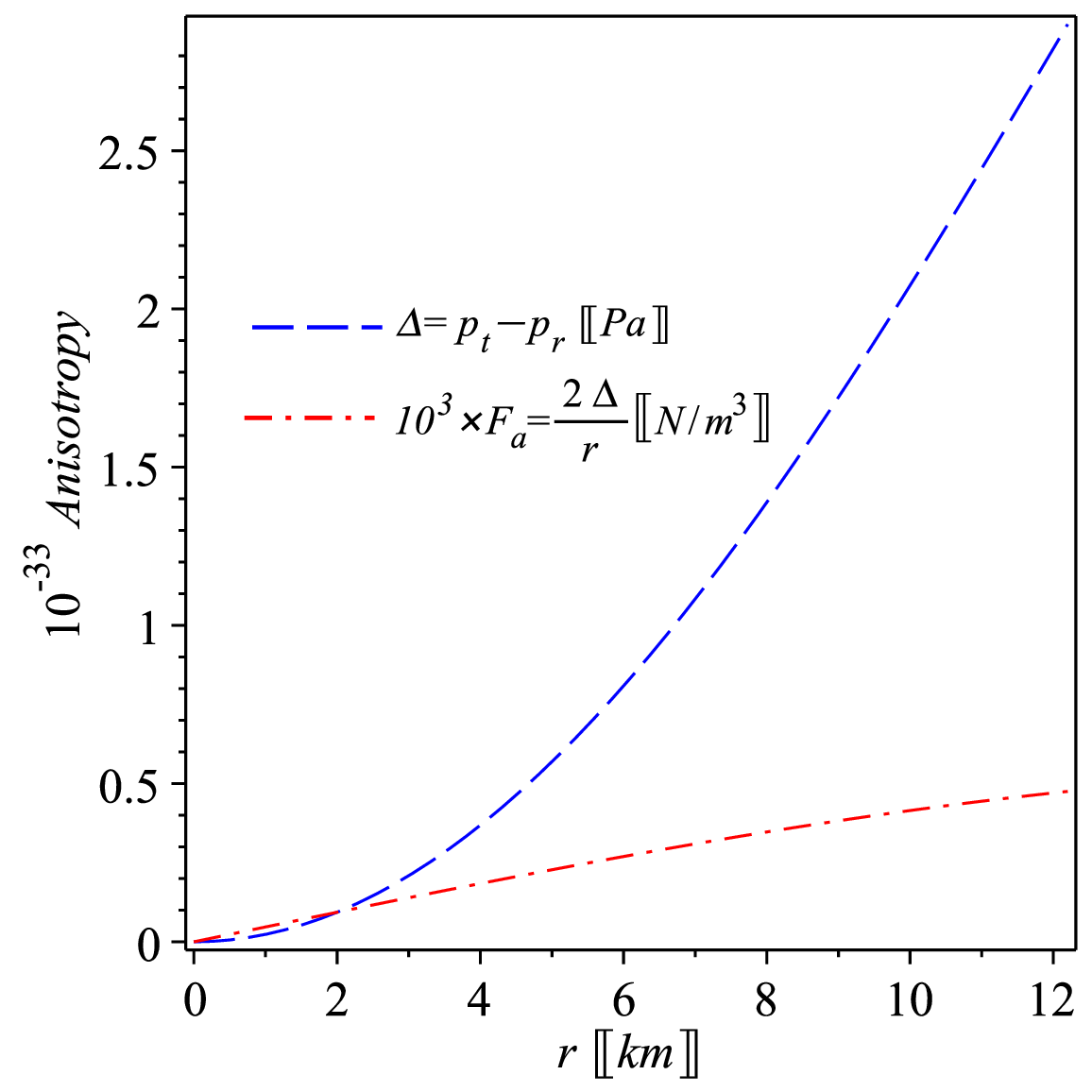}}
\caption{Matter distribution and anisotropy profiles for the pulsar \textit{U1724}.
Panels~\subref{fig:density}--\subref{fig:tangpressure} display $\rho$ together
with $p_r$ and $p_t$   profiles obtained from Eq.~\eqref{sol},
for the parameter values $\alpha_{1}=\pm 0.023$ and $\beta_{1}=\pm 0.001$.
The curves demonstrate that both the density and pressure components remain finite throughout the stellar interior
and decrease smoothly toward the stellar surface.
Panel~\subref{fig:anisotf} shows the variation of
$\Delta(r)$ inside the star for $\alpha_{1}=0,\pm 0.023$ and $\beta_{1}=0,\pm 0.001$.}
\label{Fig:dens_press}
\end{figure*}

As stated in Section~\ref{Sec:Model}, the functional forms given in Eqs.\eqref{nu} and\eqref{mu} are adopted in place of explicit equations of state (EoSs) to complete the system of equations~\eqref{9}--\eqref{11}.
Nevertheless, it can be demonstrated that these chosen forms inherently connect the pressure components with the energy density.
To proceed, we define the dimensionless variable $\epsilon = r / R_s$ and expand the solutions of Eqs.~\eqref{sol} as a power series up to the third order, i.e., $O(\epsilon^3)$. From this expansion, one obtains the following relations:
\begin{equation}\label{eq:KB_EoS}
    p_r(\rho)\approx c_1 \rho+c_2\,, \qquad  p_t(\rho) \approx c_3 \rho+c_4\,.
\end{equation}
Here, \[c_1, ..., c_4 , are \, fixed \, by   \{\epsilon_1,b_0,b_1,b_2\} \, as \, given \, in \, appendix \, \ref{Sec:App_1}.\] We can rewrite Eq.(\ref{eq:KB_EoS}) as:
\begin{equation}\label{eq:KB_EoS2}
    p_r(\rho)\approx v_r^2(\rho-\rho_{1})\,, \qquad  p_t(\rho) \approx v_t^2 (\rho-\rho_{2})\,, \quad \mbox {with} \quad v_r^2=c_1,   \rho_1=-c_2/c_1,   v_t^2=c_3 \quad \mbox{and} \quad  \rho_2=-c_4/c_3.
\end{equation}
 Here   $\rho_1$ represents the surface $\rho_s$ that fulfils  the boundary condition $p_r(\rho_s)=0$.
\subsection{Zeldovich condition}
A fundamental requirement for ensuring the stability of a compact star was formulated by
Zeldovich~\citep{1971reas.book.....Z}, stating:
\begin{equation}\label{eq:Zel}
    {\frac{{p}_r(0)}{c^2{\rho}(0)}\leq 1.}
\end{equation}
The forms of energy-density and pressure given by Eqs. \eqref{sol}, we get
\begin{align}
 & {c^2 {\rho}(r \to 0)=-{\frac {4{{b_0}}^{2} \left( 24{{b_0}}^{2}{\alpha_1}{ b_1}+48{{b_0}}^{2}{\alpha_1}{b_1}{\beta_1}-6{b_2}{{b_0}}^{2}-32{{b_0}}^{2}{b_2}{\beta_1}-3{b_1 }-21{b_1}{\beta_1} \right) }{{R}^{2}{\kappa}^{2} \left( 1+10 {\beta_1}+16{{\beta_1}}^{2} \right)  \left( {b_1}+2{b_2}{{b_0}}^{2} \right) }}}\,, \nonumber\\
 & {p_r(r\to0) =4{\frac {{b_0}^{2} \left( 48{b_0}^{2}\alpha_1b_1\beta_1+24{b_0}^{2}\alpha_1b_1+3b_1 \beta_1-2b_2{b_0}^{2} \right) }{{R}^{2}{\kappa}^{2 } \left( 1+10\beta_1+16{\beta_1}^{2} \right)  \left( b_1+2b_2{b_0}^{2} \right) }}}\,.
\end{align}
Employing the numerical results obtained for the pulsar \textit{U1724} in
Subsection~\ref{Sec:obs_const}, we evaluate the Zeldovich stability criterion~\eqref{eq:Zel}
for both parameter configurations.
For $\alpha_{1}=-0.023$ and $\beta_{1}=-0.001$, the ratio between the central pressure
and energy density is $\frac{p_{r}(0)}{c^{2}\rho(0)} = 0.044 < 1$,
while for $\alpha_{1}=0.023$ and $\beta_{1}=0.001$ we obtain
$\frac{p_{r}(0)}{c^{2}\rho(0)} \approx 0.11 < 1$.
Hence, in both scenarios, the Zeldovich condition for stellar stability is clearly satisfied.
\subsection{Causality of the model}\label{Sec:causality}

\begin{figure*}
\centering
\subfigure[~$v_r$]{\label{fig:vr}\includegraphics[scale=0.28]{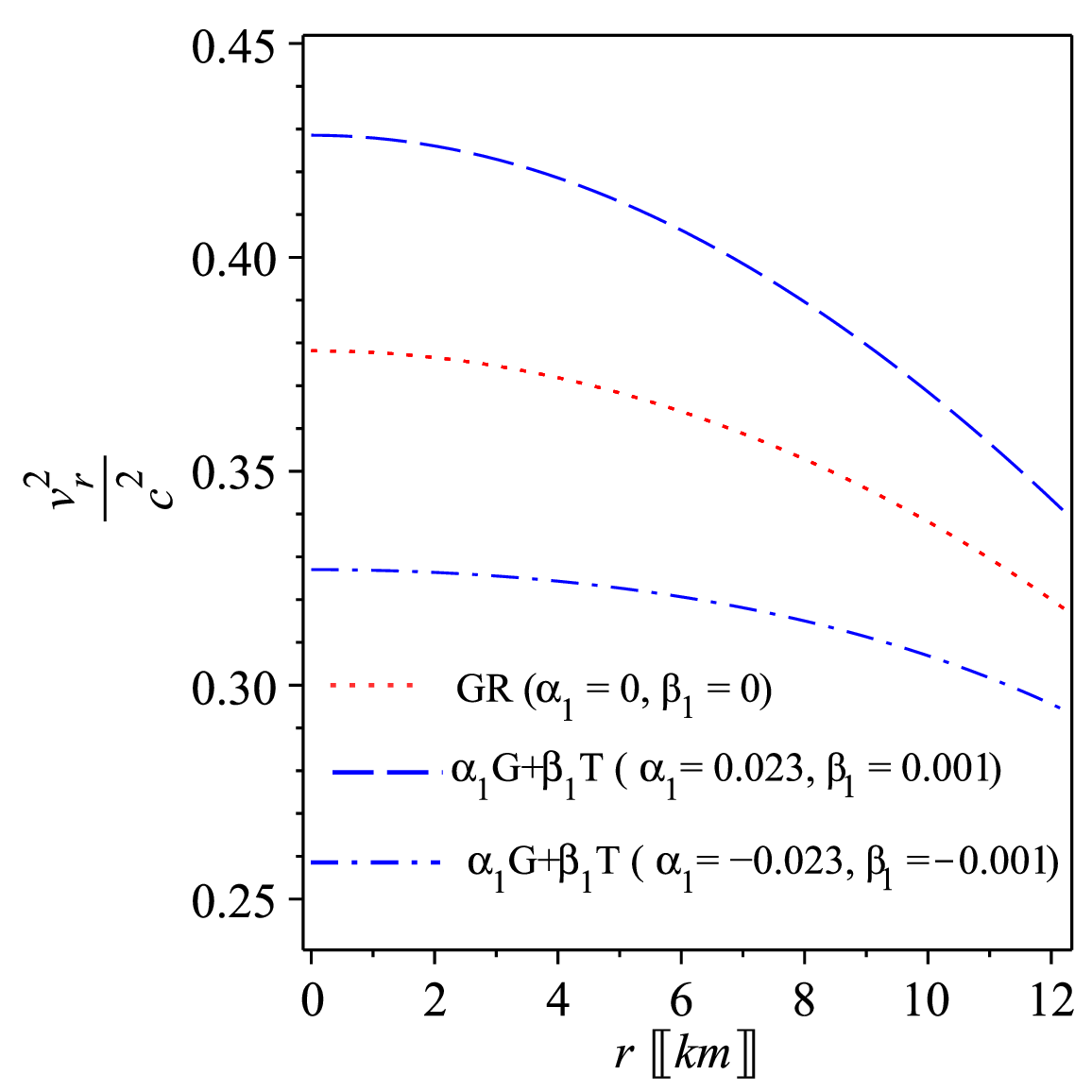}}\hspace{0.2cm}
\subfigure[~$v_t$]{\label{fig:vt}\includegraphics[scale=.28]{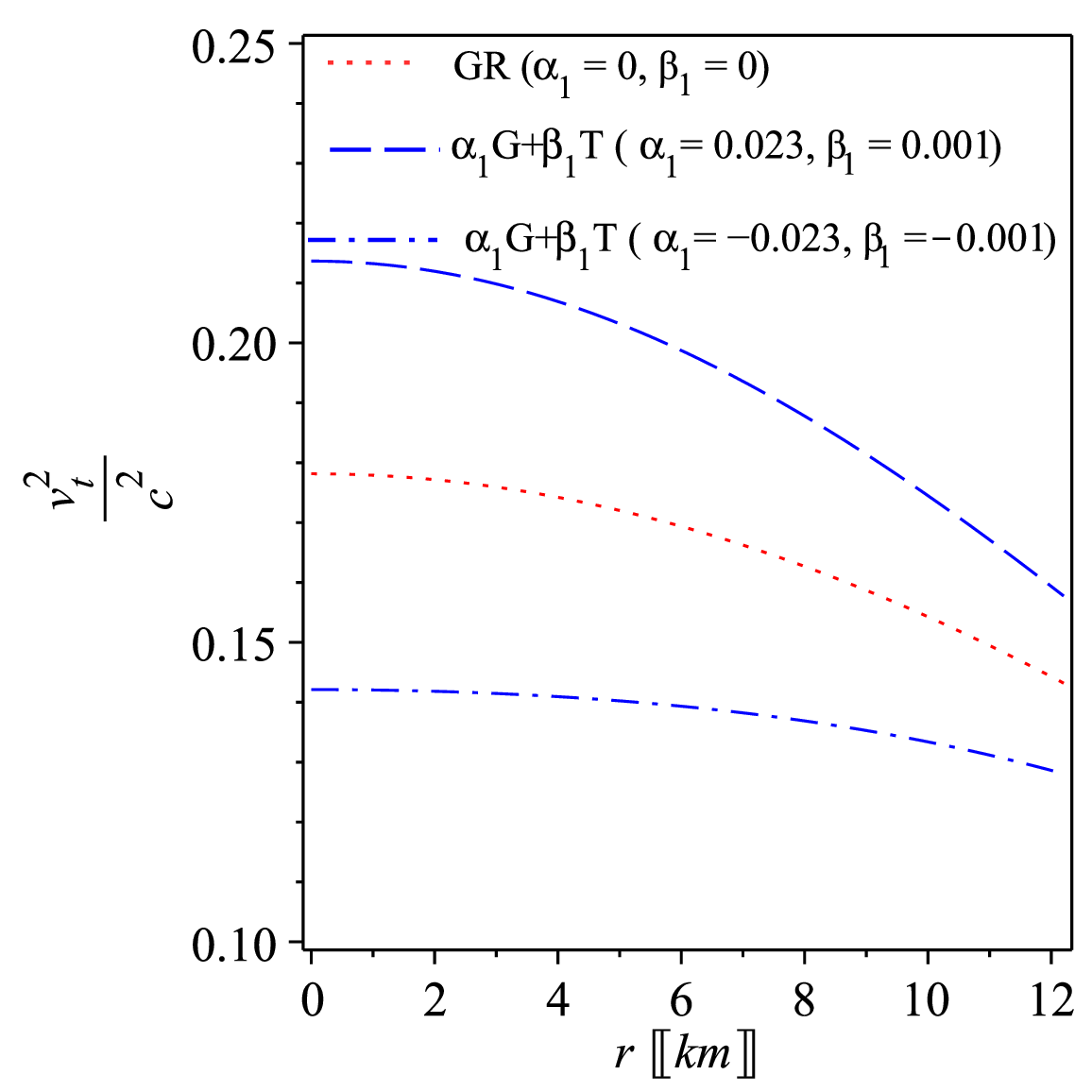}}\hspace{0.2cm}
\subfigure[~$\frac{1}{c^2}(v_t{}^2-v_r{}^2)$]{\label{fig:vt-vr}\includegraphics[scale=.28]{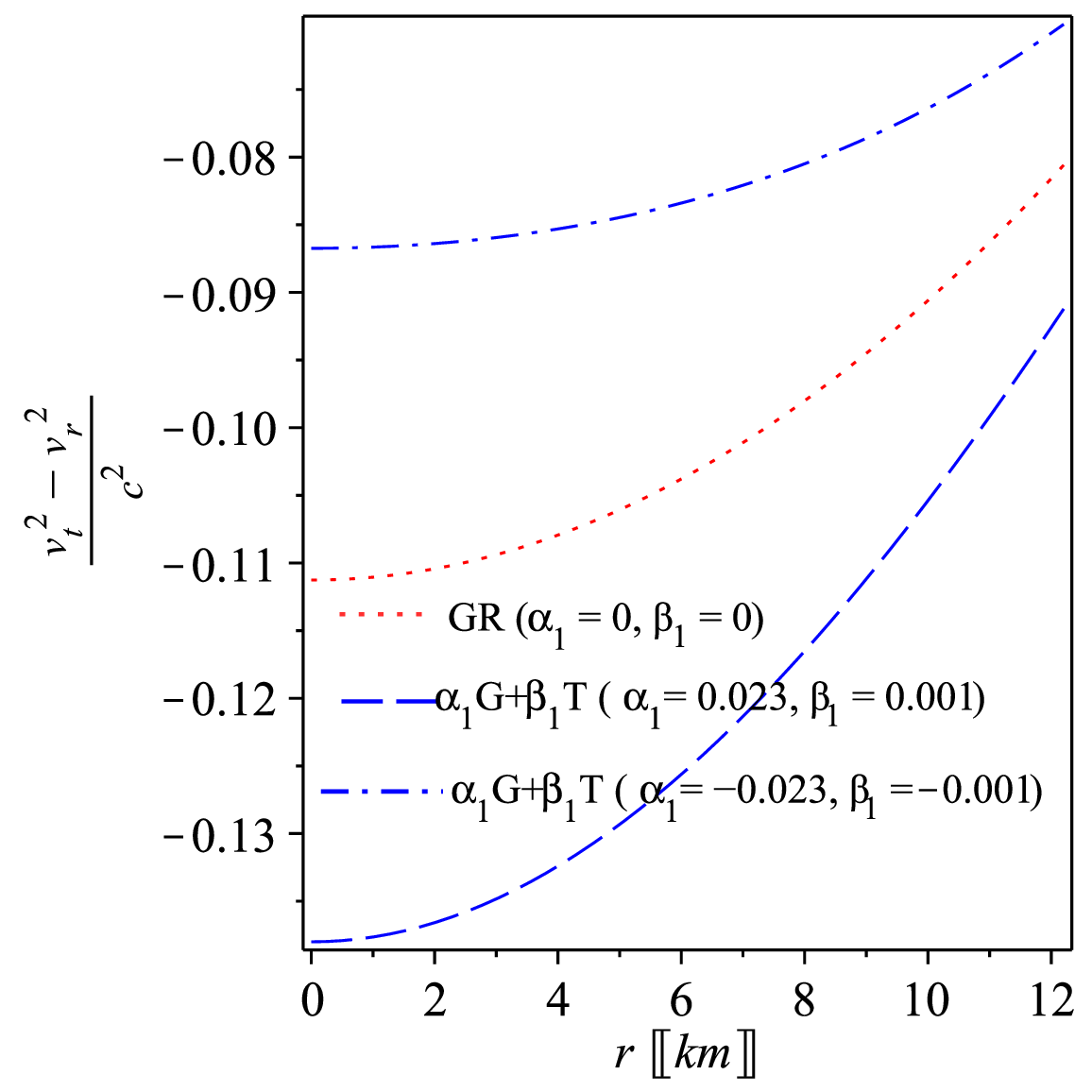}}
\caption{Variation of the sound velocity within the matter of pulsar  U1724 for $\epsilon_1 = 0,~ \pm 0.03$.
Subfigures~\subref{fig:vr} and~\subref{fig:vt} depict how $v_r$ and $v_t$ propagates as described by Eq.~\eqref{eq:sound_speed}.
Subfigure~\subref{fig:vt-vr} indicates that the solution satisfies the stability requirement of the highly anisotropic regime, given by $(v_t^2 - v_r^2)/c^2 < 0$.}
\label{Fig:Stability}
\end{figure*}

From the induced equations of state given in Eq.~\eqref{eq:KB_EoS2},
the squared sound speeds in the radial and tangential directions correspond to the respective slopes of these linear relations, namely,
\begin{equation}\label{eq:sound_speed}
  v_r^2 =  \frac{ d{ p}_r}{d { \rho}}=  \frac{p'_r}{{ \rho'}}, \quad
  v_t^2 = \frac{d{  p}_t}{d{   \rho}}= \frac{p'_t}{{ \rho'}}.
\end{equation}
Using Eqs.~\eqref{sol} provided in Appendix~A,
the corresponding expressions for the density and pressure gradients are derived
and presented in Appendix~\ref{Sec:App_3} (see Eqs.~\eqref{eq:dens_grad}--\eqref{eq:pt_grad}).
The behavior of $v_r^2$ and $v_t^2$
within the pulsar \textit{U1724} is illustrated in
Figs.\ref{Fig:Stability}\subref{fig:vr} and\subref{fig:vt}
for various choices of the model parameters $\alpha_{1}$ and $\beta_{1}$.
The results confirm that the conditions
$0 \leq v_{r}^{2}/c^{2} \leq 1$ and $0 \leq v_{t}^{2}/c^{2} \leq 1$
are satisfied throughout  stellar interior,
indicating that both causality and stability requirements are fulfilled.
Moreover, Fig.~\ref{Fig:Stability}\subref{fig:vt-vr}
demonstrates that the inequality
$-1 < (v_{t}^{2}-v_{r}^{2})/c^{2} < 0$
holds at all points inside the star,
consistent with the stability criterion for anisotropic fluid configurations
as discussed in Ref.~\citep{Herrera:1992lwz}.
\subsection{Astrophysical Bounds on $M$ and $R$ of  Pulsar U1724}\label{Sec:obs_const}

In this study, we employ the observational estimates of$M$ and $R$ of the pulsar
\textit{U1724}, namely $M = (1.81 \pm 0.27)\,M_{\odot}$ and $R = (12.2 \pm 1.4)\,\mathrm{km}$~\citep{Roupas:2020mvs},
to place constraints on the linear modified gravity parameters $\alpha_{1}$ and $\beta_{1}$.
\begin{figure*}
\centering
\subfigure[~The Mass function]{\label{Fig:Mass}\includegraphics[scale=0.45]{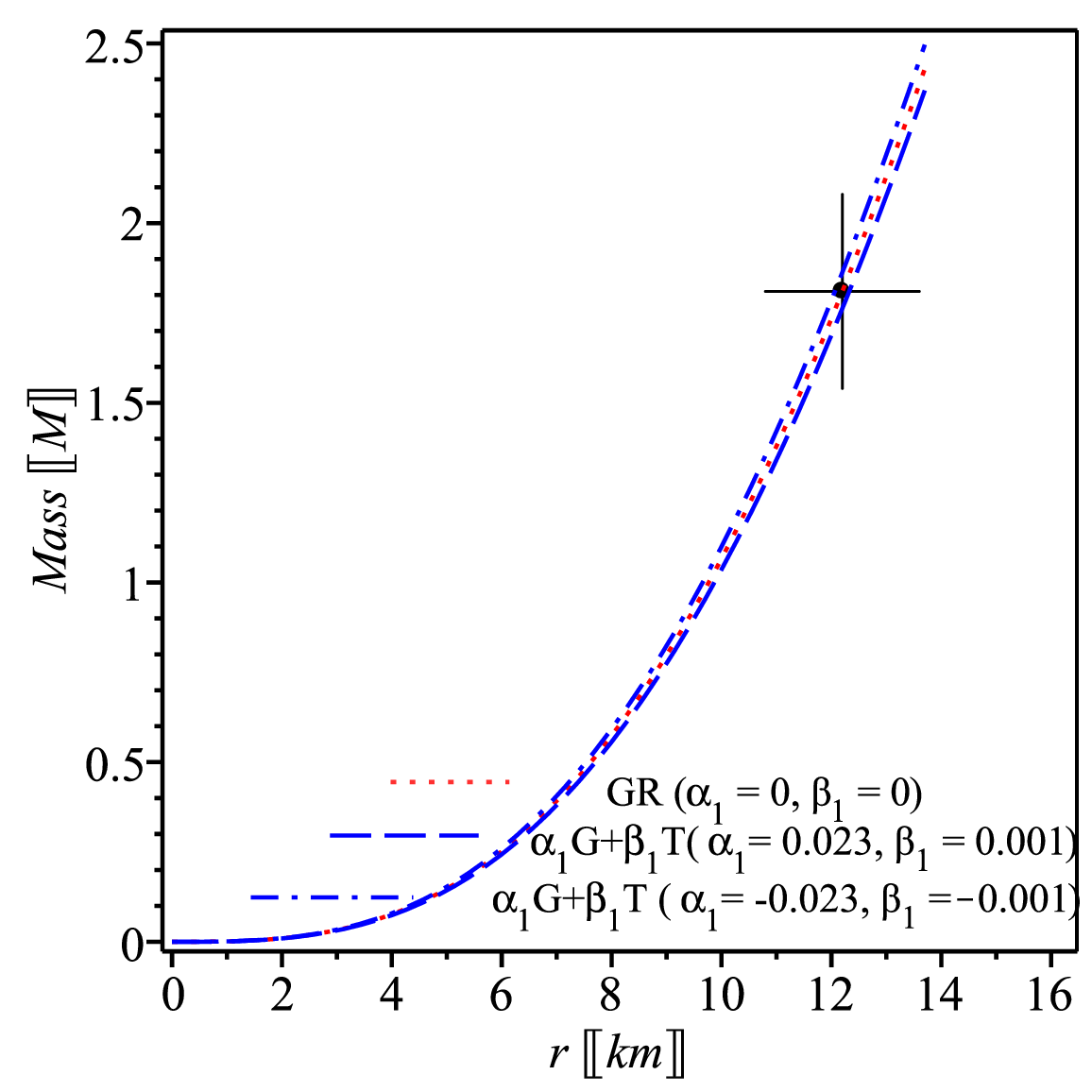}}
\caption{The mass given by Eq.~\eqref{Mf3} associated to pulsar ${\textit U1724}$.}
\label{Fig:Mass1}
\end{figure*}
The matter content can be expressed as:
 \begin{align}\label{Mf3}
     {m(r)} =  4\pi\int_{0}^{r} \rho(\eta) \eta^2 d\eta \,.
 \end{align}
Recalling the density profile presented in Eq.~\eqref{sol}, for $f(\mathcal{R,G,T}) =\mathcal{R}+ \alpha \mathcal{G} + \beta \mathcal{T}$,  yields  Fig. \ref{Fig:Mass1}.
\begin{itemize}
    \item For $\alpha_1=\beta_1=0$,  we get\{$C=0.442$, $b_0 =-0.3664450606$,  $b_1 =0.4382981879$, $b_2 =-0.386212747$\}.
    \item For $\alpha_1=0.023$ and $\beta_1=0.001$, a gravitational mass of $M = 2.38\, M_\odot$ is obtained at  $R_s \approx 13.756\, \text{km}$, yielding a compactness $C = 0.434$. Consequently, the numerical values  are fixed as
 \{$b_0 =-0.3664450606$, $b_1 =0.4157730032$, $b_2 =-0.289330509$\}.
    \item For $\alpha_1=-0.023$ and $\beta_1=-0.001$, a gravitational mass of $M \approx 2.555\, M_\odot$ is obtained at a radius $R_s \approx 13.83\, \text{km}$, yielding a compactness $C = 0.451$. Consequently, the numerical values   are fixed as
 \{ $b_0 =-0.3664450606$, $b_1 =0.4633594655$, $b_2 =-0.494002881$\}.
\end{itemize}
{This places constraints on the model parameters as $0 \leq |\alpha| \leq 2.3\, \text{km}^2$ and $0 \leq |\beta| \leq 2.08 \times 10^{-46}\, \text{N}^{-1}$. In general, as illustrated in Fig.~\ref{Fig:Mass1}, the linear gravity term in the theory $f(\mathcal{R}, \mathcal{G}, \mathcal{T}) = \mathcal{R} + \alpha \mathcal{G} + \beta \mathcal{T}$ affects the pulsar's mass.

For $\alpha_1 < 0$ and $\beta_1 < 0$, the model predicts a mass greater than that obtained in General Relativity (GR), i.e., the $\alpha_1 = 0$ and $\beta_1 = 0$ case, at the same radius (corresponding to a smaller stellar size for the same mass). Conversely, for $\alpha_1 > 0$ and $\beta_1 > 0$, the model yields a mass smaller than the GR prediction.
}
\subsection{Geometric sector}\label{Sec:geom}
\begin{figure}
\centering
\subfigure[~Matching solutions]{\label{fig:Junction}\includegraphics[scale=0.38]{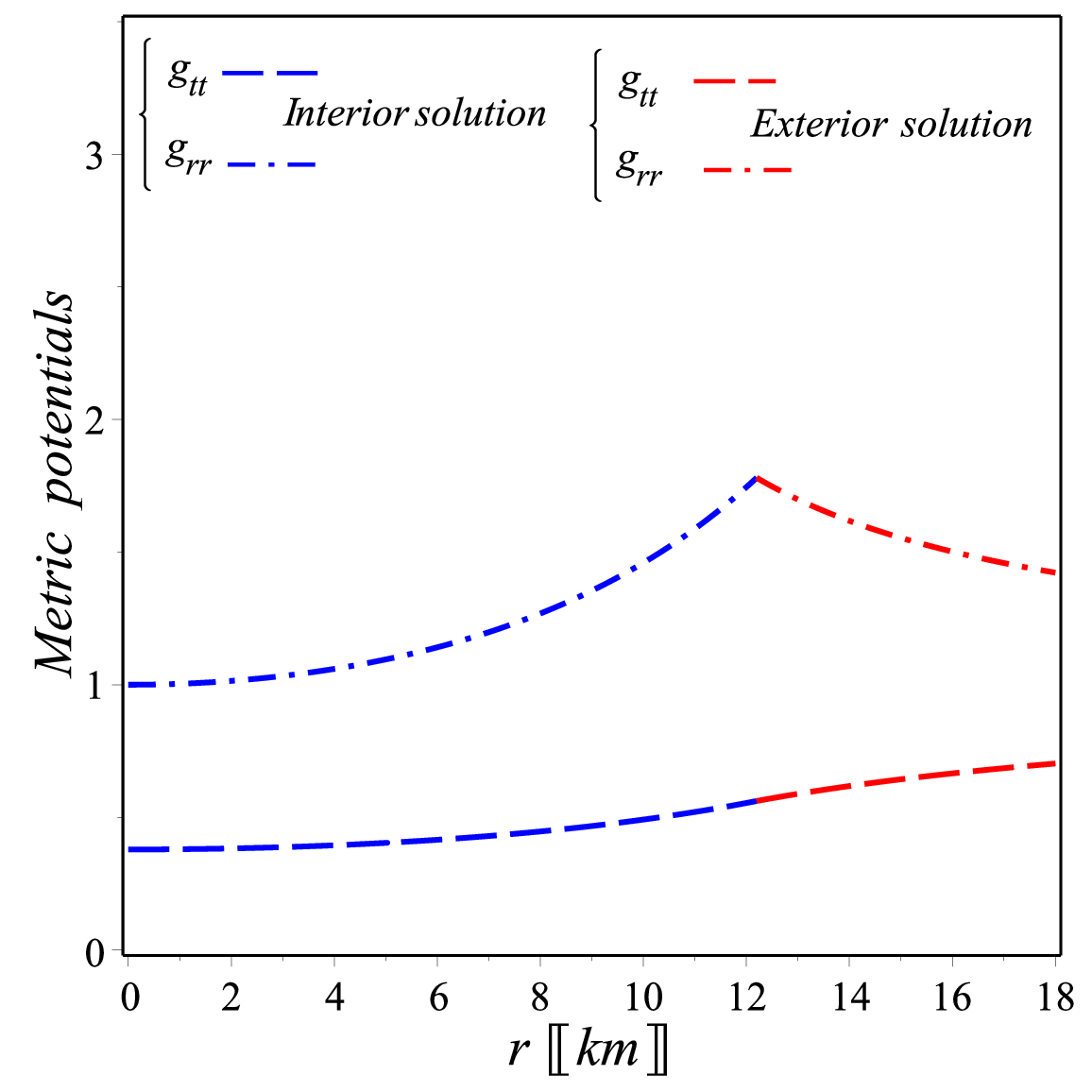}}\hspace{1cm}
\subfigure[~Gravitational red-shift]{\label{fig:redshift}\includegraphics[scale=0.38]{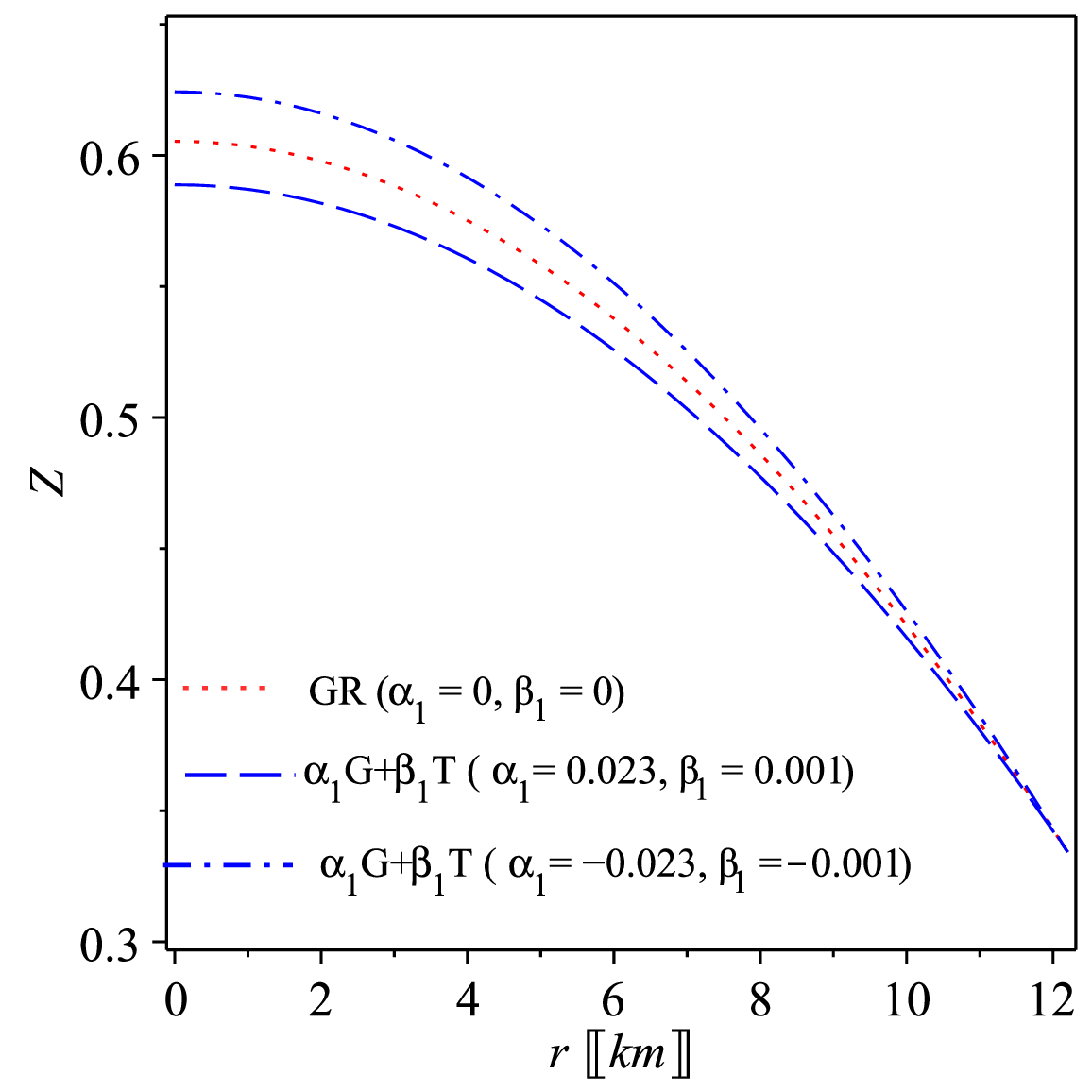}}
\caption{The spacetime structure associated with the pulsar \textit{U1724}.
\subref{fig:Junction}~illustrates the interior metric components $g_{tt}$ and $g_{rr}$, modeled through the KB prescription, together with the external Schwarzschild vacuum geometry. The figure confirms that both potentials remain regular throughout the star and transition smoothly to the outer spacetime at the boundary.
\subref{fig:redshift}~presents the corresponding gravitational redshift profile, derived from Eq.~\eqref{eq:redshift}, for the parameter sets $\alpha_1 = \pm 0.023$ and $\beta_1 = \pm 0.001$. The central value of the redshift is about $Z_s \simeq 0.59$, gradually declining to approximately $0.33$ at the stellar surface in all examined cases.}
\label{Fig:Matching}
\end{figure}

It is essential that   $\textit{g}_{tt}$ and $\textit{g}_{rr}$ remain free of singularities throughout the stellar interior. The functional forms introduced in Eqs.~\eqref{nu} and \eqref{mu} guarantee this condition, as both potentials are finite and well-behaved at the stellar center. One finds
$g_{tt}(r=0) = \left( \frac{b_1 + 2 b_2 b_0^{2} R^{2}}{4 b_0^{2} R^{2}} \right)^{2} \neq 0$
and $g_{rr}(r=0) = 1$. The radial profiles of these functions within the pulsar  are depicted in Fig.~\ref{Fig:Matching}\subref{fig:Junction}.

In addition, Fig.~\ref{Fig:Matching}\subref{fig:Junction} illustrates the seamless junction  of our pulsar, employing the numerically obtained model parameters listed in the figure caption.

In addition, the expression for the gravitational redshift, derived from the chosen metric potentials, is formulated within the theoretical framework of modified gravity described by $f(\mathcal{R}, \mathcal{G}, \mathcal{T}) = \mathcal{R} + \alpha\, \mathcal{G} + \beta\, \mathcal{T}$.
\begin{equation}\label{eq:redshift}
    Z(r)=\frac{1}{\sqrt{-g_{tt}}}-1=\frac{1}{\left( \frac{b_1+2b_2{b_0}^{2}{R}^{2
}-2b_2{b_0}^{4}{r}^{2}}{4{b_0}^{2} \left( {R}^{2}-{b_0}^{2}{r}^{2} \right)} \right)}-1.
\end{equation}
Figure~\ref{Fig:Matching}\subref{fig:redshift} illustrates how the gravitational redshift varies across the interior of the pulsar \textit{U1724} for several choices of the coupling parameters $\alpha_1$ and $\beta_1$. In the reference case of General Relativity, where $\alpha_1 = \beta_1 = 0$, the redshift at the core is approximately $Z(0) \simeq 0.61$, reducing to $Z_{R_s} \simeq 0.33$ at the stellar surface.

When the coupling constants take positive values, $\alpha_1 = 0.023$ and $\beta_1 = 0.001$, the central redshift reaches $Z(0) \simeq 0.5888$, slightly below the GR prediction, and gradually declines outward to $Z_{R_s} \simeq 0.33$. This surface value remains comfortably within the accepted theoretical limit $Z_{R_s} = 2$ \citep{Buchdahl:1959zz,Ivanov:2002xf,Barraco:2003jq,Boehmer:2006ye}.

For the opposite case, $\alpha_1 = -0.023$ and $\beta_1 = -0.001$, the central redshift rises to $Z(0) \simeq 0.624$, which is marginally higher than the GR estimate, and again decreases monotonically toward the boundary, where it approaches $Z_{R_s} \simeq 0.33$, still below the upper limit.


\subsection{Energy conditions}\label{Sec:Energy-conditions}

It is advantageous to recast Eq.~\eqref{5} into the form shown below:
\begin{equation}\label{eq:fR_MG}
    G_{\mu\nu}=\kappa\left(\mathfrak{T}_{\mu\nu}+\mathfrak{T}_{\mu\nu}^{geom}\right)=\kappa \bar{\mathfrak{T}}_{\mu\nu},
\end{equation}
The quantity $G_{\mu\nu} = R_{\mu\nu} - \tfrac{1}{2} g_{\mu\nu} R$ corresponds to the Einstein tensor, whereas the additional contribution arises from the modified gravitational framework characterized by $f(\mathcal{R}, \mathcal{G}, \mathcal{T}) = \mathcal{R} + \alpha\, \mathcal{G} + \beta\, \mathcal{T}$~\cite{Shamir:2017rjz}.
\begin{align}
 &   \mathfrak{T}_{\mu\nu}^{geom}=\frac{1}{{{f_{\cal R}}\left( {{\cal R},{\cal G},{\cal T}} \right)}}\left[\right. {\nabla _\rho }{\nabla _\sigma }{f_{\cal R}}\left( {{\cal R},{\cal G},{\cal T}} \right) - \left( {{{\cal T}_{\rho \sigma }} + {\Theta _{\rho \sigma }}} \right){f_T}\left( {{\cal R},{\cal G},{\cal T}} \right)
 + \frac{1}{2}{g_{\rho \sigma }}(f\left( {{\cal R},{\cal G},{\cal T}} \right) + \mathcal{R}{f_{\cal R}}\left( {{\cal R},{\cal G},{\cal T}} \right))\nonumber\\
& - {g_{\rho \sigma }}\Box{f_{\cal R}}\left( {{\cal R},{\cal G},{\cal T}} \right) - (2\mathcal{R}{\mathcal{R}_{\rho \sigma }} - 4\mathcal{R}_\rho ^\xi {\mathcal{R}_{\xi \sigma }} - 4{\mathcal{R}_{\rho \xi \sigma \eta }}{\mathcal{R}^{\xi \eta }} + 2\mathcal{R}_\rho ^{\xi \eta \delta }{R_{\sigma \xi \eta \delta }}){f_{\cal G}}\left( {{\cal R},{\cal G},{\cal T}} \right) - (2\mathcal{R}{g_{\rho \sigma }}{\nabla ^2} - 2\mathcal{R}{\nabla _\rho }{\nabla _\sigma } \\
& - 4{g_{\rho \sigma }}{\mathcal{R}^{\xi \eta }}{\nabla _\xi }{\nabla _\eta }- 4{\mathcal{R}_{\rho \sigma }}{\nabla ^2} + 4\mathcal{R}_\rho ^\xi {\nabla _\sigma }{\nabla _\xi } + 4\mathcal{R}_\sigma ^\xi {\nabla _\rho }{\nabla _\xi } + 4{\mathcal{R}_{\rho \xi \sigma \eta }}{\nabla ^\xi }{\nabla ^\eta })\left. {{f_{\cal G}}\left( {{\cal R},{\cal G},{\cal T}} \right)} \right].\nonumber
\end{align}
Expressed in mixed indices, the total energy-momentum distribution is written as $\bar{\mathfrak{T}}_{\mu}{^\nu} = \mathrm{diag}(-\bar{\rho} c^{2},\, \bar{p}_{r},\, \bar{p}_{t},\, \bar{p}_{t})$.
\begin{figure*}
\centering
\subfigure[~WEC \& NEC ]{\label{fig:Cond1}\includegraphics[scale=0.27]{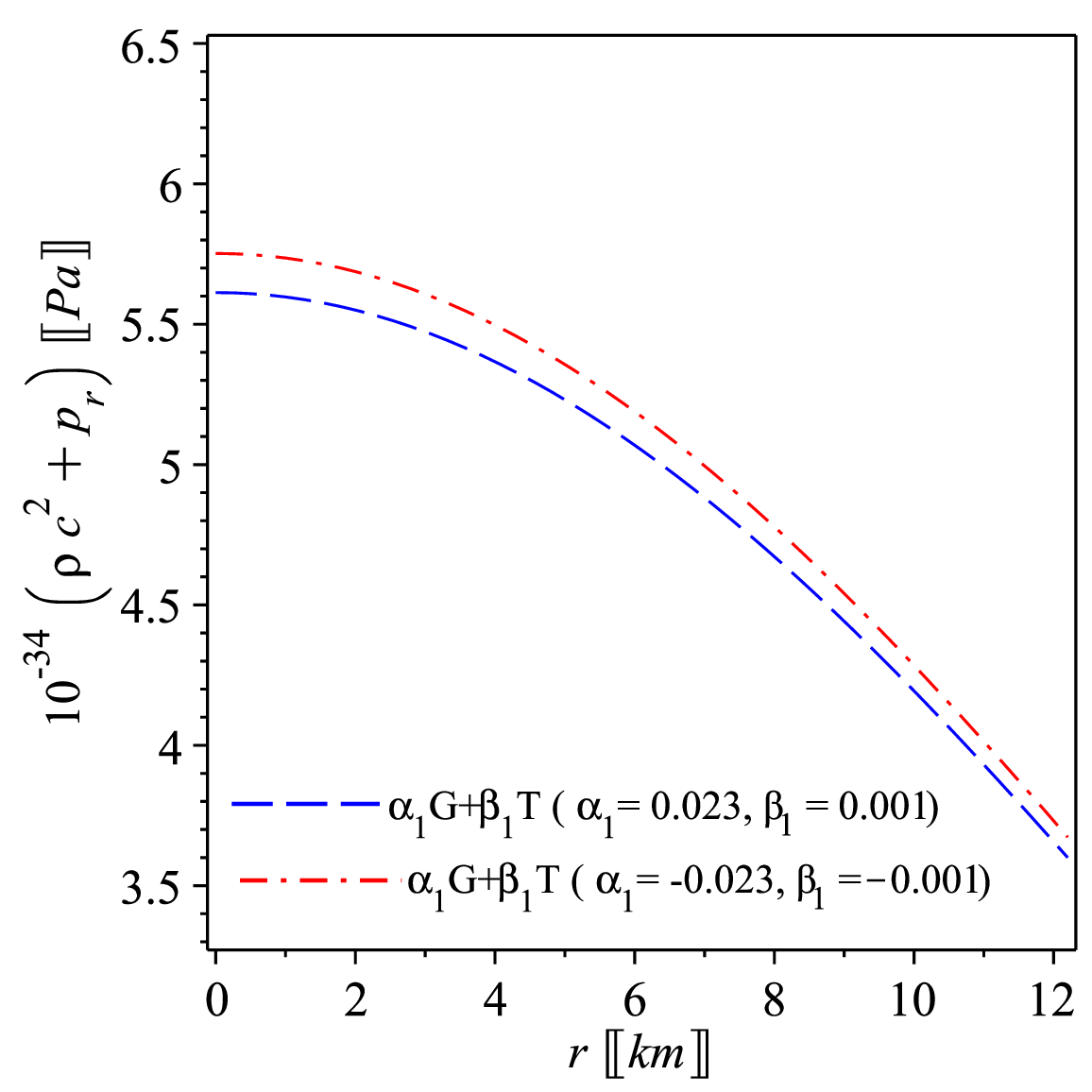}}\hspace{0.2cm}
\subfigure[~WEC \&NEC]{\label{fig:Cond2}\includegraphics[scale=0.27]{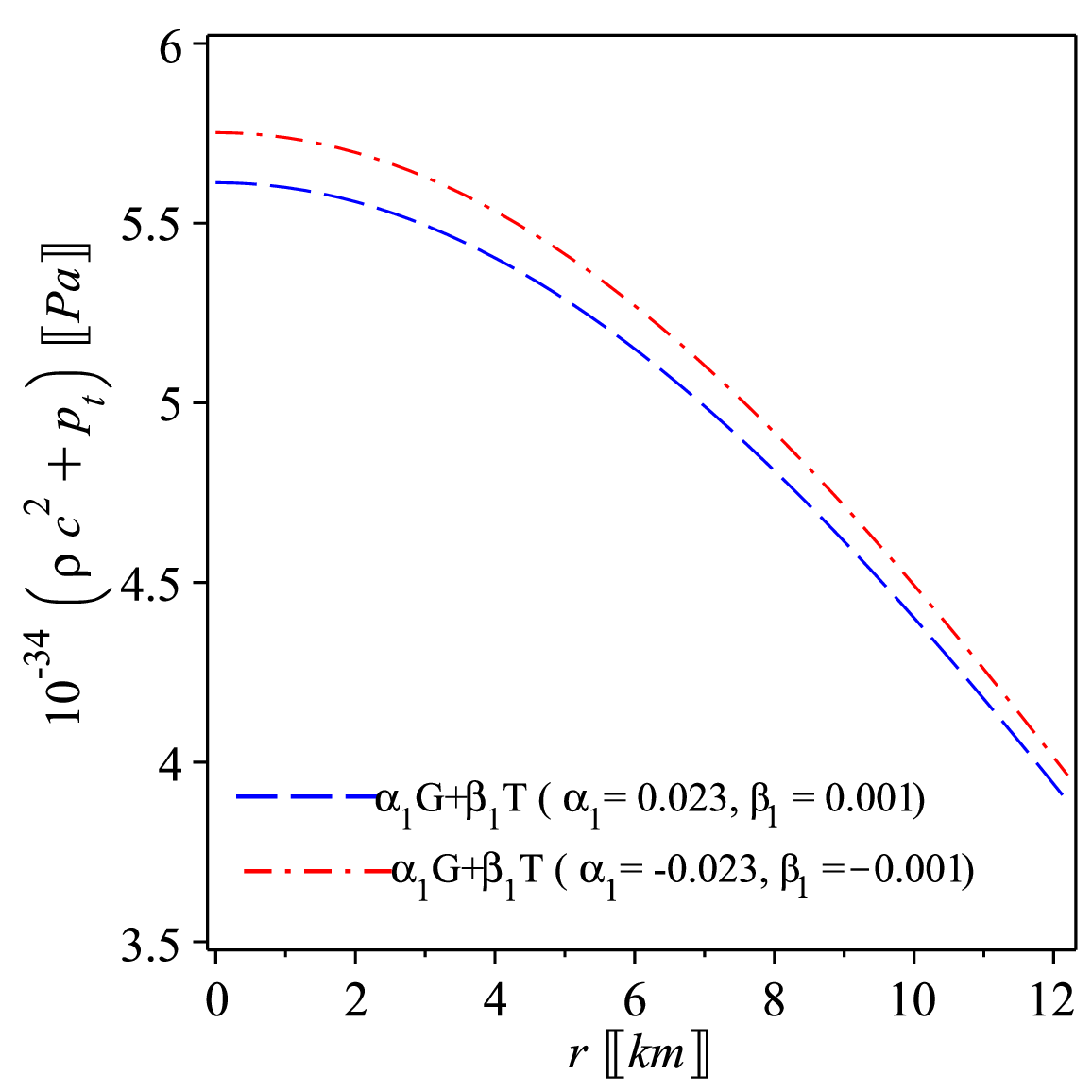}}\hspace{0.2cm}
\subfigure[~SEC]{\label{fig:Cond3}\includegraphics[scale=.27]{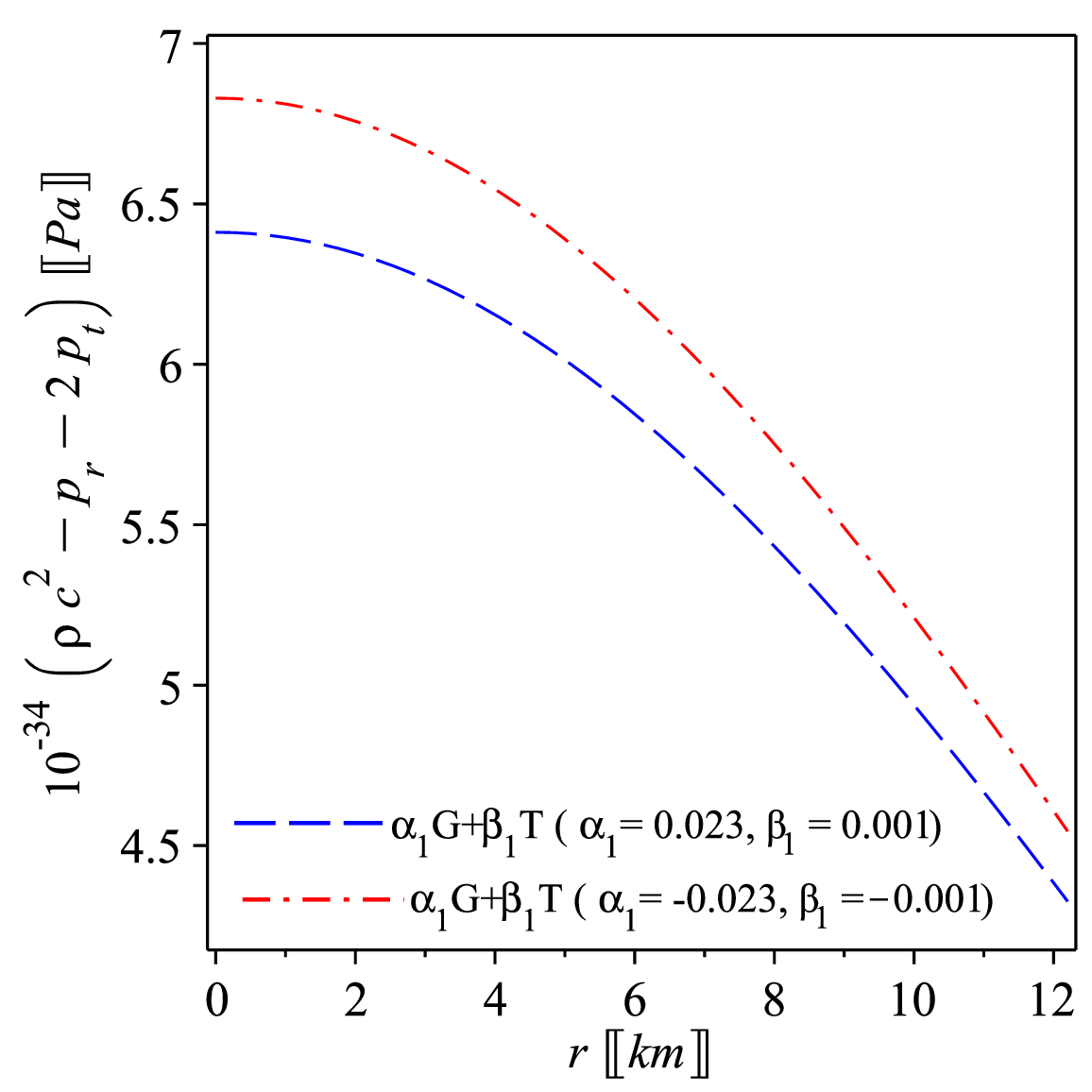}}\\
\subfigure[~DEC]{\label{fig:DEC}\includegraphics[scale=.27]{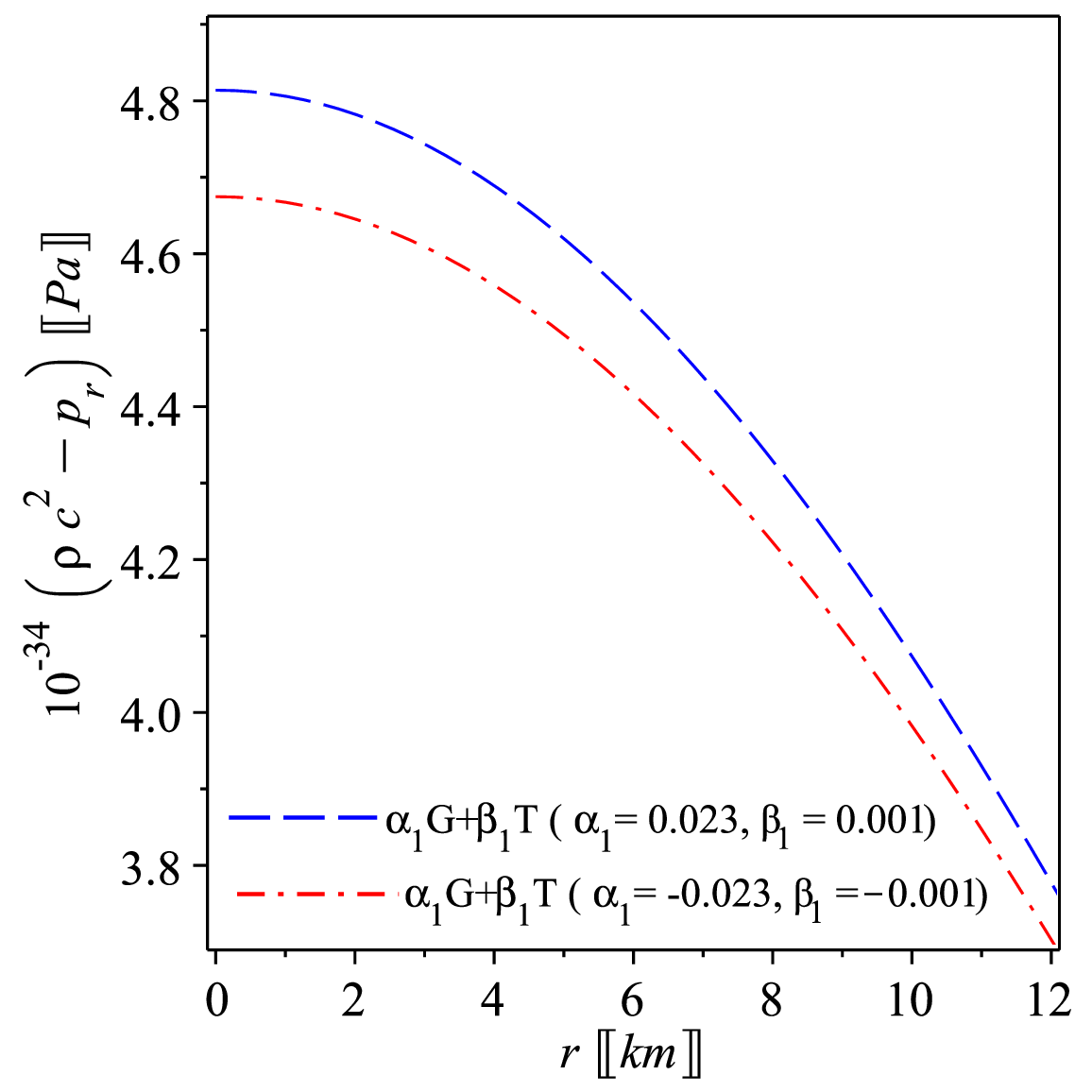}}\hspace{0.2cm}
\subfigure[~DEC ]{\label{fig:DEC}\includegraphics[scale=.27]{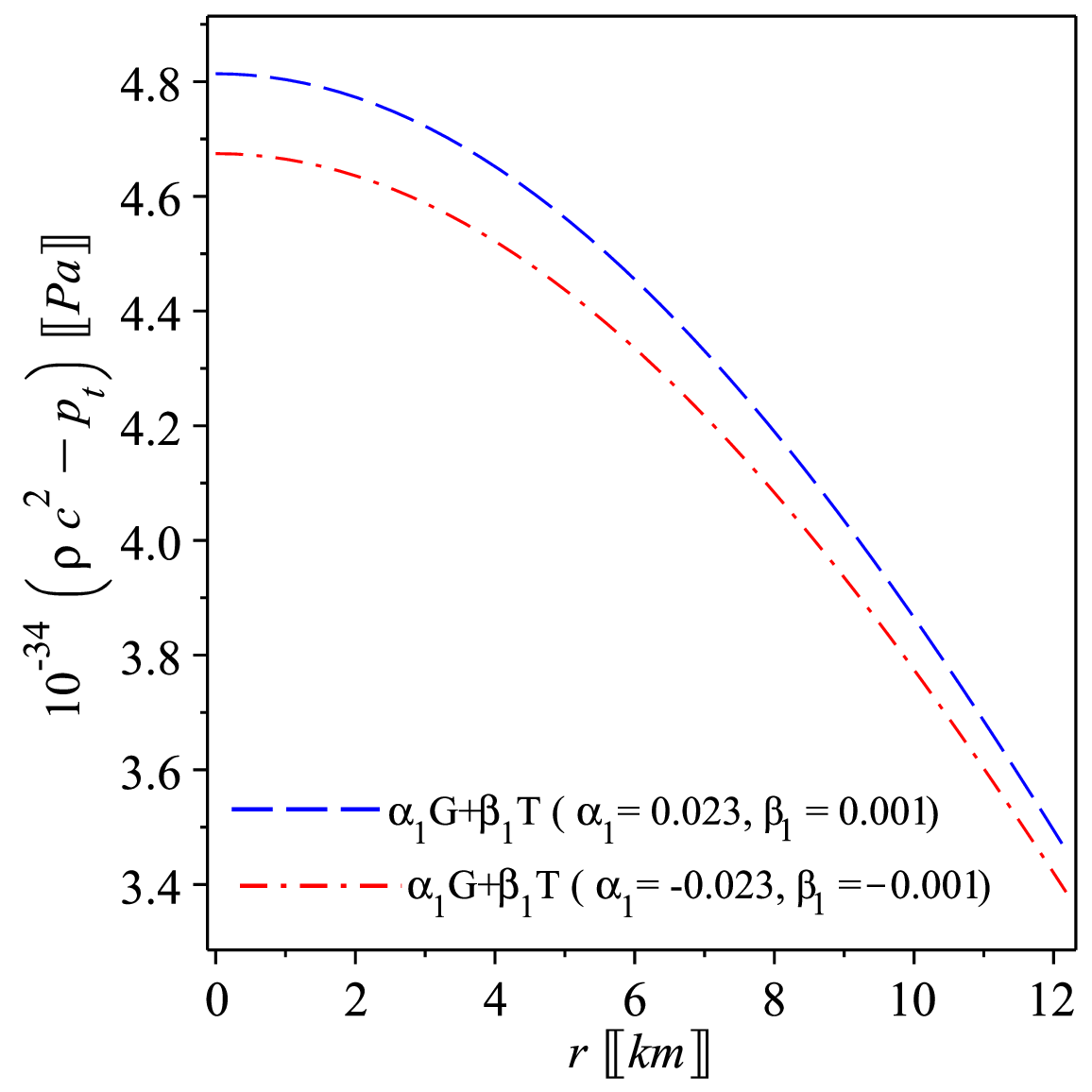}}
\caption{All energy conditions associated with the effective matter tensor $\bar{\mathfrak{T}}_{\mu\nu}$, defined in Subsection~\ref{Sec:Energy-conditions}, are verified to hold for the pulsar model J0740+6620, as illustrated in the plots.
}
\label{Fig:EC}
\end{figure*}

Drawing on the Raychaudhuri equation together with the focusing theorem, one concludes that the trace of the tidal field must obey the following inequality $R_{\mu\nu} u^{\mu} u^{\nu} \geq 0$ and $R_{\mu\nu} n^{\mu} n^{\nu} \geq 0$.
Within the modified gravity framework
$f(\mathcal{R}, \mathcal{G}, \mathcal{T}) = \mathcal{R} + \alpha\, \mathcal{G} + \beta\, \mathcal{T}$,
the Ricci tensor assumes the form
$R_{\mu\nu} = \kappa \left( \bar{\mathfrak{T}}_{\mu\nu} - \tfrac{1}{2} g_{\mu\nu} \bar{\mathfrak{T}} \right)$,
as  Eq.~\eqref{eq:fR_MG} shows.
On this basis, the following,
\begin{itemize}
  \item[a.] $\bar{\rho}\geq 0$, $ \bar{\rho} c^2+ \bar{p}_r > 0$ and $\bar{\rho} c^2+\bar{p}_t > 0$,    WEC.
  \item[b.] $\bar{\rho} c^2+ \bar{p}_r \geq 0$ and $\bar{\rho} c^2+  \bar{p}_t \geq 0$,  NEC .
  \item[c.] $\bar{\rho} c^2+\bar{p}_r+2\bar{p}_t \geq 0$, $\bar{\rho} c^2+\bar{p}_r \geq 0$ and $\bar{\rho} c^2+\bar{p}_t \geq 0$, SEC.
  \item[d.] $\bar{\rho}\geq 0$, $\bar{\rho} c^2-\bar{p}_r \geq 0$ and $\bar{\rho} c^2-\bar{p}_t \geq 0$, DEC,\\
\end{itemize}
represent the energy conditions.
The fulfillment of the energy conditions, computed from the total stress-energy tensor, is depicted in Fig.~\ref{Fig:EC} for both signs of $\alpha_1$ and $\beta_1$. All conditions are seen to hold for the case of the pulsar \textit{U1724}.

\subsection{Adiabatic behavior and fluid equilibrium}\label{Sec:TOV}
Within the framework of Newtonian gravity, the stellar mass cannot be restricted from above when the adiabatic index defined as the ratio of specific heats takes values greater than $4/3$, that is, $\gamma > 4/3$. Stability in the Newtonian regime therefore requires the opposite condition, $\gamma < 4/3$.

Relativistic treatments of anisotropic neutron stars, however, reveal that equilibrium against radial perturbations can still be maintained even when $\gamma > 4/3$. Accordingly, the adiabatic index is introduced following the formulation of \citep{Chandrasekhar:1964zz, chan1993dynamical} as

\begin{equation}\label{eq:adiabatic}
{\gamma}=\frac{4}{3}\left(1+\frac{{ \Delta}}{r|{  p}'_r|}\right)_{max},\qquad \qquad
{\Gamma_r}=\frac{{\rho c^2}+{p_r}}{{p_r}}{v_r^2}, \qquad \qquad
{\Gamma_t}=\frac{{\rho c^2}+{p_t}}{{p_t}}{v_t^2 .}
\end{equation}
For an isotropic configuration, where $\Delta = 0$, the adiabatic index attains the value $\gamma = 4/3$.
When a mild degree of anisotropy is present ($\Delta < 0$), the index becomes $\gamma < 4/3$, reproducing the behavior predicted by Newtonian gravity.
In cases of pronounced anisotropy ($\Delta > 0$), as examined in this work, the value increases beyond this limit, yielding $\gamma > 4/3$.

\begin{figure}
\centering
\subfigure[~$\gamma$]{\label{fig:gamar1}\includegraphics[scale=0.28]{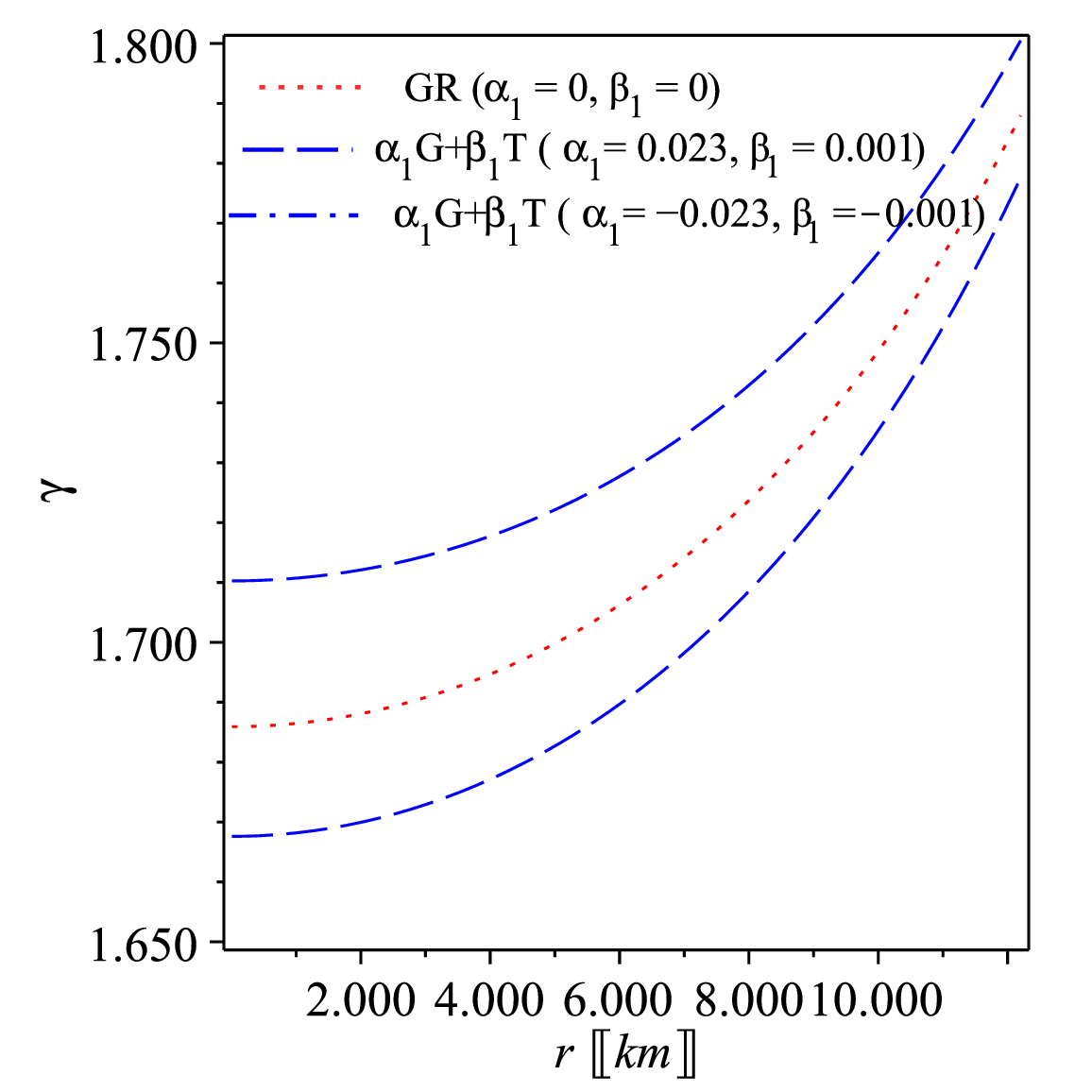}}\hspace{0.2cm}
\subfigure[~$\Gamma_r$]{\label{fig:gamar}\includegraphics[scale=0.28]{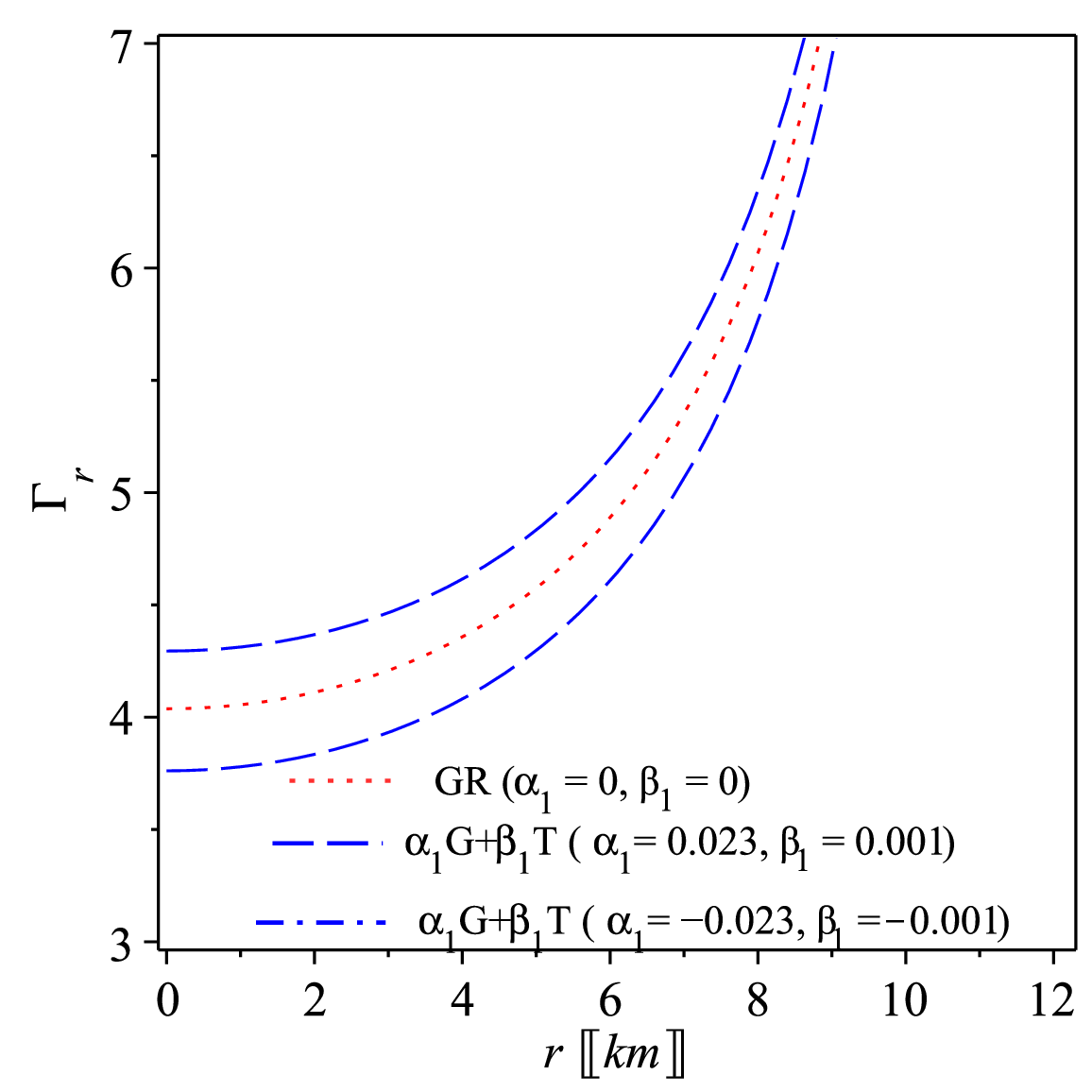}}\hspace{0.2cm}
\subfigure[~$\Gamma_t$]{\label{fig:gamar}\includegraphics[scale=0.28]{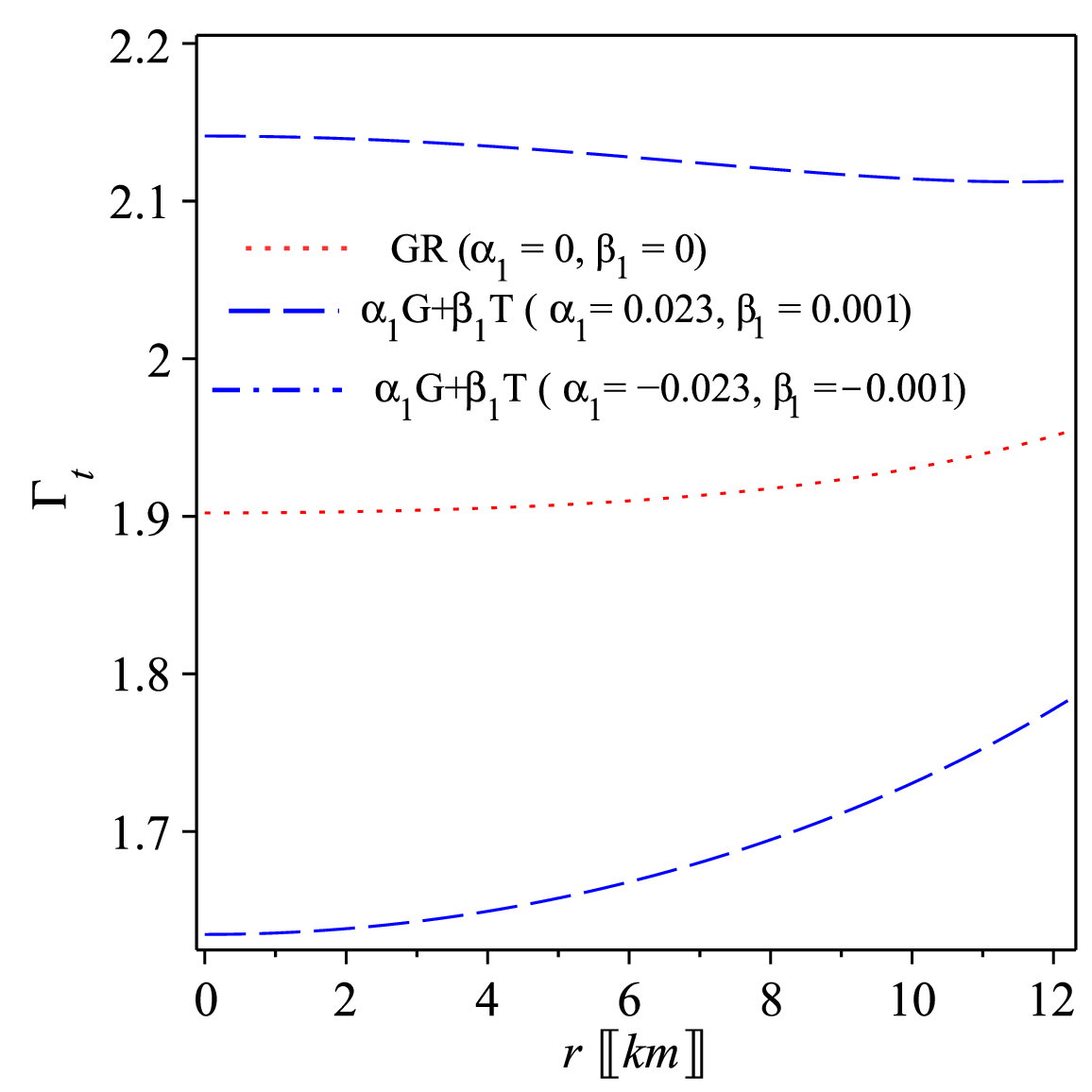}}
\caption{Variation of the adiabatic indices calculated using Eq.~\eqref{eq:adiabatic} for  \textit{U1724}.}
\label{Fig:Adiab}
\end{figure}

Equilibrium is marginal when $\Gamma = \gamma$, while stability against perturbations yields $\Gamma > \gamma$~\cite{chan1993dynamical,1975A&A....38...51H}.
Based on the field equations given in Eq.~\eqref{sol} (Appendix~A) and the gradient expressions from Appendix~C, Eqs.~\eqref{eq:pr_grad}--\eqref{eq:pt_grad}, the analysis indicates that our model produces a stable anisotropic description of the pulsar \textit{U1724}.
This stability persists for both parameter sets $\alpha_1$ and $\beta_1$, as illustrated in Fig.~\ref{Fig:Adiab}.

The state of hydrostatic equilibrium for the current configuration is determined by employing the TolmanOppenheimer-Volkoff equation, suitably modified within  $f(\mathcal{R}, \mathcal{G}, \mathcal{T}) = \mathcal{R} + \alpha\, \mathcal{G} + \beta\, \mathcal{T}$ gravity, expressed as:
\begin{equation}\label{eq:RS_TOV}
{\mathit F_a}+{\mathit F_g}+{\mathit F_h}+{\mathit F_G=0}\,.
\end{equation}

\begin{figure}
\centering
\subfigure[~TOV ($\alpha_1=\beta_1=0$, GR)]{\label{fig:GRTOV}\includegraphics[scale=0.23]{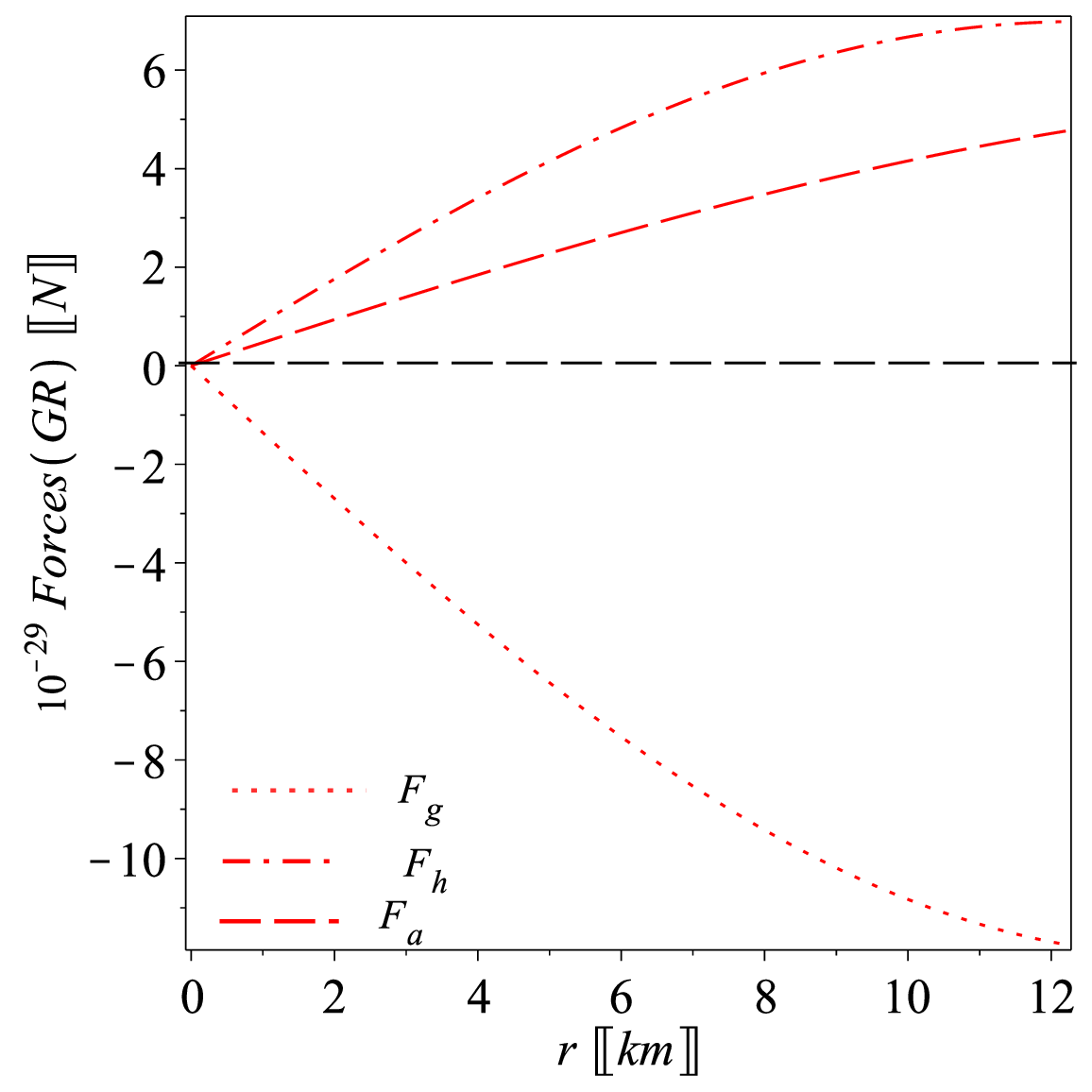}}\hspace{0.2cm}
\subfigure[~TOV  ($\alpha_1>0$, $\beta_1>0$, $\left( \mathcal{R,G,T} \right)$) ]{\label{fig:FRp}\includegraphics[scale=0.23]{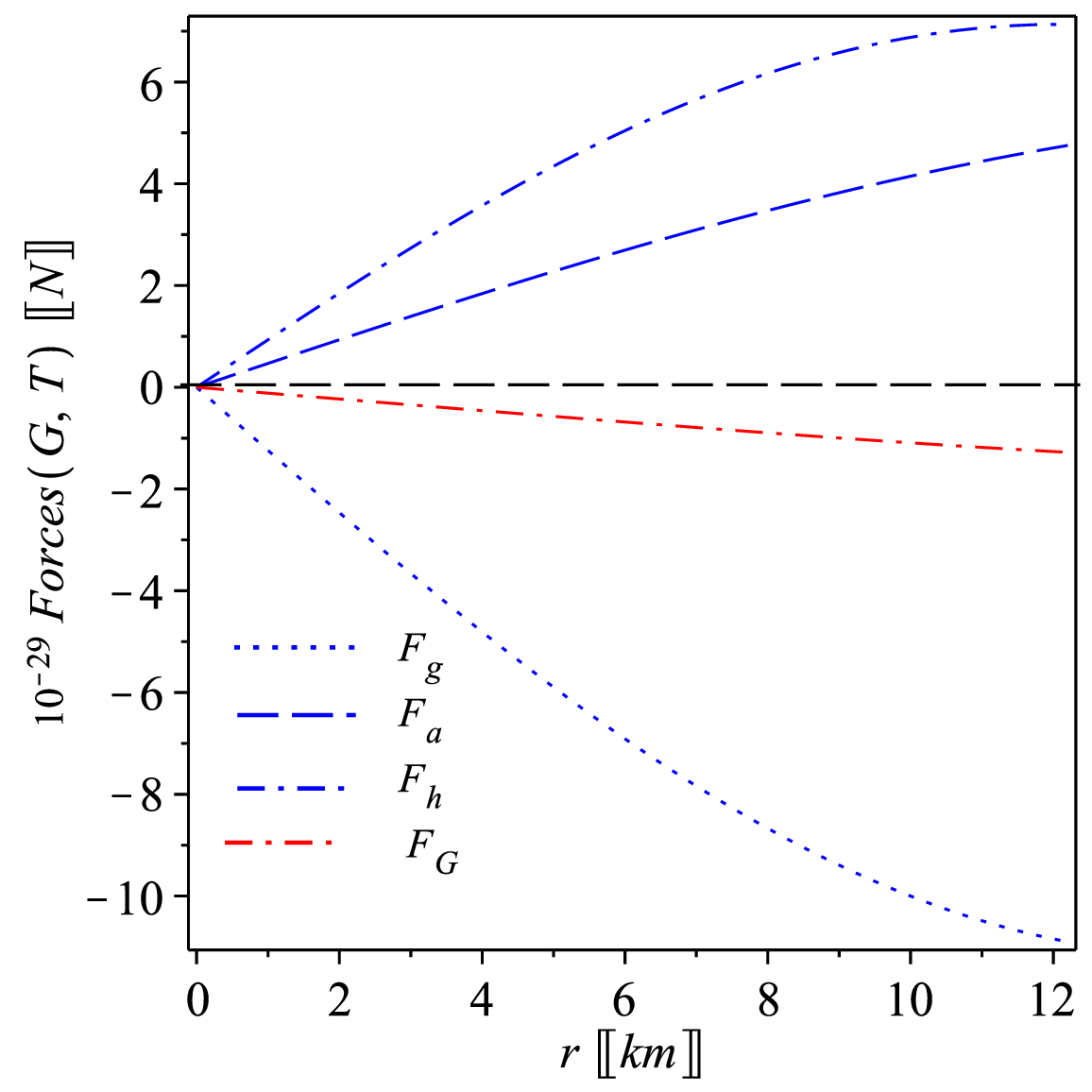}}\hspace{0.3cm}
\subfigure[~TOV  ($\alpha_1<0$, $\beta_1<0$, $\left( \mathcal{R,G,T} \right)$)]{\label{fig:FRn}\includegraphics[scale=0.23]{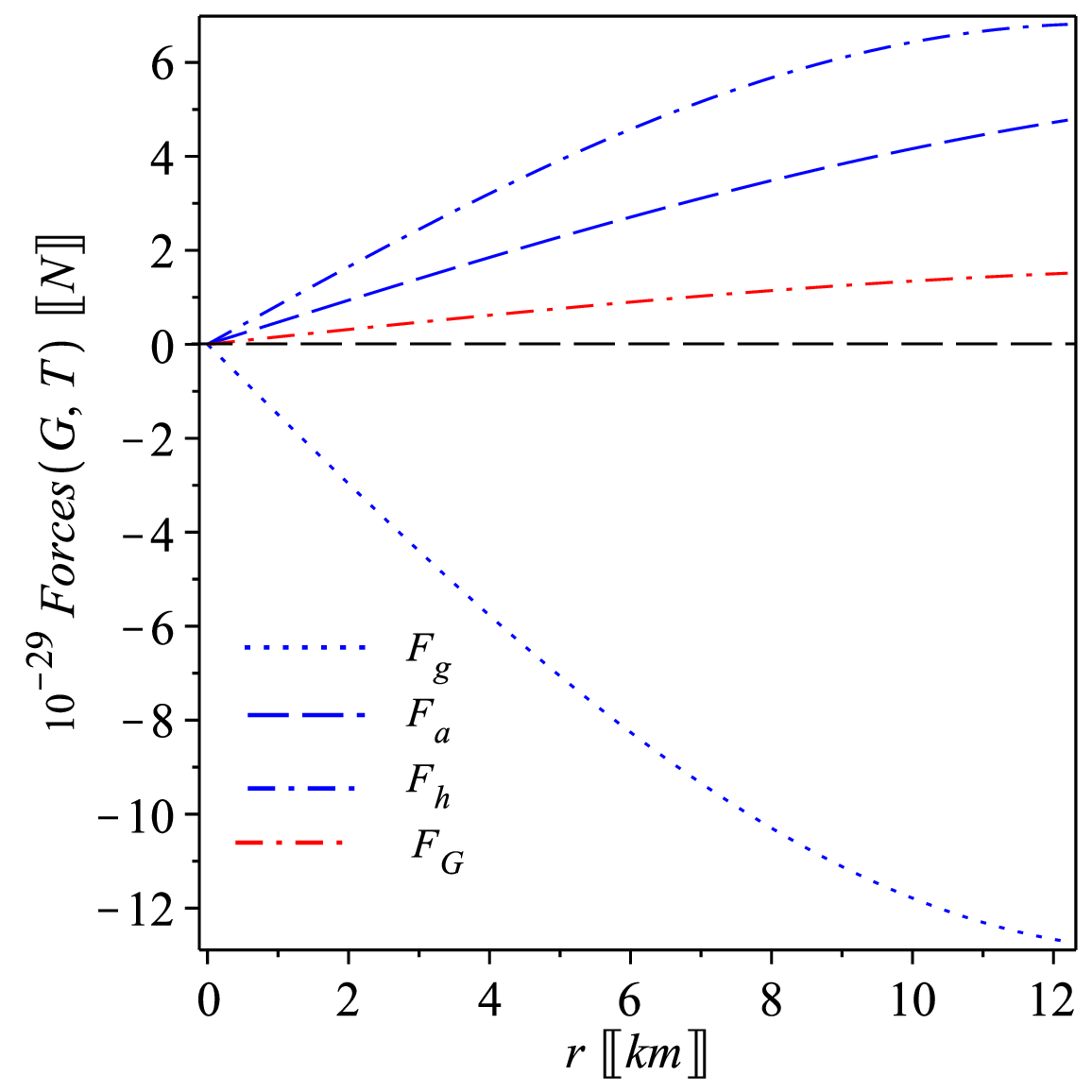}}
\caption{TOV constraints given by Eq.~\eqref{eq:Forces}.}
\label{Fig:TOV}
\end{figure}

Their explicit forms are given by:

\begin{eqnarray}\label{eq:Forces}
  {\mathit F_a} =&\frac{ 2{\mathit  \Delta}}{\mathit r} ,\qquad
  {\mathit F_g} = -\frac{{\mathit  M_g}}{r}({\mathit  \rho c^2}+{\mathit p_r})e^{\epsilon/2} ,\qquad\nonumber\\
  {  F_h} =&-{\mathit  p'_t} ,\qquad
  { F_R} =[\frac{\beta_1}{3(1+\beta_1)}+ 2\alpha_1]({  c^2 \rho}'-{ p}'_r-2{  p}'_{t}).
\end{eqnarray}
The quantity $\epsilon := \mu - \nu$ is introduced in the formulation of $F_g$.
The gravitational mass $M_g$ of an isolated system, defined on the three-space ${\mathit V}$ $(t = \text{constant})$, is obtained through the Tolman mass expression~\citep{1930PhRv...35..896T}$\,$ generalized to the $f(\mathcal{R}, \mathcal{G}, \mathcal{T})$ gravitational framework.
\begin{eqnarray}\label{eq:grav_mass}
{\mathit M_g(r)}&=&{\int_{\mathit V}}\Big(\mathbb{{\bar{\mathfrak T}}}{^r}{_r}+\mathbb{\bar{\mathfrak T}}{^\theta}{_\theta}+\bar{\mathfrak{T}}{^\phi}{_\phi}-\bar{\mathfrak{T}}{^t}{_t}\Big)\sqrt{-g}\,dV\nonumber\\
&=&e^{-\mu}(e^{\mu/2})'  e^{\nu/2} r =\frac{1}{2} r \mu' e^{-\epsilon/2}.
\end{eqnarray}
The gravitational force is obtained as
${\mathit F_g} = -\dfrac{b_0\, r}{R_s^2}\,(\rho c^2 + p_r)$.
Based on the field equations of Eq.~\eqref{sol} and the corresponding pressure gradients given by Eqs.~\eqref{eq:pr_grad}--\eqref{eq:pt_grad}, the linear $f(\mathcal{R}, \mathcal{G}, \mathcal{T})$ framework is shown to satisfy the equilibrium relation \eqref{eq:RS_TOV}.
This outcome confirms that the pulsar \textit{U1724} remains stable for both parameter sets $\alpha_1$, $\beta_1$, and in the general relativistic limit $\alpha_1 = \beta_1 = 0$, as illustrated in Fig.~\ref{Fig:TOV}.

When the anisotropy parameter becomes positive ($\Delta > 0$), a repulsive contribution emerges that acts outward, opposing the inward-directed gravitational pull.
This outward pressure effectively enlarges the stellar configuration, enabling the star to sustain a higher mass and compactness without losing stability.

Figures~\ref{Fig:TOV}\subref{fig:FRp} and~\subref{fig:FRn} reveal that the additional term introduced by the linear form of the $f(\mathcal{R}, \mathcal{G}, \mathcal{T})$ gravity theory, defined as $\mathcal{R} + \alpha\,\mathcal{G} + \beta\,\mathcal{T}$, modifies the internal balance depending on the signs of $\alpha_1$ and $\beta_1$.
For positive values of these parameters, the correction tends to enhance the collapse, while for negative values, it exerts a counteracting influence that partially offsets gravity.

This trend aligns with the results derived previously for the pulsar \textit{U1724} in Subsection~\ref{Sec:obs_const}, corresponding to the parameter sets
$\{\alpha_1 = -0.023,\, \beta_1 = -0.001,\, C = 0.451,\, M \approx 2.21\,M_\odot,\, R_s \approx 13.2~\text{km}\}$
and
$\{\alpha_1 = 0.023,\, \beta_1 = 0.001,\, C = 0.434,\, M \approx 2.199\,M_\odot,\, R_s \approx 13.22~\text{km}\}$.

\section{Mass-Radius Characteristics and the Equation of State
}\label{Sec:EoS_MR}

The physical composition of neutron star interiors remains an open question in modern astrophysics, as the matter in their cores reaches densities several times higher than the nuclear saturation threshold far beyond the range that can be reproduced in laboratory conditions.
Despite the uncertainty in the true equation of state (EoS) describing such dense matter, measurements of stellar masses and radii provide valuable empirical bounds, allowing certain models to be constrained or excluded.
Consequently, observational data serve to narrow the possible mass-radius relations associated with a given EoS.

In this work, rather than prescribing a specific EoS, we employ the metric functions introduced in Eqs.~\eqref{nu} and~\eqref{mu}.
These expressions naturally connect the energy density and pressures, giving rise to effective EoS relations as shown in Eq.~\eqref{eq:KB_EoS2}, which remain most accurate near the stellar center due to their power-series construction.
To test their applicability, we compute the evolution of the pressure and density from the core to the stellar boundary for both positive and negative choices of the parameters $\alpha_1$ and $\beta_1$.

\begin{figure}[th!]
\centering
\subfigure[~EoS along the radial axis]{\label{fig:RfEoSp}\includegraphics[scale=0.45]{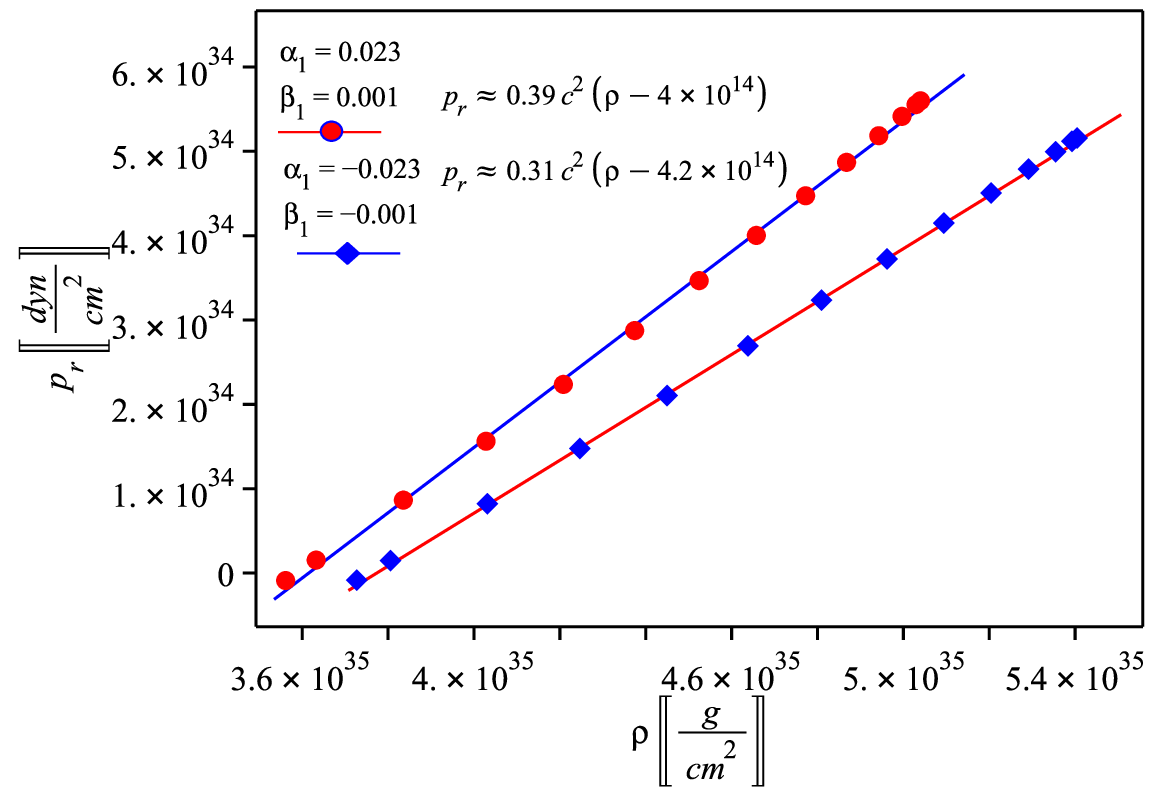}}
\subfigure[~EoS along  the tangential axis]{\label{fig:TEoSn}\includegraphics[scale=0.45]{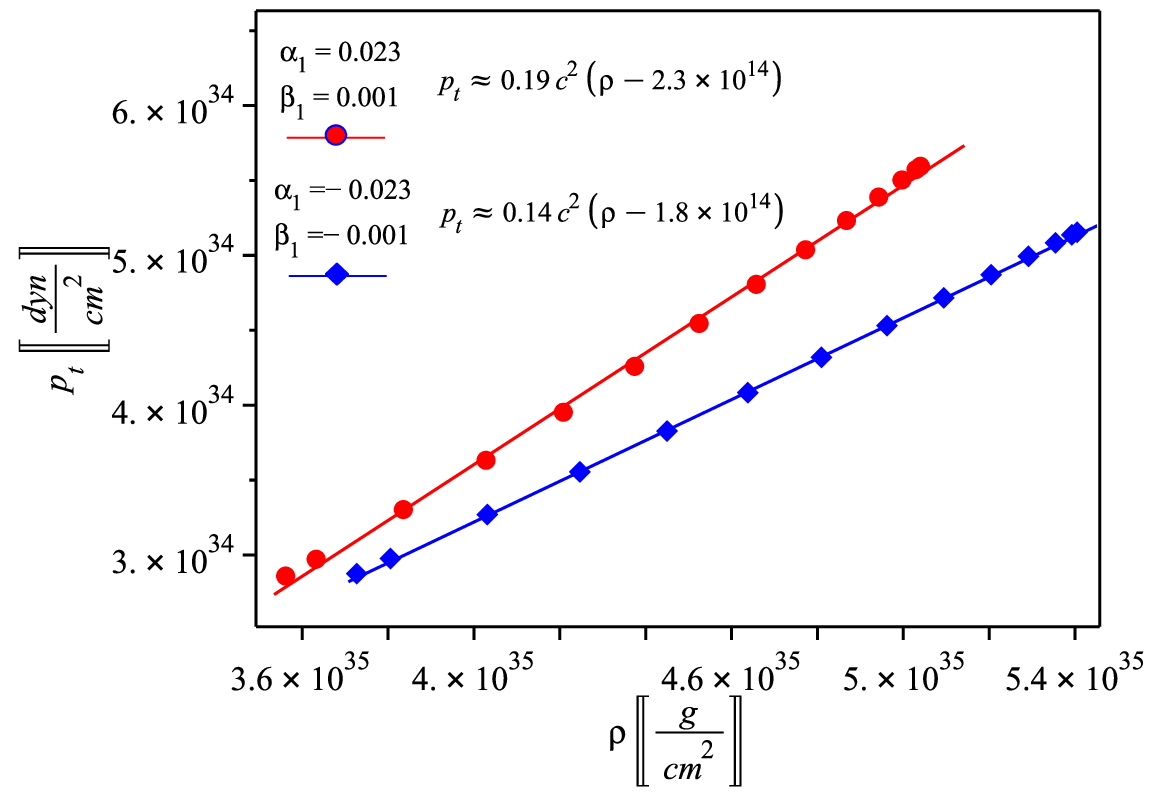}}
\caption{Best-fitting equations of state (EoSs) for \textit{U1724}.
\subref{fig:RfEoSp}~Using Eqs.~\eqref{sol}, a series of data points connecting $\rho$, $p_r$, and $p_t$  are derived for $\alpha_1 = 0.023$ and $\beta_1 = 0.001$.}
\label{Fig:EoS}
\end{figure}

Adopting the numerical parameters reported in Subsection~\ref{Sec:obs_const} for the pulsar \textit{U1724}, together with the linear $f(\mathcal{R}, \mathcal{G}, \mathcal{T}) = \mathcal{R} + \alpha\,\mathcal{G} + \beta\,\mathcal{T}$ field equations [Eqs.~\eqref{sol}], we generate the corresponding sequences, which are illustrated in Fig.~\ref{Fig:EoS}.

For negative $\alpha_1$ and $\beta_1$, the linear approximation slightly modifies the sound-speed profiles and surface density,
whereas for positive parameters, the results almost coincide with the induced EoS relations.
This suggests that in the latter configuration, a second-order polynomial of the form
$p_{r,t}(\rho) \approx \tilde{c}_0 + \tilde{c}_1\rho + \tilde{c}_2\rho^2$
may yield a more accurate, non-linear description of the matter distribution.

\begin{figure*}[t]
\centering
\subfigure[~$C$]{\label{fig:Comp}\includegraphics[scale=0.4]{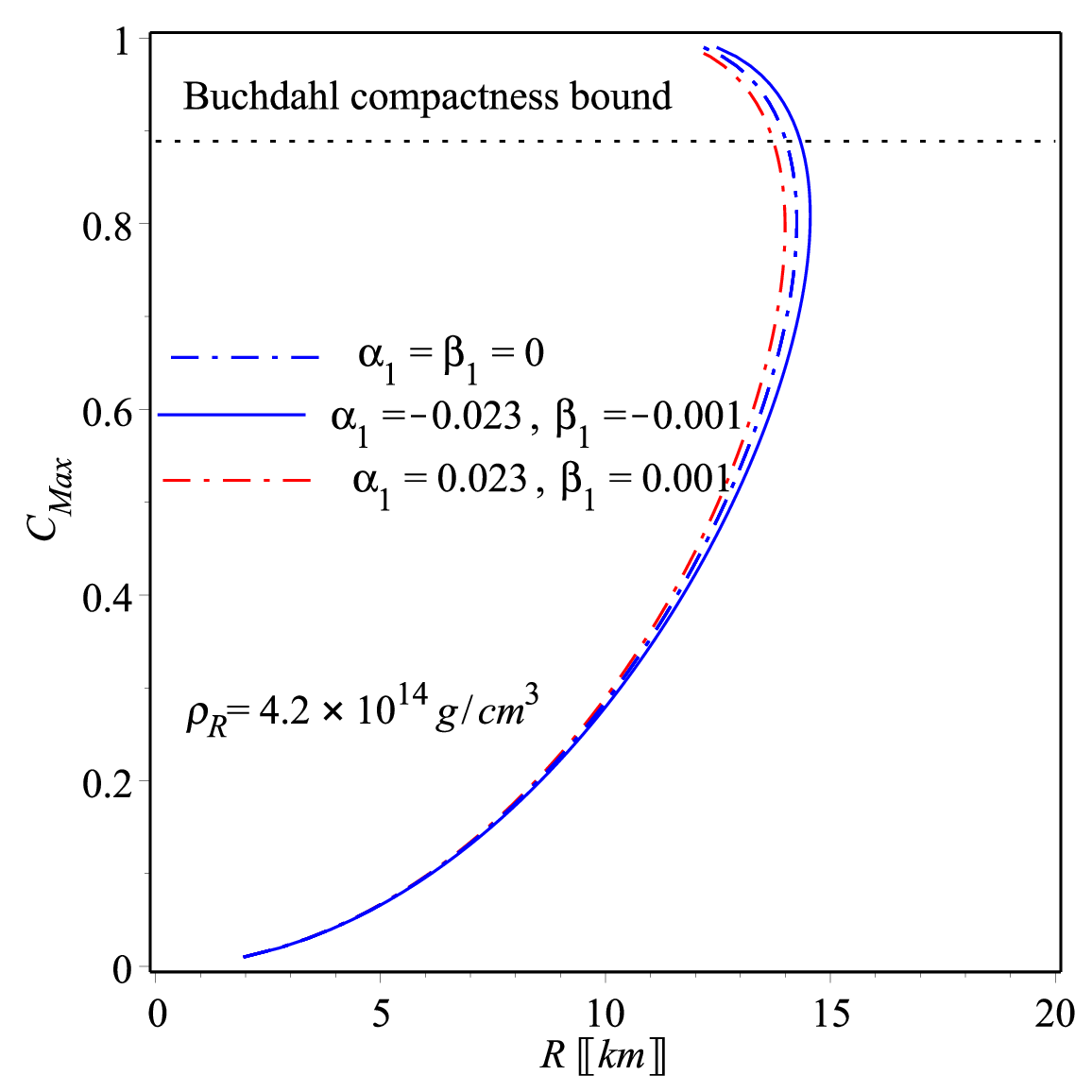}}\hspace{0.5cm}
\subfigure[~$M-R$ diagram]{\label{fig:MR}\includegraphics[scale=.4]{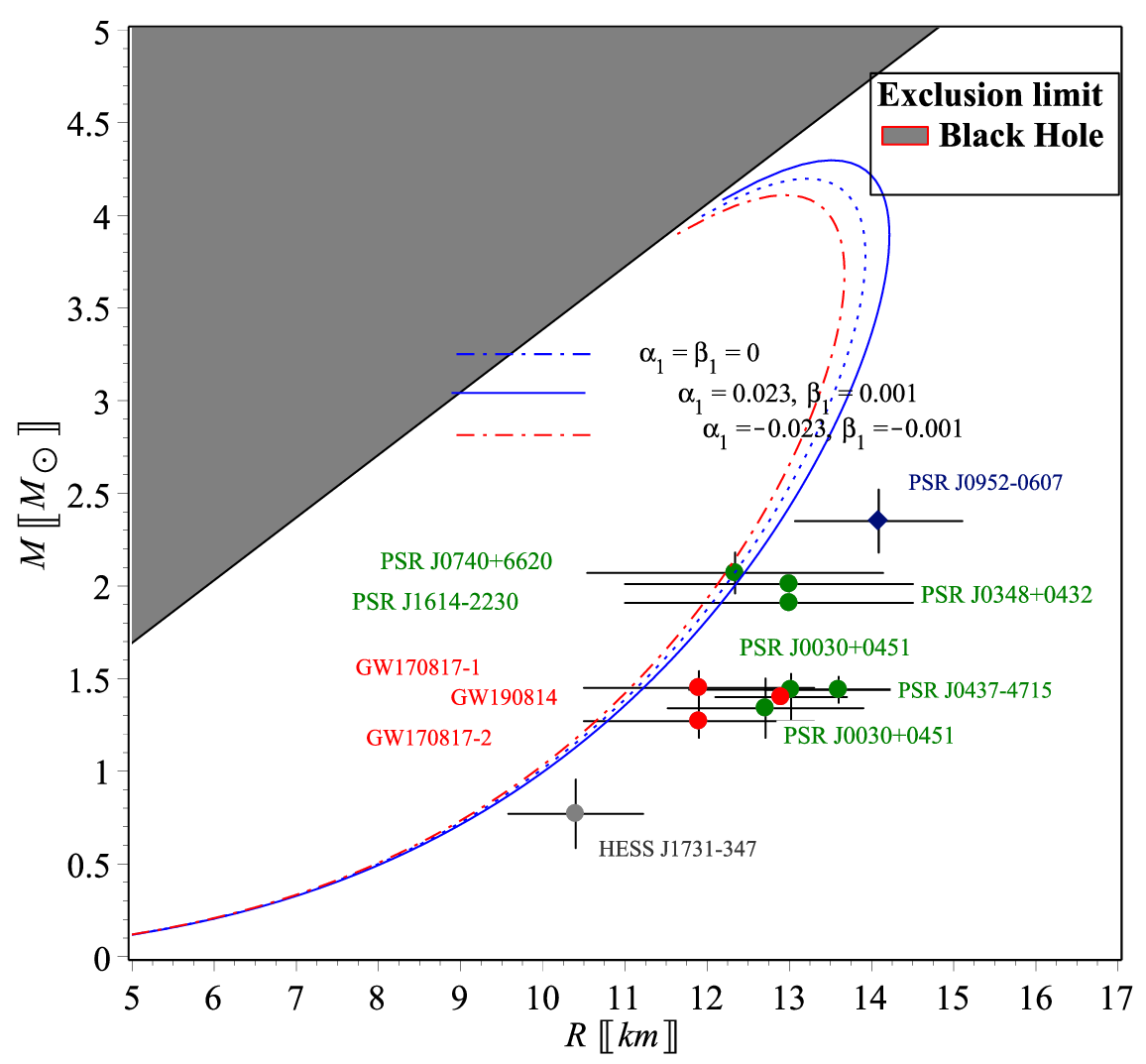}}
\caption{\subref{fig:Comp}~Compactness-radius (CR) diagram.
The horizontal solid and dashed lines indicate the theoretical compactness bounds corresponding to the black hole (BH) ($C = 1$) and Buchdahl ($C = 8/9$) limits, respectively.
The CR profiles are obtained from the best-fitting equations of state shown in Fig.~\ref{Fig:EoS}, together with additional curves generated for surface densities of $\rho_s = 3.2 \times 10^{14}$~g/cm$^3$ and $\rho_s = 2.9 \times 10^{14}$~g/cm$^3$.
When $\epsilon_1 = 0.03$, the $C$ approaches the BH threshold ($C \rightarrow 1$), resembling the trend found in general relativity~\citep{Roupas:2020mvs}.
For $\epsilon_1 = -0.03$, however, the maximum compactness remains below the Buchdahl upper bound, emphasizing a distinct behavior of the quadratic gravity model compared to GR. \subref{fig:MR}~Mass-radius (MR) diagram.
The shaded regions mark the boundaries imposed by the black-hole and Buchdahl limits (gray and orange zones, respectively).
For positive $\epsilon$, the MR trajectories extend beyond the Buchdahl constraint, forming nonphysical segments where the maximum mass lies near this boundary (solid markers).
For negative $\epsilon$, on the other hand, the sequences remain entirely within the Buchdahl bound, and the corresponding maxima are represented by open markers.
}
\label{Fig:CompMR}
\end{figure*}

Buchdahl established a fundamental restriction on the compactness of static, stable stars, setting an upper threshold of $C < 8/9$~\cite{PhysRev.116.1027}.
This classical result was originally formulated within general relativity under the assumptions of isotropy (or weak anisotropy) and spherical symmetry.
Later studies have shown that once any of these idealized conditions are relaxed, the compactness can exceed Buchdahl's limit.

In more realistic situations, where strong pressure anisotropy is present, models within general relativity allow the compactness to approach the black-hole boundary ($C \rightarrow 1$)~\cite{Alho:2022bki}.
Even so, other physical requirements, such as energy and stability constraints, typically impose stricter upper limits~\cite{Alho:2021sli,Roupas:2020mvs,Raposo:2018rjn,Cardoso:2019rvt}.
Comparable findings have been reported in theories involving non-minimal interactions between matter and geometry~\citep{Nashed:2022zyi,ElHanafy:2022kjl,2023arXiv230514953E}.
For this reason, reexamining the compactness constraint within the framework of the linear $f(\mathcal{R}, \mathcal{G}, \mathcal{T})$ gravity model considered here is particularly meaningful.

In this work, the compactness constraint for the pulsar \textit{U1724} is examined within the framework of the linear $f(\mathcal{R}, \mathcal{G}, \mathcal{T})$ theory.
For the parameter combination $\alpha_1 = 0.023$, $\beta_1 = 0.001$, the upper compactness limit is found to be $C \lesssim 0.995$,
while for $\alpha_1 = -0.023$, $\beta_1 = -0.001$, it reaches $C \lesssim 0.999$.
Both estimates slightly modify the classical general relativistic Buchdahl bound.
A notable trend appears for negative $\alpha_1$ and $\beta_1$, where the permissible compactness increases.
This behavior supports the interpretation that, in the linear $f(\mathcal{R}, \mathcal{G}, \mathcal{T})$ model, an additional repulsive term emerges in the Tolman-Oppenheimer-Volkoff equilibrium equation,
partially offsetting gravity and thereby allowing configurations of greater mass and compactness.

Figure~\ref{Fig:CompMR}\subref{fig:MR} presents the mass-radius (MR) curves derived from the optimal equations of state obtained previously in this section, corresponding to both signs of the parameters $\alpha_1$ and $\beta_1$.
The stellar mass is evaluated through the matching relation~\eqref{eq:bo}, expressed as
\[
M = \frac{c^{2} R}{2 G}\left(1 - e^{-b_{2}}\right).
\]
When the parameters take the values $\alpha_1 = -0.023$ and $\beta_1 = -0.001$, with  $\rho_s = 4.2 \times 10^{14}$~g/cm$^3$, the configuration attains $M \simeq 4.24\,M_\odot$ at a radius of $R \simeq 13.45$~km.
For the opposite case, $\alpha_1 = 0.023$, $\beta_1 = 0.001$, and $\rho_s = 4.0 \times 10^{14}$~g/cm$^3$, the model predicts a slightly higher peak of $M \simeq 4.46\,M_\odot$ occurring at $R \simeq 14.05$~km.
These calculated values closely reproduce the physical characteristics observed for the pulsar \textit{U1724}.

\newpage

\section{Pulsars  data}
In addition to ${\textit U1724}$, a similar analysis is developed for other twenty one pulsars cover a wide range of masses from $0.8 M_\odot$ to heavy pulsars of mass $2.01 M_\odot$. In Table \ref{Table1}, we give the observed masses and radii for each associated with the corresponding model parameters \{$b_0$, $b_1$, $b_2$\} assuming that the  parameters $\alpha=-0.023$ and $\beta=-0.001$ because the values of this parameters are the one that reproduce radial velocity that is not violating the conjectured conformal upper limit.
\begin{table*}
\caption{Observed mass-radius of twenty pulsars and the corresponding model parameters ($\alpha=-0.023$,\, $\beta=-0.001$).}
\label{Table1}
\begin{tabular*}{\textwidth}{@{\extracolsep{\fill}}llcccccc@{}}
\hline
{{Pulsar}}   &Ref.& observed mass ($M_{\odot}$) &  observed radius [{km}]& estimated mass ($M_{\odot}$) &  {$b_0$}    & {$b_1$}     & {$b_2$}   \\
\hline
Her X-1 &\cite{Abubekerov:2008inw}         &  $0.85\pm 0.15$    &  $8.1\pm 0.41$   &$0.905$&  $-0.298$    & $0.323$    & $-0.34$     \\
RX J185635-3754 &\cite{Pons:2001px}     &  $0.9\pm 0.2$      &  $6$             &$0.949$&  $-0.369$    & $0.469$    & $-0.5$     \\
LMC X-4 &\cite{Rawls:2011jw}         &  $1.04\pm 0.09$    &  $8.301\pm 0.2$  &$1.103$&  $-0.33$    & $0.389$    & $-0.41$     \\
GW170817-2  &\cite{LIGOScientific:2018cki}     &  $1.27\pm 0.09$    &  $11.9\pm 1.4$   &$1.351$&  $-0.3$    & $0.33$    & $-0.347$     \\
EXO 1785-248 &\cite{Ozel:2008kb}    &  $1.3\pm 0.2$      &  $8.849\pm 0.4$  &$1.372$&  $-0.364$    & $0.459$    & $-0.489$     \\
PSR J0740+6620 &\cite{Raaijmakers:2019qny}  &  $1.34\pm 0.16$    &  $12.71\pm 1.19$ &$1.426$&  $-0.299$    & $0.325$    & $-0.342$     \\
M13 &\cite{Webb:2007tc}            &  $1.38\pm 0.2$     &  $9.95\pm 0.27$  &$1.459$&  $-0.351$    & $0.432$    & $-0.459$     \\
LIGO    &\cite{LIGOScientific:2020zkf}         &  $1.4$             &  $12.9\pm 0.8$   &$1.489$&  $-0.3$    & $0.335$    & $-0.353$     \\
X7  &\cite{Rybicki:2005id}             &  $1.4$             &  $14.5\pm 1.8$   &$1.492$&  $-0.284$    & $0.297$    & $-0.312$     \\
PSR J0037-4715 &\cite{Reardon:2015kba}   &  $1.44\pm 0.07$    &  $13.6\pm 0.9$   &$1.532$&  $-0.3$    & $0.326$    & $-0.344$     \\
PSR J0740+6620 &\cite{Miller:2019cac}  &  $1.44\pm 0.16$    &  $13.02\pm 1.24$ &$1.531$&  $-0.31$    & $0.341$    & $-0.36$     \\
GW170817-1  &\cite{LIGOScientific:2018cki}     &  $1.45\pm 0.09$    &  $11.9\pm 1.4$   &$1.539$&  $-0.325$    & $0.378$    & $-0.4$     \\
4U 1820-30 &\cite{Guver:2010td}      &  $1.46\pm 0.2$     &  $11.1\pm 1.8$   &$1.546$&  $-0.341$    & $0.41$    & $-0.433$     \\
Cen X-3 &\cite{Naik:2011qc}         &  $1.49\pm 0.49$    &  $9.178\pm 0.13$ &$1.566$&  $-0.388$    & $0.51$    & $-0.546$     \\
4U 1608-52  &\cite{1996IAUC.6331....1M}     &  $1.57\pm 0.3$     &  $9.8\pm 1.8$    &$1.651$&  $-0.385$    & $0.502$    & $-0.538$     \\
KS 1731-260 &\cite{Ozel:2008kb}     &  $1.61\pm 0.37$    &  $10\pm 2.2$     &$1.692$&  $-0.386$    & $0.505$    & $-0.541$     \\
EXO 1745-268  &\cite{Ozel:2008kb}   &  $1.65\pm 0.25$    &  $10.5\pm 1.8$   &$1.736$&  $-0.38$    & $0.492$    & $-0.53$     \\
Vela X-1 &\cite{Rawls:2011jw}        &  $1.77\pm 0.08$    &  $9.56\pm 0.08$  &$1.845$&  $-0.424$    & $0.547$    & $-0.51$     \\
4U 1724-207 &\cite{Ozel:2008kb}     &  $1.81\pm 0.27$    &  $12.2\pm 1.4$   &$1.909$&  $-0.367$    & $0.463$    & $-0.494$     \\
SAX J1748.9-2021 &\cite{Ozel:2008kb} &  $1.81\pm 0.3$     &  $11.7\pm 1.7$   &$1.906$&  $-0.376$    & $0.484$    & $-0.518$     \\
PSR J1614-2230 &\cite{Demorest:2010bx}  &  $1.97\pm 0.04$    &  $13\pm 2$       &$2.076$&  $-0.371$    & $0.474$    & $-0.506$     \\
PSR J0348+0432 &\cite{Antoniadis:2013pzd}   &  $2.01\pm 0.04$    &  $13\pm 2$       &$2.117$&  $-0.376$    & $0.484$    & $-0.517$     \\
\hline
\end{tabular*}
\end{table*}
\newpage
\section{Conclusion}\label{Sec:Conclusion}

The investigation into stellar phenomena within the framework of modified gravitational theories, specifically those encompassing the functional form $\left( \mathcal{R,G,T} \right)$, has yielded intriguing insights into the nature of gravity and its effects on celestial bodies. These alternative theories of gravity have emerged as a response to certain limitations and open questions posed by the traditional GR.

The $\left( \mathcal{R,G,T} \right)$ class of modified gravitational theories extends beyond the confines of Einstein's field equations by introducing additional terms that depend on the curvature scalar $\mathcal{R}$, the Gauss-Bonnet scalar $\mathcal{G}$, and the trace of the energy-momentum tensor $\mathcal{T}$. This comprehensive approach acknowledges the intricate interplay between spacetime curvature and the distribution of matter and energy, providing a nuanced perspective on gravitational dynamics.

One of the key motivations behind exploring $\left( \mathcal{R,G,T} \right)$ theories is their potential to address long-standing challenges and discrepancies observed in various astrophysical and cosmological phenomena. These theories offer the flexibility to explain cosmic acceleration, dark matter, and dark energy without resorting to ad hoc concepts. By altering the gravitational field equations, researchers can emulate the effects of dark energy through modifications to gravity itself, thus potentially resolving some of the most perplexing enigmas of the universe.

In the context of stellar studies, $\left( \mathcal{R,G,T} \right)$ theories introduce novel aspects to the behavior of compact objects, such as NSs and white dwarfs. The modified field equations can give rise to distinct predictions for the internal structure, stability, and evolution of these celestial bodies. This prompts a comprehensive examination of the equations of state, hydrostatic equilibrium, and energy-momentum conservation within these modified gravitational frameworks. This was our aim in this study to apply the field equations of $\left( \mathcal{R,G,T} \right)$ to a spherically symmetric spacetime. Using such philosophy we got a system of non-linear differential equations, three equations, in five unknowns, two of the ansatzs of the metric potentials, the energy-density $\rho$, radial and tangential pressures, $p_r$ and $p_t$ respectively. To close such system we  employ the customary form for the metric potential $g_{rr}$ in interior solutions. Furthermore, we introduce an adjustment by nullifying the contribution of the metric potential $g_{tt}$ to the anisotropy, in alignment with standard practices. Using  these constrains we succeeded to derive the explicate forms of  $\rho$, $p_r$ and $p_t$.

One particularly intriguing avenue of exploration is the effect of f(R,G,T) modifications on the maximum mass of neutron stars. General Relativity predicts an upper limit to the mass of neutron stars, known as the Tolman-Oppenheimer-Volkoff (TOV) limit. However, in $\left( \mathcal{R,G,T} \right)$ theories, this limit may be altered due to the modified gravitational dynamics. Understanding how these modifications influence neutron star properties can have profound implications for our comprehension of stellar stability and the ultimate fate of massive stars. Through the use of the junction condition, matching the interior solution on the boundary of the star with the exterior solution given by Schwarzschild solution and by imposing the vanishing of the radial pressure on the surface of the pulsar,   we derived the values of the parameters  appeared in this interior solution. Through the  use of the limits on $R$ and $M$ of   ${\textit U1724}$ we determined the  parameters of the linear form of $\left( \mathcal{R,G,T} \right)=\mathcal{R}+ \alpha_1 \mathcal{G} + \beta_1 \mathcal{T}$ to be at most $\alpha_1=\pm0.023$ and $\beta_1=\pm0.001$.

By using the numerical values of the linear form of $\left( \mathcal{R,G,T} \right)=\mathcal{R}+ \alpha_1 \mathcal{G} + \beta_1 \mathcal{T}$ we assured the reliability of the derived model through several tests on matter and geometric sectors. We show that the maximum densities at the core of the star ${\textit U1724}$ are ${\rho_\text{core}\approx 6.02\times 10^{14}}$ g/cm$^{3}$ for for negative values of the model parameters $\alpha_1$ and $\beta_1$ and ${\rho_\text{core}\approx 5.61\times 10^{14}}$ g/cm$^{3}$ for positive values of the model parameters $\alpha_1$ and $\beta_1$.  Interestingly, we have determined that the maximum squared sound speeds within the core of the neutron star are $v_r{}^2=0.24c^2$ in the radial direction and $v_t{}^2=0.15c^2$ in the tangential direction. These values adhere to the proposed conformal bound on sound speed within the core and hold true throughout the entire interior of the neutron star, in contrast to the case in GR. This stands in contrast to approaches involving hadronic equations of state (EoS) or the Gaussian process non-parametric EoS (utilizing NICER+XMM observations of pulsar ${\textit U1724}$), where there is a significant violation of the speed of sound.



Despite the model's capacity to reduce the speed of sound compared to cases involving GR or the linear form of $\left( \mathcal{R,G,T} \right)$ gravity with positive value of the model parameter, the sound speed satisfies the conjectured upper bound of the maximum sound speed $c_s^2=c^2/3$. This outcome ensures that a matter-geometry non-minimal coupling scenario, as proposed by \cite{ElHanafy:2022kjl,2023arXiv230514953E,Nashed:2023pxd},   offer a more suitable framework for addressing the issue of the conjectured sound speed.
\appendix
\section{The form of $\rho$, $p_r$ and $p_t$}\label{Sec:App_1}
Using Eqs. (\ref{nu}) and (\ref{mu}) in Eqs. (\ref{9}), (\ref{10}), and (\ref{11}) we get the energy-density, radial and tangential pressures in the form:
\begin{align}\label{sol}
&\rho  =\frac{128{b_0}^{2} }{c^2{R}^{16}} \bigg\{{\frac {55}{8}}b_1 \left( \frac{{\kappa}^{2}}2+\beta \right) {R}^{2}\alpha{r}^{10}{b_0}^{12}-{\frac {9}{8}}b_1 \left( \frac{{\kappa}^{2}}2+\beta \right) \alpha{r}^{12}{b_0}^{14 }+\frac{3}4{R}^{4} \left[ b_2 \left( \frac{3{\kappa}^{2}}{16}+\beta \right) {R}^{4}-{\frac {47 }{2}}b_1 \left( \frac{{\kappa}^{2}}2+\beta \right) \alpha \right] {r}^{8}{b_0}^{10}\nonumber\\
&-{\frac {37}{12}} \left[ b_2 \left( \frac{3}{16}{\kappa}^{2}+\beta \right) {R}^{4}-{\frac {585}{74}}b_1 \left( \frac{1}2{\kappa}^{2}+\beta \right) \alpha \right] {R}^ {6}{r}^{6}{b_0}^{8}+{\frac {29{R}^{8}{r}^{4}{b_0}^{6}}{6}} \bigg[ b_2 \left( \frac{3}{16}{\kappa}^{2}+\beta \right) {R}^{4}-{\frac {51b_1}{464}}  \bigg\{  \left( \beta+{\frac {9}{68}}{\kappa}^{2} \right) {r}^{2}+ {\frac {616}{17}} \nonumber\\
&\times\alpha\left( \frac{{\kappa}^{2}}2+\beta \right)  \bigg\}  \bigg] -\frac{7}2 \left[ b_2  \left( \frac{3{\kappa}^{2}}{16}+\beta \right) {R}^{4}-{\frac {157b_1}{336}}  \left\{  \left( \beta+{\frac {21{\kappa}^{2}}{157}} \right) {r}^{2}+{\frac {792}{157}} \left( \frac{{\kappa}^{2}}2+\beta \right) \alpha \right\}  \right] {R}^{10}{r}^{2}{b_0}^{4}+ \bigg\{ b_2 \left( \frac{3{\kappa}^{2}}{16}+\beta \right) {R}^{4}\nonumber\\
&-{ \frac {55b_1}{32}} \left[  \left( {\frac {3}{22}}{\kappa}^{2}+\beta \right) {r}^{2}+{\frac {48}{55}} \left( \frac{1}2{\kappa}^{2}+\beta \right) \alpha \right]  \bigg\} {Rn}^{12}{b_0}^{2}+{ \frac {21b_1{R}^{14} }{32}}\left( \beta+\frac{1}7{\kappa}^{2} \right)  \bigg\} \bigg[{\kappa}^{4}+10{ \kappa}^{2}\beta+16{\beta}^{2} \bigg] ^{-1}\nonumber\\
& \left( b_1+2b_2{b_0}^{2}{Rn}^{2}-2b_2{b_0}^{4}{r}^{2} \right) ^{-1}\,,\\
&p_r=-8 \bigg\{110b_1 \left(\frac{{ \kappa}^{2}}2+\beta \right) {R}^{2}\alpha{r}^{10} {b_0}^{12} -18b_1 \left(\frac{{ \kappa}^{2}}2+\beta \right) \alpha{r}^{12}{b_0}^{14}-4{Rn}^{4} \bigg\{ b_2 \left( \beta-\frac{{\kappa}^ {2}}{16} \right) {R}^{4}+{\frac {141}{2}}b_1 \left( \frac{{ \kappa}^{2}}2+\beta \right) \alpha \bigg\} {r}^{8}{b_0}^{10}\nonumber\\
&+{ \frac {44}{3}}{R}^{6} \left[ b_2 \left(\beta  -{\frac {15}{176}} {\kappa}^{2}\right) {R}^{4}+{\frac {585}{22}}b_1 \left( \frac{{\kappa}^{2}}2+\beta \right) \alpha \right] {r}^{6}{b_0}^{8}-{\frac {56{R}^{8 }{r}^{4}{b_0}^{6}}{3}} \bigg\{ b_2 \left(\beta -{\frac {15}{112}} {\kappa}^{2} \right) {R}^{4}-\frac{3b_1}{16} \bigg[ \left( {\frac {3}{28}}{\kappa}^{2}+\beta \right) {r}^{2}\nonumber\\
&-88 \left( \frac{{\kappa}^{2}}2+\beta \right) \alpha \bigg]  \bigg\}+8{R}^{10} \left( b_2 \left( \beta-{ \frac {5}{16}}{\kappa}^{2} \right) {R}^{4}-{\frac {59}{48}} \left(  \left( \beta+{\frac {6}{59}}{\kappa}^{2} \right) {r}^{2}-{ \frac {792}{59}} \left( \frac{{\kappa}^{2}}2+\beta \right) \alpha \right) b_1 \right) {r}^{2}{b_0}^{4}+ {R}^{12}{b_0}^{ 2}\bigg\{ {R}^{4}b_2{\kappa}^{2}\nonumber\\
&+\frac{17}2b_1 \left(  \left( \beta+{\frac {3} {34}}{\kappa}^{2} \right) {r}^{2}-{\frac {48}{17}} \left( \frac{{ \kappa}^{2}}2+\beta \right) \alpha \right)  \bigg\} -\frac{3}2{R}^{14}b_1\beta \bigg\} {b_0}^{2}{R}^{-16} \left( {\kappa}^{4}+10{\kappa}^{2}\beta+16{\beta}^{2} \right) ^{- 1} \left( b_1+2b_2{b_0}^{2}{R}^{2}-2b_2{b_0}^{4}{r}^{2} \right) ^{-1}\,,\\
&p_t  =-8 \bigg\{ 110b_1  \left( \frac{{\kappa}^{2}}2+\beta \right)  {R}^{2}\alpha{r}^{10} {b_0}^{12}-18b_1  \left( \frac{{\kappa}^{2}}2+\beta \right)  \alpha{r}^{12}{b_0}^{14}+2 \left( {R}^{4}b_2-141b_1\alpha \right)  \left( \frac{{\kappa}^{2}}2+\beta \right)  {R}^{4}{r}^{8}{b_0}^{10}-{\frac {22{r}^{6}}{3}}{ b_0}^{8}{R}^{6} \bigg\{ b_2 \nonumber\\
&\times\left( \beta+{ \frac {6}{11}}{\kappa}^{2} \right) {R}^{4}-{\frac {585}{11}}b_1  \left( \frac{{\kappa}^{2}}2+\beta \right)  \alpha \bigg\} +{\frac {28}{3}} \left\{ b_2 \left( {\frac {9}{ 14}}{\kappa}^{2}+\beta \right) {R}^{4}+{\frac {3}{56}}b_1 \left[ {r}^{2}\beta-616 \left( \frac{{\kappa}^{2}}2+\beta \right)  \alpha \right]  \right\} {R}^{8}{r}^{4}{b_0}^{6} \nonumber\\
& -4{R}^{10}{r}^{2 }{b_0}^{4}\bigg\{ b_2\left( \beta+{\kappa}^{2} \right) {R}^{4}+{ \frac {11}{24}} \left( {r}^{2}\beta-72 \left( \frac{{\kappa}^{2}}2+\beta \right) \alpha \right) b_1 \bigg\} + \left\{ {R}^{4}b_2{\kappa}^{2}+5/2b_1 \left[ {r}^{ 2}\beta-{\frac {48}{5}} \left( \frac{{\kappa}^{2}}2+\beta \right)  \alpha \right]  \right\} {R}^{12}{b_0}^{2} \nonumber\\
& -\frac{3}2{R}^{14}b_1 \beta \bigg\} {b_0}^{2}{R}^{-16} \left( {\kappa}^{4}+10{ \kappa}^{2}\beta+16{\beta}^{2} \right) ^{-1} \left( b_1+2b_2{b_0}^{2}{R}^{2}-2b_2{b_0}^{4}{r}^{2} \right) ^{-1}\,.
\end{align}
The metric ansatz is selected so that all physical quantities remain finite and well behaved throughout the stellar interior.
In our analysis, the observational estimates of the mass and radius of the pulsar $U1724$ are employed to constrain the parameters $\alpha$ and $\beta$ appearing in the model.

For convenience, it is useful to introduce dimensionless combinations of these constants.
We define
\[
\alpha_{1} \equiv \frac{\alpha}{R^{2}},
\]
where the scale $R$ is chosen to match the radius of a typical neutron star, taken here as $\ell = 10~\text{km}$.
In a similar way, we introduce
\[
\beta_{1} \equiv \frac{\beta}{\kappa^{2}} .
\]

Substituting the functions given in Eqs.~\eqref{mu} and \eqref{nu} into the system of equations \eqref{9}--\eqref{11} leads to the following set of expressions:

\begin{align}\label{sol}
&  \rho=-{b_0}^{2}\bigg\{  252{R}^{12}b_2{b_0}^ {4}{r}^{2}-348{R}^{10}b_2{b_0}^{6}{r}^{4}+222{R}^{8}b_2{b_0}^{8}{r}^{6}-54{R}^{6}b_2{b_0}^{10}{r} ^{8}-384{R}^{14}b_2{b_0}^{2}\beta_1+90{R}^{12}{b_0}^{2}b_1{r}^{2}+27{R}^{8}{b_0}^{6}{r}^{6}b_1\nonumber\\
&-84{R}^{10}{b_0}^{4}{r}^{4}b_1-252{R}^{14}b_1\beta_1-72{R}^{14}b_2{b_0}^{2}+432{b_0}^{14}b_1{r}^{12}\alpha_1{L}^{2}\beta_1
-3168{R}^{10}{b_0}^{4}b_1\alpha_1{L}^{2}\beta_1{r}^{2}+7392{R}^ {8}{b_0}^{6}b_1\alpha_1{L}^{2}\beta_1{r}^{4}\nonumber\\
&- 9360{R}^{6}{b_0}^{8}b_1\alpha_1{L}^{2}{r}^{6}\beta_1+6768{R}^{4}{b_0}^{10}b_1{r}^{8}\alpha_1{L}^ {2}\beta_1-2640{R}^{2}{b_0}^{12}b_1{r}^{10}\alpha_1{L}^{2}\beta_1+288{R}^{12}{b_0}^{2}b_1\alpha_1{L}^{2}+1344{R}^{12}b_2{b_0}^{4}\beta_1{r }^{2}\nonumber\\
&-1856{R}^{10}b_2{b_0}^{6}\beta_1{r}^{4}+1184 {R}^{8}b_2{b_0}^{8}{r}^{6}\beta_1-288{R}^{6}b_2{b_0}^{10}\beta_1{r}^{8}+660{R}^{12}{b_0}^{2}b_1{r}^{2}\beta_1+204{R}^{8}{b_0}^{6}b_1{r}^{6}\beta_1-628{R}^{10}{b_0}^{4}b_1{r}^{ 4}\beta_1\nonumber\\
&+ 216{b_0}^{14}b_1{r}^{12}\alpha_1{L}^{2}+576{R}^{ 12}{b_0}^{2}b_1\alpha_1{L}^{2}\beta_1-1584{R}^ {10}{b_0}^{4}b_1\alpha_1{L}^{2}{r}^{2}+3696{R}^{8} {b_0}^{6}b_1\alpha_1{L}^{2}{r}^{4}-4680{R}^{6}{b_0}^{8}b_1\alpha_1{L}^{2}{r}^{6}-36{R}^{14}b_1\nonumber\\
&+3384{R}^{4}{b_0}^{10}b_1{r}^{8}\alpha_1{L}^{2}-1320{R}^{2}{b_0 }^{12}b_1{r}^{10}\alpha_1{L}^{2} \bigg\}\bigg\{3{R}^{14}{\kappa}^{2} \left( 1+10\beta_1+16{\beta_1}^{2} \right)\left( b_1{R}^{2}+2b_2{b_0}^ {2}{R}^{2}-2b_2{b_0}^{4}{r}^{2} \right) {c}^{2}\bigg\}^{-1}\,, \nonumber\\
& { p_r}={b_0}^{2} \bigg\{ 60{R}^{12}b_2{b_0}^{4 }{r}^{2}-60{R}^{10}b_2{b_0}^{6}{r}^{4}+30{R}^{8}b_2{b_0}^{8}{r}^{6}-6{R}^{6}b_2{b_0}^{10}{r}^{8}- 18{R}^{12}{b_0}^{2}b_1{r}^{2}+24{R}^{10}{b_0}^{4} {r}^{4}b_1-9{R}^{8}{b_0}^{6}{r}^{6}b_1+\nonumber\\
&36{R}^{14}b_1\beta_1-24{R}^{14}b_2{b_0}^{2}+432{b_0}^{14}b_1{r}^{12}\alpha_1{L}^{2}\beta_1-3168{R}^ {10}{b_0}^{4}b_1\alpha_1{L}^{2}\beta_1{r}^{2}+ 7392{R}^{8}{b_0}^{6}b_1\alpha_1{L}^{2}\beta_1 {r}^{4}-9360{R}^{6}{b_0}^{8}b_1\alpha_1{L}^{2}{r }^{6}\beta_1\nonumber\\
&+6768{R}^{4}{b_0}^{10}b_1{r}^{8}\alpha_1{L}^{2}\beta_1-2640{R}^{2}{b_0}^{12}b_1{r}^ {10}\alpha_1{L}^{2}\beta_1+288{R}^{12}{b_0}^{2}b_1\alpha_1{L}^{2}-192{R}^{12}b_2{b_0}^{4}\beta_1{r}^{2}+448{R}^{10}b_2{b_0}^{6}\beta_1{r}^ {4}\nonumber\\
&-352{R}^{8}b_2{b_0}^{8}{r}^{6}\beta_1+96{R}^{6} b_2{b_0}^{10}\beta_1{r}^{8}-204{R}^{12}{b_0}^ {2}b_1{r}^{2}\beta_1+236{R}^{10}{b_0}^{4}b_1{ r}^{4}\beta_1-84{R}^{8}{b_0}^{6}b_1{r}^{6}\beta_1+216{b_0}^{14}b_1{r}^{12}\alpha_1{L}^{2}\nonumber\\
&+576{R} ^{12}{b_0}^{2}b_1\alpha_1{L}^{2}\beta_1-1584{R }^{10}{b_0}^{4}b_1\alpha_1{L}^{2}{r}^{2}+3696{R}^{ 8}{b_0}^{6}b_1\alpha_1{L}^{2}{r}^{4}-4680{R}^{6}{b_0}^{8}b_1\alpha_1{L}^{2}{r}^{6}+3384{R}^{4}{b_0}^{10}b_1{r}^{8}\alpha_1{L}^{2}\nonumber\\
&-1320{R}^{2}{b_0 }^{12}b_1{r}^{10}\alpha_1{L}^{2} \bigg\}\bigg\{3{R}^{14}{ \kappa}^{2} \left( 1+10\beta_1+16{\beta_1}^{2} \right) \left( b_1{R}^{2}+2b_2{b_0}^{2}{R}^{2}-2b_2{b_0}^{4}{r}^{2} \right) \bigg\}^{-1}\,,\nonumber\\
& {  p_t}=4{b_0}^{2} \bigg\{ 24{R}^{12}b_2{b_0}^{4 }{r}^{2}-36{R}^{10}b_2{b_0}^{6}{r}^{4}+24{R}^{8}b_2{b_0}^{8}{r}^{6}-6{R}^{6}b_2{b_0}^{10}{r}^{8}+ 9{R}^{14}b_1\beta_1-6{R}^{14}b_2{b_0}^{2}+ 108{b_0}^{14}b_1{r}^{12}\alpha_1{L}^{2}\beta_1-\nonumber\\
& 792{R}^{10}{b_0}^{4}b_1\alpha_1{L}^{2}\beta_1 {r}^{2}+1848{R}^{8}{b_0}^{6}b_1\alpha_1{L}^{2}\beta_1{r}^{4}-2340{R}^{6}{b_0}^{8}b_1\alpha_1 {L}^{2}{r}^{6}\beta_1+1692{R}^{4}{b_0}^{10}b_1{r}^ {8}\alpha_1{L}^{2}\beta_1-660{R}^{2}{b_0}^{12}b_1{r}^{10}\alpha_1{L}^{2}\beta_1\nonumber\\
&+72{R}^{12}{b_0}^{2} b_1\alpha_1{L}^{2}+24{R}^{12}b_2{b_0}^{4}\beta_1{r}^{2}-56{R}^{10}b_2{b_0}^{6}\beta_1{ r}^{4}+44{R}^{8}b_2{b_0}^{8}{r}^{6}\beta_1-12{R}^{ 6}b_2{b_0}^{10}\beta_1{r}^{8}-15{R}^{12}{b_0} ^{2}b_1{r}^{2}\beta_1\nonumber\\
&+11{R}^{10}{b_0}^{4}b_1{ r}^{4}\beta_1-3{R}^{8}{b_0}^{6}b_1{r}^{6}\beta_1 +54{b_0}^{14}b_1{r}^{12}\alpha_1{L}^{2}+144{R}^{ 12}{b_0}^{2}b_1\alpha_1{L}^{2}\beta_1-396{R}^{ 10}{b_0}^{4}b_1\alpha_1{L}^{2}{r}^{2}+924{R}^{8}{b_0}^{6}b_1\alpha_1{L}^{2}{r}^{4}\nonumber\\
&-1170{R}^{6}{b_0}^{8}b_1\alpha_1{L}^{2}{r}^{6}+846{R}^{4}{b_0}^ {10}b_1{r}^{8}\alpha_1{L}^{2}-330{R}^{2}{b_0}^{12} b_1{r}^{10}\alpha_1{L}^{2} \bigg\}\bigg\{3{R}^{14}{\kappa}^{2} \left( 1+10\beta_1+16{\beta_1}^{2} \right)\nonumber\\
&  \left( b_1{R}^{2}+2b_2{b_0}^{2}{R}^{2}-2b_2{b_0 }^{4}{r}^{2} \right) \bigg\}^{-1}\,.
\end{align}
Pressure anisotropy inside the stellar matter can be characterized by the difference
\[
\Delta = p_{t}-p_{r},
\]
and its contribution to the internal force balance may be written in the form
\[
F_{a}(r) = \frac{2\,\Delta}{r}.
\]
Since both pressures coincide at the center for a physically admissible configuration, the quantity $F_{a}$ necessarily approaches zero as $r\to 0$.

The sign of $\Delta$ determines the behaviour of the matter distribution throughout the region $0<r\leq R_{s}$.
A positive value of $\Delta$ indicates that the tangential pressure exceeds the radial pressure everywhere, corresponding to a strongly anisotropic distribution of stresses.
If $\Delta$ is negative, the radial pressure is locally greater than the tangential component, which represents the opposite, weak-anisotropy regime.
\section{The KB solution and the corresponding EoSs}\label{Sec:App_2}

In the adopted framework, the KB construction prescribes how the density determines the two pressure components.
Once this dependence is applied, one obtains the set of relations displayed in Eqs.~\eqref{eq:KB_EoS}, which effectively play the role of the system's equations of state.
The constants that enter these expressions are not arbitrary; they arise directly from the model parameters that are enumerated in the subsequent section.
\begin{align}
&c_1=\left\{  \left[\alpha_1 \left( \frac{1}2+\beta_1 \right) b_1+{\frac {2}{27}} \left(  \beta_1-\frac{3}{16} \right) b_2 \right] b_2{b_0}^{4}+{\frac {11}{18}} \left( \alpha_1 \left( \frac{1}2+\beta_1 \right) b_1+\frac{1}{6} \left( \beta_1-\frac{1}{22} \right) b_2 \right) b_1{b_0}^{2}+{\frac {
17}{432}} \left( {\frac {3}{34}}+\beta_1 \right) {b_1}^{2}
 \right\}\nonumber\\
 & {c}^{2} \left\{  \left[\alpha_1 \left( \frac{1}2+\beta_1
 \right) b_1-{\frac {10}{27}} \left( \beta_1+\frac{3}{16} \right) b_2 \right] b_2{b_0}^{4}+{\frac {11}{18}} \left( \alpha_1 \left( \frac{1}2+\beta_1 \right) b_1-{\frac {15}{22}}
 \left( \frac{1}6+\beta_1 \right) b_2 \right) b_1{b_0
}^{2}-{\frac {55}{432}}{b_1}^{2} \left( {\frac {3}{22}}+\beta_1 \right)  \right\} ^{-1}
,\label{eq:A1}\\[5pt]
&c_2=\frac{{4b_0}^{2}}3 \left\{  \left( \alpha_1 \left( \frac{1}2+\beta_1 \right) b_1+\frac{2}9 \left( \beta_1+\frac{3}{16} \right) b_2
 \right) b_2{b_0}^{4}+\frac{5b_1}{4} \left( \alpha_1
 \left( \frac{1}2+\beta_1 \right) b_1+{\frac {17}{90}} \left( \beta_1+{\frac {9}{68}} \right) b_2 \right) {b_0}^{2}+\frac{1}{
12}\left( \beta_1+\frac{3}{16} \right)  \right\}\nonumber\\
 &\times \bigg\{  \left( \frac{{b_1}^{2} }
2+\beta_1 \right){R}^{2}{\kappa}^{2}  \left( \alpha_1 \left(\frac{1}2+\beta_1 \right) b_1-{\frac {10}{27}}
 \left( \beta_1+\frac{3}{16} \right) b_2 \right) b_2{b_0}
^{4}+{\frac {11}{18}} \left( \alpha_1 \left( \frac{1}2+\beta_1
 \right) b_1-{\frac {15}{22}} \left( \frac{1}6+\beta_1 \right) b_2 \right) b_1{b_0}^{2}\nonumber\\
 &-{\frac {55}{432}}{b_1}
^{2} \left( {\frac {3}{22}}+\beta_1 \right)  \bigg\} ^{-1},\label{eq:A2}\\[5pt]
&c_3=- {c}^{2}\left\{  \left( \alpha_1 \left( \frac{1}2+\beta_1 \right) -\frac{1}{27}b_2 \left( \beta_1+\frac{3}4 \right)  \right) b_2{
b_0}^{4}+{\frac {11}{18}}b_1 \left( \alpha_1
 \left( \frac{1}2+\beta_1 \right) b_1-{\frac {1}{66}}b_2
 \left( \beta_1+2 \right)  \right) {b_0}^{2}+{\frac {5}{432}}
{b_1}^{2}\beta_1 \right\}\times \nonumber\\
 &\left\{  \left( \alpha_1 \left( \frac{1}2+\beta_1 \right) b_1-{\frac {10}{27}}
 \left( \frac{3}{16}+\beta_1 \right) b_2 \right) b_2{b_0}
^{4}+{\frac {11}{18}} \left( \alpha_1 \left(\frac{1}2+\beta_1
 \right) b_1-{\frac {15}{22}} \left( \frac{1}6+\beta_1 \right) b_2 \right) b_1{b_0}^{2}-{\frac {55}{432}} \left( {
\frac {3}{22}}+\beta_1 \right) {b_1}^{2} \right\} ^{-1},\label{eq:A3}\\[5pt]
&c_4=2 {b_0}^{2}\left\{  \left( \alpha_1 \left( \frac{1}2+\beta_1 \right) b_1-{\frac {2}{27}} \left( \frac{3}{16}+\beta_1 \right) b_2
 \right) b_2{b_0}^{4}+b_1 \left( \alpha_1
 \left( \frac{1}2+\beta_1 \right) b_1-{\frac {43}{432}} \left( \beta_1+{\frac {9}{43}} \right) b_2 \right) {b_0}^{2}-{
\frac {5}{288}}{b_1}^{2}\beta_1 \right\}\nonumber\\
 &  \bigg\{{
\kappa}^{2} \left( \frac{1}2+\beta_1 \right){R}^{2}
 \left( \alpha_1 \left( \frac{1}2+\beta_1 \right) b_1-{\frac
{10}{27}} \left( \frac{3}{16}+\beta_1 \right) b_2 \right) b_2
{b_0}^{4}+{\frac {11}{18}} \left( \alpha_1 \left( \frac{1}2+\beta_1\right) b_1-{\frac {15}{22}} \left( \frac{1}6+\beta_1
 \right) b_2 \right) b_1{b_0}^{2}\nonumber\\
 & -{\frac {55}{432}}
 \left( {\frac {3}{22}}+\beta_1 \right) {b_1}^{2} \bigg\} ^{-
1}.\label{eq:A4}
\end{align}
The relations derived earlier allow every term appearing in Eq.~\eqref{eq:KB_EoS2} to be rewritten purely as functions of the model parameters.
By reformulating the expressions in this way, we arrive at
\[
v_r^2 = c_1,\qquad
\rho_1 = \rho_s = -\frac{c_2}{c_1},\qquad
v_t^2 = c_3,\qquad
\rho_2 = -\frac{c_4}{c_3}.
\]

\section{The derivatives of $\rho$, $p_r$ and $p_t$}\label{Sec:App_3}
From the solutions for the density and pressure components obtained within the quadratic $f(R)$ framework (see Eqs.~\eqref{sol}),
we can next examine how these functions change throughout the stellar interior.
Differentiating them with respect to the radial coordinate yields the following expressions.
\begin{align}\label{eq:dens_grad}
&\rho'=3456 \bigg\{ \left( \frac{1}2+\beta_1 \right)[\frac{5\alpha_1b_1 }6 b_2{r}^{12}{b_0}^{16}-{\frac {137\alpha_1b_1{R}^{2} }{27}} b_2{r }^{10}{b_0}^{14}]+{\frac {349{R}^{2} {r}^{8}{b_0}^{12}}{27}} \bigg[  \left( \alpha_1 \left( \frac{1}2+\beta_1 \right) b_1-{\frac {9 b_2 }{349}} \left( \beta_1+\frac{3}{16} \right)\right) b_2{R}^{2}\nonumber\\
&-{ \frac {27}{698}}\alpha_1{b_1}^{2} \left( \frac{1}2+\beta_1 \right){r}^{2} \bigg]-{\frac {53}{3}} \left(  \left( \alpha_1 \left( \frac{1}2+\beta_1 \right) b_1 -{\frac {110}{1431}} \left( \beta_1+\frac{3}{16} \right) b_2 \right) b_2{R}^{2}-{\frac {275}{1908}}\alpha_1{b_1}^{2} \left( \frac{1}2+\beta_1 \right) {r}^{2} \right) {R}^{4}{r}^{6}{b_0}^{10}\nonumber\\
&+{\frac {739}{54}} \left(  \left( \alpha_1 \left( \frac{1}2+\beta_1 \right) b_1-{\frac {338}{2217}} \left( \beta_1+\frac{3}{16} \right) b_2 \right) b_2{R}^{2}-{\frac { 282}{739}} \left( \alpha_1 \left( \frac{1}2+\beta_1 \right)  b_1-{\frac {41}{564}} \left( \beta_1+{\frac {27}{164}} \right) b_2 \right) b_1{r}^{2} \right) {R}^{6}{r}^{4}{b_0}^{8}\nonumber\\
&-{\frac {154}{27}} \left( b_2 \left( \alpha_1 \left( \frac{1}2+\beta_1 \right)  b_1-{\frac {58}{231}} \left( \beta_1+\frac{3}{16} \right) b_2 \right) {R}^{2}-{\frac {585}{616}} b_1 \left( \alpha_1 \left( \frac{1}2+\beta_1 \right)  b_1-{\frac {29}{135}} \left( {\frac {249}{1508}}+\beta_1 \right) b_2 \right) {r}^{2} \right) {R}^{8}{r}^{2}{b_0}^{6}+ \bigg\{\nonumber\\
&  b_2\left( \alpha_1 \left( \frac{1}2+\beta_1 \right) b_1-{\frac {10}{27}} \left( \beta_1+\frac{3}{16} \right) b_2 \right) {R}^{4}-{\frac {77b_1 }{27}}\left( \alpha_1 \left(\frac{1}2+\beta_1 \right) b_1-{\frac {389}{924}} \left( {\frac {129}{778}}+\beta_1 \right) b_2 \right) {r}^{2 }{R}^{2}-{\frac {17{b_1}^{2}}{144}} \left( \beta_1+{\frac {9}{68}} \right){r}^{4} \bigg\}\nonumber\\
&  {R}^{8}{b_0}^{4}+{\frac { 11}{18}} \left(  \left( \alpha_1 \left( 1/2+\beta_1 \right) b_1-{\frac {15}{22}} \left( \beta_1+1/6 \right) b_2 \right) {R}^{2}+{\frac {157}{396}} \left( {\frac {21}{157}}+ \beta_1 \right) b_1{r}^{2} \right) b_1{R}^{10}{b_0}^{2}-{\frac {55}{432}} \left( {\frac {3}{22}}+\beta_1 \right) {b_1}^{2}{R}^{12} \bigg\} \nonumber\\
& {b_0}^{4}r{R}^{-12}{ \kappa}^{-2} \left( 1+10\beta_1+16{\beta_1}^{2} \right) ^{ -1} \left( b_1{R}^{2}+2b_2{b_0}^{2}{R}^{2}-2b_2{b_0}^{4}{r}^{2} \right) ^{-2}{c}^{-2},
\end{align}
\begin{align}\label{eq:pr_grad}
  & p'_r=
-2r{b_0}^{4}\bigg\{35472{R}^{8}{b_0}^{8}b_1\alpha_1{r}^{4}b_2-13152{b_0}^{14}b_1 {r}^{10}\alpha_1b_2{R}^{2}-45792{R}^{6}{b_0}^{10}b_1\alpha_1{r}^{6}b_2
+33504{R}^{4}{b_0}^{12}{ b_1}{r}^{8}\alpha_1b_2\nonumber\\
& -1840{R}^{10}b_1\beta_1b_2{b_0}^{4}{r}^{2}+2032{R}^{8}{b_0}^{6}b_1{r}^{4}\beta_1b_2-720{R}^{6}{b_0}^{8}b_1{r}^{6}\beta_1
b_2-14784{R}^{10}{b_0}^{6}
b_1\alpha_1b_2{r}^{2}-2592{b_0}^{12}{b_1}^{2} {r}^{10}\alpha_1\beta_1{R}^{2}\nonumber\\
&+4320{b_0}^{16}b_1{r}^{12}\alpha_1\beta_1b_2+5184{R}^{12}{b_0}^{4}b_1\alpha_1\beta_1b_2
-14784{R}^{10}{ b_0}^{4}{b_1}^{2}\alpha_1\beta_1{r}^{2}+28080{ R}^{8}{b_0}^{6}{b_1}^{2}\alpha_1{r}^{4}\beta_1- 27072{R}^{6}{b_0}^{8}{b_1}^{2}{r}^{6}\alpha_1\beta_1\nonumber\\
&+13200{R}^{4}{b_0}^{10}{b_1}^{2}{r}^{8}\alpha_1 \beta_1+384{R}^{12}{b_2}^{2}{b_0}^{4}\beta_1+ 1584{R}^{12}{b_0}^{2}{b_1}^{2}\alpha_1-24{R}^{12}b_1b_2{b_0}^{2}-48{R}^{10}{b_0}^{2}{b_1} ^{2}{r}^{2}+27{R}^{8}{b_0}^{4}{b_1}^{2}{r}^{4}\nonumber\\
&-300{R}^{8 }{b_2}^{2}{b_0}^{8}{r}^{4}+168{R}^{6}{b_2}^{2}{b_0}^{10}{r}^{6}+240{R}^{10}{b_2}^{2}{b_0}^{6}{r}^{2}-36 {R}^{4}{b_2}^{2}{b_0}^{12}{r}^{8}-1296{b_0}^{12}{b_1}^{2}{r}^{10}\alpha_1{R}^{2}+2160{b_0}^{16}b_1{r}^{12}\alpha_1b_2+\nonumber\\
&14040{R}^{8}{b_0}^{6}{
b_1}^{2}\alpha_1{r}^{4}-13536{R}^{6}{b_0}^{8}{b_1}^{2} {r}^{6}\alpha_1+6600{R}^{4}{b_0}^{10}{b_1}^{2}{r}^{8}\alpha_1+528{R}^{12}b_1\beta_1b_2{b_0}^{ 2}-472{R}^{10}{b_0}^{2}{b_1}^{2}{r}^{2}\beta_1+252{R }^{8}{b_0}^{4}{b_1}^{2}{r}^{4}\beta_1\nonumber\\
&-1792{R}^{10}{b_2}^{2}{b_0}^{6}\beta_1{r}^{2}+3008{R}^{8}{b_2} ^{2}{b_0}^{8}\beta_1{r}^{4}-2176{R}^{6}{b_2}^{2}{b_0}^{10}{r}^{6}\beta_1+576{R}^{4}{b_2}^{2}{b_0}^{ 12}\beta_1{r}^{8}-7392{R}^{10}{b_0}^{4}{b_1}^{2}\alpha_1{r}^{2}+12{R}^{8}{b_0}^{6}b_1{r}^{4}b_2\nonumber\\
&+2592{R}^{12}{b_0}^{4}b_1\alpha_1b_2+24{R}^{10}b_1b_2{b_0}^{4}{r}^{2} -12{R}^{6}{b_0}^{8}b_1{r}^{6}b_2+3168{R}^{12}{b_0}^{2}{b_1}^{2}\alpha_1\beta_1+204{R}^{12}{b_1}^{2}\beta_1 -72{R}^{12}{b_2}^{2}{b_0}^{4}+18{R}^{12}{b_1}^{2}\nonumber\\
&- 26304{b_0}^{14}b_1{r}^{10}\alpha_1\beta_1b_2{R}^{2}-29568{R}^{10}{b_0}^{6}b_1\alpha_1 \beta_1b_2{r}^{2}+70944{R}^{8}{b_0}^{8}b_1\alpha_1\beta_1{r}^{4}b_2-91584{R}^{6}{b_0}^{10 }b_1\alpha_1{r}^{6}\beta_1b_2\nonumber\\
&+67008{R}^{4}{b_0}^{12}b_1{r}^{8}\alpha_1\beta_1b_2 \bigg\}\bigg\{3{R}^{12}{\kappa}^{2} \left( 1+10\beta_1+16{\beta_1}^{2} \right)  \left( b_1{R}^{2}+2b_2{b_0}^ {2}{R}^{2}-2b_2{b_0}^{4}{r}^{2} \right) ^{2}\bigg\}^{-1},
\end{align}
\begin{align}\label{eq:pt_grad}
&p'_t=-8{b_0}^{4}r\bigg\{ 8868{R}^{8}{b_0}^{8}b_1\alpha_1{r}^{4}b_2-3288{b_0}^{14}b_1{ r}^{10}\alpha_1b_2{R}^{2}-11448{R}^{6}{b_0}^{10}b_1\alpha_1{r}^{6}b_2+8376{R}^{4}{b_0}^{12}{ b_1}{r}^{8}\alpha_1b_2+68{R}^{10}b_1\beta_1b_2{b_0}^{4}{r}^{2}\nonumber\\
&-92{R}^{8}{b_0}^{6}b_1{r}^{4}\beta_1b_2+36{R}^{6}{b_0}^{8}b_1{ r}^{6}\beta_1b_2-3696{R}^{10}{b_0}^{6}b_1{ \alpha_1}b_2{r}^{2}-648{b_0}^{12}{b_1}^{2}{r}^{ 10}\alpha_1\beta_1{R}^{2}+1080{b_0}^{16}b_1{ r}^{12}\alpha_1\beta_1b_2\nonumber\\
&+1296{R}^{12}{b_0}^{4 }b_1\alpha_1\beta_1b_2-3696{R}^{10}{b_0 }^{4}{b_1}^{2}\alpha_1\beta_1{r}^{2}+7020{R}^{8}{{ b_0}}^{6}{b_1}^{2}\alpha_1{r}^{4}\beta_1-6768{R}^{ 6}{b_0}^{8}{b_1}^{2}{r}^{6}\alpha_1\beta_1+3300{ R}^{4}{b_0}^{10}{b_1}^{2}{r}^{8}\alpha_1\beta_1\nonumber\\
&-48 {R}^{12}{b_2}^{2}{b_0}^{4}\beta_1+396{R}^{12}{b_0}^{2}{b_1}^{2}\alpha_1-24{R}^{12}b_1b_2{{ b_0}}^{2}-216{R}^{8}{b_2}^{2}{b_0}^{8}{r}^{4}+144{R}^ {6}{b_2}^{2}{b_0}^{10}{r}^{6}+144{R}^{10}{b_2}^{2}{{ b_0}}^{6}{r}^{2}\nonumber\\
&-36{R}^{4}{b_2}^{2}{b_0}^{12}{r}^{8}- 324{b_0}^{12}{b_1}^{2}{r}^{10}\alpha_1{R}^{2}+540{ b_0}^{16}b_1{r}^{12}\alpha_1b_2+3510{R}^{8}{{ b_0}}^{6}{b_1}^{2}\alpha_1{r}^{4}-3384{R}^{6}{b_0 }^{8}{b_1}^{2}{r}^{6}\alpha_1+1650{R}^{4}{b_0}^{10}{{ b_1}}^{2}{r}^{8}\alpha_1\nonumber\\
&-12{R}^{12}b_1\beta_1{ b_2}{b_0}^{2}-22{R}^{10}{b_0}^{2}{b_1}^{2}{r}^{2 }\beta_1+9{R}^{8}{b_0}^{4}{b_1}^{2}{r}^{4}\beta_1+ 224{R}^{10}{b_2}^{2}{b_0}^{6}\beta_1{r}^{2}-376{R} ^{8}{b_2}^{2}{b_0}^{8}\beta_1{r}^{4}+272{R}^{6}{{ b_2}}^{2}{b_0}^{10}{r}^{6}\beta_1\nonumber\\
&-72{R}^{4}{b_2}^{2 }{b_0}^{12}\beta_1{r}^{8}+648{R}^{12}{b_0}^{4}b_1\alpha_1b_2-1848{R}^{10}{b_0}^{4}{b_1}^{2} \alpha_1{r}^{2}+72{R}^{10}b_1b_2{b_0}^{4}{r }^{2}-72{R}^{8}{b_0}^{6}b_1{r}^{4}b_2+24{R}^{6}{{ b_0}}^{8}b_1{r}^{6}b_2\nonumber\\
&+792{R}^{12}{b_0}^{2}{{ b_1}}^{2}\alpha_1\beta_1-36{R}^{12}{b_2}^{2}{b_0}^{4}-6576{b_0}^{14}{ b_1}{r}^{10}\alpha_1\beta_1b_2{R}^{2}-7392{R }^{10}{b_0}^{6}b_1\alpha_1\beta_1b_2{r} ^{2}+17736{R}^{8}{b_0}^{8}b_1\alpha_1\beta_1 {r}^{4}b_2\nonumber\\
&+15{R}^{12}{b_1}^{2}\beta_1-22896{R}^{6}{b_0}^{10}b_1\alpha_1{ r}^{6}\beta_1b_2+16752{R}^{4}{b_0}^{12}b_1{r} ^{8}\alpha_1\beta_1b_2 \bigg\}\bigg\{8{R}^{12}{\kappa}^{2} \left( 1+10\beta_1+16{\beta_1}^{2} \right)  \left( b_1{R}^{2}+2b_2{b_0}^{2}{R}^{2}\right.\nonumber\\
&\left.-2b_2{b_0 }^{4}{r}^{2} \right) ^{2}\bigg\}^{-1}.
\end{align}


\begin{thebibliography}{131}
\expandafter\ifx\csname natexlab\endcsname\relax\def\natexlab#1{#1}\fi
\expandafter\ifx\csname bibnamefont\endcsname\relax
  \def\bibnamefont#1{#1}\fi
\expandafter\ifx\csname bibfnamefont\endcsname\relax
  \def\bibfnamefont#1{#1}\fi
\expandafter\ifx\csname citenamefont\endcsname\relax
  \def\citenamefont#1{#1}\fi
\expandafter\ifx\csname url\endcsname\relax
  \def\url#1{\texttt{#1}}\fi
\expandafter\ifx\csname urlprefix\endcsname\relax\def\urlprefix{URL }\fi
\providecommand{\bibinfo}[2]{#2}
\providecommand{\eprint}[2][]{\url{#2}}

\bibitem[{\citenamefont{Perlmutter
  et~al.}(1997)}]{SupernovaCosmologyProject:1997czu}
\bibinfo{author}{\bibfnamefont{S.}~\bibnamefont{Perlmutter}}
  \bibnamefont{et~al.} (\bibinfo{collaboration}{Supernova Cosmology Project}),
  \bibinfo{journal}{Bull. Am. Astron. Soc.} \textbf{\bibinfo{volume}{29}},
  \bibinfo{pages}{1351} (\bibinfo{year}{1997}), \eprint{astro-ph/9812473}.

\bibitem[{\citenamefont{Riess et~al.}(1998)}]{SupernovaSearchTeam:1998fmf}
\bibinfo{author}{\bibfnamefont{A.~G.} \bibnamefont{Riess}} \bibnamefont{et~al.}
  (\bibinfo{collaboration}{Supernova Search Team}), \bibinfo{journal}{Astron.
  J.} \textbf{\bibinfo{volume}{116}}, \bibinfo{pages}{1009}
  (\bibinfo{year}{1998}), \eprint{astro-ph/9805201}.

\bibitem[{\citenamefont{Tegmark et~al.}(2004)}]{SDSS:2003eyi}
\bibinfo{author}{\bibfnamefont{M.}~\bibnamefont{Tegmark}} \bibnamefont{et~al.}
  (\bibinfo{collaboration}{SDSS}), \bibinfo{journal}{Phys. Rev. D}
  \textbf{\bibinfo{volume}{69}}, \bibinfo{pages}{103501}
  (\bibinfo{year}{2004}), \eprint{astro-ph/0310723}.

\bibitem[{\citenamefont{Spergel et~al.}(2007)}]{WMAP:2006bqn}
\bibinfo{author}{\bibfnamefont{D.~N.} \bibnamefont{Spergel}}
  \bibnamefont{et~al.} (\bibinfo{collaboration}{WMAP}),
  \bibinfo{journal}{Astrophys. J. Suppl.} \textbf{\bibinfo{volume}{170}},
  \bibinfo{pages}{377} (\bibinfo{year}{2007}), \eprint{astro-ph/0603449}.

\bibitem[{\citenamefont{De~Felice and Tsujikawa}(2010)}]{DeFelice:2010aj}
\bibinfo{author}{\bibfnamefont{A.}~\bibnamefont{De~Felice}} \bibnamefont{and}
  \bibinfo{author}{\bibfnamefont{S.}~\bibnamefont{Tsujikawa}},
  \bibinfo{journal}{Living Rev. Rel.} \textbf{\bibinfo{volume}{13}},
  \bibinfo{pages}{3} (\bibinfo{year}{2010}), \eprint{1002.4928}.

\bibitem[{\citenamefont{Carroll et~al.}(2004)\citenamefont{Carroll, Duvvuri,
  Trodden, and Turner}}]{Carroll:2003wy}
\bibinfo{author}{\bibfnamefont{S.~M.} \bibnamefont{Carroll}},
  \bibinfo{author}{\bibfnamefont{V.}~\bibnamefont{Duvvuri}},
  \bibinfo{author}{\bibfnamefont{M.}~\bibnamefont{Trodden}}, \bibnamefont{and}
  \bibinfo{author}{\bibfnamefont{M.~S.} \bibnamefont{Turner}},
  \bibinfo{journal}{Phys. Rev. D} \textbf{\bibinfo{volume}{70}},
  \bibinfo{pages}{043528} (\bibinfo{year}{2004}), \eprint{astro-ph/0306438}.

\bibitem[{\citenamefont{Nojiri and Odintsov}(2003)}]{Nojiri:2003ft}
\bibinfo{author}{\bibfnamefont{S.}~\bibnamefont{Nojiri}} \bibnamefont{and}
  \bibinfo{author}{\bibfnamefont{S.~D.} \bibnamefont{Odintsov}},
  \bibinfo{journal}{Phys. Rev. D} \textbf{\bibinfo{volume}{68}},
  \bibinfo{pages}{123512} (\bibinfo{year}{2003}), \eprint{hep-th/0307288}.

\bibitem[{\citenamefont{Capozziello et~al.}(2006)\citenamefont{Capozziello,
  Nojiri, Odintsov, and Troisi}}]{Capozziello:2006dj}
\bibinfo{author}{\bibfnamefont{S.}~\bibnamefont{Capozziello}},
  \bibinfo{author}{\bibfnamefont{S.}~\bibnamefont{Nojiri}},
  \bibinfo{author}{\bibfnamefont{S.~D.} \bibnamefont{Odintsov}},
  \bibnamefont{and} \bibinfo{author}{\bibfnamefont{A.}~\bibnamefont{Troisi}},
  \bibinfo{journal}{Phys. Lett. B} \textbf{\bibinfo{volume}{639}},
  \bibinfo{pages}{135} (\bibinfo{year}{2006}), \eprint{astro-ph/0604431}.

\bibitem[{\citenamefont{Amendola
  et~al.}(2007{\natexlab{a}})\citenamefont{Amendola, Polarski, and
  Tsujikawa}}]{Amendola:2006kh}
\bibinfo{author}{\bibfnamefont{L.}~\bibnamefont{Amendola}},
  \bibinfo{author}{\bibfnamefont{D.}~\bibnamefont{Polarski}}, \bibnamefont{and}
  \bibinfo{author}{\bibfnamefont{S.}~\bibnamefont{Tsujikawa}},
  \bibinfo{journal}{Phys. Rev. Lett.} \textbf{\bibinfo{volume}{98}},
  \bibinfo{pages}{131302} (\bibinfo{year}{2007}{\natexlab{a}}),
  \eprint{astro-ph/0603703}.

\bibitem[{\citenamefont{Nojiri and
  Odintsov}(2006{\natexlab{a}})}]{Nojiri:2006gh}
\bibinfo{author}{\bibfnamefont{S.}~\bibnamefont{Nojiri}} \bibnamefont{and}
  \bibinfo{author}{\bibfnamefont{S.~D.} \bibnamefont{Odintsov}},
  \bibinfo{journal}{Phys. Rev. D} \textbf{\bibinfo{volume}{74}},
  \bibinfo{pages}{086005} (\bibinfo{year}{2006}{\natexlab{a}}),
  \eprint{hep-th/0608008}.

\bibitem[{\citenamefont{Santos et~al.}(2007)\citenamefont{Santos, Alcaniz,
  Reboucas, and Carvalho}}]{Santos:2007bs}
\bibinfo{author}{\bibfnamefont{J.}~\bibnamefont{Santos}},
  \bibinfo{author}{\bibfnamefont{J.~S.} \bibnamefont{Alcaniz}},
  \bibinfo{author}{\bibfnamefont{M.~J.} \bibnamefont{Reboucas}},
  \bibnamefont{and} \bibinfo{author}{\bibfnamefont{F.~C.}
  \bibnamefont{Carvalho}}, \bibinfo{journal}{Phys. Rev. D}
  \textbf{\bibinfo{volume}{76}}, \bibinfo{pages}{083513}
  (\bibinfo{year}{2007}), \eprint{0708.0411}.

\bibitem[{\citenamefont{Nojiri and Odintsov}(2007)}]{Nojiri:2007jr}
\bibinfo{author}{\bibfnamefont{S.}~\bibnamefont{Nojiri}} \bibnamefont{and}
  \bibinfo{author}{\bibfnamefont{S.~D.} \bibnamefont{Odintsov}},
  \bibinfo{journal}{Phys. Lett. B} \textbf{\bibinfo{volume}{652}},
  \bibinfo{pages}{343} (\bibinfo{year}{2007}), \eprint{0706.1378}.

\bibitem[{\citenamefont{Tsujikawa}(2008)}]{Tsujikawa:2007xu}
\bibinfo{author}{\bibfnamefont{S.}~\bibnamefont{Tsujikawa}},
  \bibinfo{journal}{Phys. Rev. D} \textbf{\bibinfo{volume}{77}},
  \bibinfo{pages}{023507} (\bibinfo{year}{2008}), \eprint{0709.1391}.

\bibitem[{\citenamefont{Nojiri and Odintsov}(2004)}]{Nojiri:2004bi}
\bibinfo{author}{\bibfnamefont{S.}~\bibnamefont{Nojiri}} \bibnamefont{and}
  \bibinfo{author}{\bibfnamefont{S.~D.} \bibnamefont{Odintsov}},
  \bibinfo{journal}{Phys. Lett. B} \textbf{\bibinfo{volume}{599}},
  \bibinfo{pages}{137} (\bibinfo{year}{2004}), \eprint{astro-ph/0403622}.

\bibitem[{\citenamefont{Allemandi et~al.}(2005)\citenamefont{Allemandi,
  Borowiec, Francaviglia, and Odintsov}}]{Allemandi:2005qs}
\bibinfo{author}{\bibfnamefont{G.}~\bibnamefont{Allemandi}},
  \bibinfo{author}{\bibfnamefont{A.}~\bibnamefont{Borowiec}},
  \bibinfo{author}{\bibfnamefont{M.}~\bibnamefont{Francaviglia}},
  \bibnamefont{and} \bibinfo{author}{\bibfnamefont{S.~D.}
  \bibnamefont{Odintsov}}, \bibinfo{journal}{Phys. Rev. D}
  \textbf{\bibinfo{volume}{72}}, \bibinfo{pages}{063505}
  (\bibinfo{year}{2005}), \eprint{gr-qc/0504057}.

\bibitem[{\citenamefont{Bertolami et~al.}(2007)\citenamefont{Bertolami,
  Boehmer, Harko, and Lobo}}]{Bertolami:2007gv}
\bibinfo{author}{\bibfnamefont{O.}~\bibnamefont{Bertolami}},
  \bibinfo{author}{\bibfnamefont{C.~G.} \bibnamefont{Boehmer}},
  \bibinfo{author}{\bibfnamefont{T.}~\bibnamefont{Harko}}, \bibnamefont{and}
  \bibinfo{author}{\bibfnamefont{F.~S.~N.} \bibnamefont{Lobo}},
  \bibinfo{journal}{Phys. Rev. D} \textbf{\bibinfo{volume}{75}},
  \bibinfo{pages}{104016} (\bibinfo{year}{2007}), \eprint{0704.1733}.

\bibitem[{\citenamefont{Nojiri et~al.}(2017)\citenamefont{Nojiri, Odintsov, and
  Oikonomou}}]{Nojiri:2017ncd}
\bibinfo{author}{\bibfnamefont{S.}~\bibnamefont{Nojiri}},
  \bibinfo{author}{\bibfnamefont{S.}~\bibnamefont{Odintsov}}, \bibnamefont{and}
  \bibinfo{author}{\bibfnamefont{V.}~\bibnamefont{Oikonomou}},
  \bibinfo{journal}{Physics Reports} \textbf{\bibinfo{volume}{692}},
  \bibinfo{pages}{1} (\bibinfo{year}{2017}).

\bibitem[{\citenamefont{Nojiri and Odintsov}(2011)}]{Nojiri:2010wj}
\bibinfo{author}{\bibfnamefont{S.}~\bibnamefont{Nojiri}} \bibnamefont{and}
  \bibinfo{author}{\bibfnamefont{S.~D.} \bibnamefont{Odintsov}},
  \bibinfo{journal}{Phys. Rept.} \textbf{\bibinfo{volume}{505}},
  \bibinfo{pages}{59} (\bibinfo{year}{2011}), \eprint{1011.0544}.

\bibitem[{\citenamefont{Nashed}(2018{\natexlab{a}})}]{Nashed:2018oaf}
\bibinfo{author}{\bibfnamefont{G.}~\bibnamefont{Nashed}}, \bibinfo{journal}{The
  European Physical Journal Plus} \textbf{\bibinfo{volume}{133}},
  \bibinfo{pages}{1} (\bibinfo{year}{2018}{\natexlab{a}}).

\bibitem[{\citenamefont{Nashed}(2018{\natexlab{b}})}]{Nashed:2018efg}
\bibinfo{author}{\bibfnamefont{G.}~\bibnamefont{Nashed}},
  \bibinfo{journal}{International Journal of Modern Physics D}
  \textbf{\bibinfo{volume}{27}}, \bibinfo{pages}{1850074}
  (\bibinfo{year}{2018}{\natexlab{b}}).

\bibitem[{\citenamefont{Nashed}(2018{\natexlab{c}})}]{Nashed:2018piz}
\bibinfo{author}{\bibfnamefont{G.}~\bibnamefont{Nashed}},
  \bibinfo{journal}{Advances in High Energy Physics}
  \textbf{\bibinfo{volume}{2018}}, \bibinfo{pages}{1}
  (\bibinfo{year}{2018}{\natexlab{c}}).

\bibitem[{\citenamefont{Bertolami et~al.}(2008)\citenamefont{Bertolami, Lobo,
  and Paramos}}]{Bertolami:2008ab}
\bibinfo{author}{\bibfnamefont{O.}~\bibnamefont{Bertolami}},
  \bibinfo{author}{\bibfnamefont{F.~S.~N.} \bibnamefont{Lobo}},
  \bibnamefont{and} \bibinfo{author}{\bibfnamefont{J.}~\bibnamefont{Paramos}},
  \bibinfo{journal}{Phys. Rev. D} \textbf{\bibinfo{volume}{78}},
  \bibinfo{pages}{064036} (\bibinfo{year}{2008}), \eprint{0806.4434}.

\bibitem[{\citenamefont{Bertolami and Paramos}(2008)}]{Bertolami:2007vu}
\bibinfo{author}{\bibfnamefont{O.}~\bibnamefont{Bertolami}} \bibnamefont{and}
  \bibinfo{author}{\bibfnamefont{J.}~\bibnamefont{Paramos}},
  \bibinfo{journal}{Phys. Rev. D} \textbf{\bibinfo{volume}{77}},
  \bibinfo{pages}{084018} (\bibinfo{year}{2008}), \eprint{0709.3988}.

\bibitem[{\citenamefont{Bertolami and Sequeira}(2009)}]{Bertolami:2009cd}
\bibinfo{author}{\bibfnamefont{O.}~\bibnamefont{Bertolami}} \bibnamefont{and}
  \bibinfo{author}{\bibfnamefont{M.~C.} \bibnamefont{Sequeira}},
  \bibinfo{journal}{Phys. Rev. D} \textbf{\bibinfo{volume}{79}},
  \bibinfo{pages}{104010} (\bibinfo{year}{2009}), \eprint{0903.4540}.

\bibitem[{\citenamefont{Calcagni et~al.}(2005)\citenamefont{Calcagni,
  Tsujikawa, and Sami}}]{Calcagni:2005im}
\bibinfo{author}{\bibfnamefont{G.}~\bibnamefont{Calcagni}},
  \bibinfo{author}{\bibfnamefont{S.}~\bibnamefont{Tsujikawa}},
  \bibnamefont{and} \bibinfo{author}{\bibfnamefont{M.}~\bibnamefont{Sami}},
  \bibinfo{journal}{Class. Quant. Grav.} \textbf{\bibinfo{volume}{22}},
  \bibinfo{pages}{3977} (\bibinfo{year}{2005}), \eprint{hep-th/0505193}.

\bibitem[{\citenamefont{De~Felice et~al.}(2006)\citenamefont{De~Felice,
  Hindmarsh, and Trodden}}]{DeFelice:2006pg}
\bibinfo{author}{\bibfnamefont{A.}~\bibnamefont{De~Felice}},
  \bibinfo{author}{\bibfnamefont{M.}~\bibnamefont{Hindmarsh}},
  \bibnamefont{and} \bibinfo{author}{\bibfnamefont{M.}~\bibnamefont{Trodden}},
  \bibinfo{journal}{JCAP} \textbf{\bibinfo{volume}{08}}, \bibinfo{pages}{005}
  (\bibinfo{year}{2006}), \eprint{astro-ph/0604154}.

\bibitem[{\citenamefont{De~Felice and
  Tsujikawa}(2009{\natexlab{a}})}]{DeFelice:2008wz}
\bibinfo{author}{\bibfnamefont{A.}~\bibnamefont{De~Felice}} \bibnamefont{and}
  \bibinfo{author}{\bibfnamefont{S.}~\bibnamefont{Tsujikawa}},
  \bibinfo{journal}{Phys. Lett. B} \textbf{\bibinfo{volume}{675}},
  \bibinfo{pages}{1} (\bibinfo{year}{2009}{\natexlab{a}}), \eprint{0810.5712}.

\bibitem[{\citenamefont{Metsaev and Tseytlin}(1987)}]{Metsaev:1987zx}
\bibinfo{author}{\bibfnamefont{R.~R.} \bibnamefont{Metsaev}} \bibnamefont{and}
  \bibinfo{author}{\bibfnamefont{A.~A.} \bibnamefont{Tseytlin}},
  \bibinfo{journal}{Nucl. Phys. B} \textbf{\bibinfo{volume}{293}},
  \bibinfo{pages}{385} (\bibinfo{year}{1987}).

\bibitem[{\citenamefont{Nojiri et~al.}(2006)\citenamefont{Nojiri, Odintsov, and
  Sami}}]{Nojiri:2006je}
\bibinfo{author}{\bibfnamefont{S.}~\bibnamefont{Nojiri}},
  \bibinfo{author}{\bibfnamefont{S.~D.} \bibnamefont{Odintsov}},
  \bibnamefont{and} \bibinfo{author}{\bibfnamefont{M.}~\bibnamefont{Sami}},
  \bibinfo{journal}{Phys. Rev. D} \textbf{\bibinfo{volume}{74}},
  \bibinfo{pages}{046004} (\bibinfo{year}{2006}), \eprint{hep-th/0605039}.

\bibitem[{\citenamefont{Amendola
  et~al.}(2007{\natexlab{b}})\citenamefont{Amendola, Charmousis, and
  Davis}}]{Amendola:2007ni}
\bibinfo{author}{\bibfnamefont{L.}~\bibnamefont{Amendola}},
  \bibinfo{author}{\bibfnamefont{C.}~\bibnamefont{Charmousis}},
  \bibnamefont{and} \bibinfo{author}{\bibfnamefont{S.~C.} \bibnamefont{Davis}},
  \bibinfo{journal}{JCAP} \textbf{\bibinfo{volume}{10}}, \bibinfo{pages}{004}
  (\bibinfo{year}{2007}{\natexlab{b}}), \eprint{0704.0175}.

\bibitem[{\citenamefont{Nojiri and Odintsov}(2005)}]{Nojiri:2005jg}
\bibinfo{author}{\bibfnamefont{S.}~\bibnamefont{Nojiri}} \bibnamefont{and}
  \bibinfo{author}{\bibfnamefont{S.~D.} \bibnamefont{Odintsov}},
  \bibinfo{journal}{Phys. Lett. B} \textbf{\bibinfo{volume}{631}},
  \bibinfo{pages}{1} (\bibinfo{year}{2005}), \eprint{hep-th/0508049}.

\bibitem[{\citenamefont{De~Felice and
  Tsujikawa}(2009{\natexlab{b}})}]{DeFelice:2009aj}
\bibinfo{author}{\bibfnamefont{A.}~\bibnamefont{De~Felice}} \bibnamefont{and}
  \bibinfo{author}{\bibfnamefont{S.}~\bibnamefont{Tsujikawa}},
  \bibinfo{journal}{Phys. Rev. D} \textbf{\bibinfo{volume}{80}},
  \bibinfo{pages}{063516} (\bibinfo{year}{2009}{\natexlab{b}}),
  \eprint{0907.1830}.

\bibitem[{\citenamefont{Cognola et~al.}(2006)\citenamefont{Cognola, Elizalde,
  Nojiri, Odintsov, and Zerbini}}]{Cognola:2006eg}
\bibinfo{author}{\bibfnamefont{G.}~\bibnamefont{Cognola}},
  \bibinfo{author}{\bibfnamefont{E.}~\bibnamefont{Elizalde}},
  \bibinfo{author}{\bibfnamefont{S.}~\bibnamefont{Nojiri}},
  \bibinfo{author}{\bibfnamefont{S.~D.} \bibnamefont{Odintsov}},
  \bibnamefont{and} \bibinfo{author}{\bibfnamefont{S.}~\bibnamefont{Zerbini}},
  \bibinfo{journal}{Phys. Rev. D} \textbf{\bibinfo{volume}{73}},
  \bibinfo{pages}{084007} (\bibinfo{year}{2006}), \eprint{hep-th/0601008}.

\bibitem[{\citenamefont{Nojiri and
  Odintsov}(2006{\natexlab{b}})}]{Nojiri:2006ri}
\bibinfo{author}{\bibfnamefont{S.}~\bibnamefont{Nojiri}} \bibnamefont{and}
  \bibinfo{author}{\bibfnamefont{S.~D.} \bibnamefont{Odintsov}},
  \bibinfo{journal}{eConf} \textbf{\bibinfo{volume}{C0602061}},
  \bibinfo{pages}{06} (\bibinfo{year}{2006}{\natexlab{b}}),
  \eprint{hep-th/0601213}.

\bibitem[{\citenamefont{Yousaf et~al.}(2017)\citenamefont{Yousaf, Sharif,
  Ilyas, and Bhatti}}]{Yousaf:2017lto}
\bibinfo{author}{\bibfnamefont{Z.}~\bibnamefont{Yousaf}},
  \bibinfo{author}{\bibfnamefont{M.}~\bibnamefont{Sharif}},
  \bibinfo{author}{\bibfnamefont{M.}~\bibnamefont{Ilyas}}, \bibnamefont{and}
  \bibinfo{author}{\bibfnamefont{M.~Z.} \bibnamefont{Bhatti}},
  \bibinfo{journal}{Eur. Phys. J. C} \textbf{\bibinfo{volume}{77}},
  \bibinfo{pages}{691} (\bibinfo{year}{2017}), \eprint{1710.05717}.

\bibitem[{\citenamefont{Bhatti et~al.}(2017)\citenamefont{Bhatti, Sharif,
  Yousaf, and Ilyas}}]{Bhatti:2017fov}
\bibinfo{author}{\bibfnamefont{M.~Z.-u.-H.} \bibnamefont{Bhatti}},
  \bibinfo{author}{\bibfnamefont{M.}~\bibnamefont{Sharif}},
  \bibinfo{author}{\bibfnamefont{Z.}~\bibnamefont{Yousaf}}, \bibnamefont{and}
  \bibinfo{author}{\bibfnamefont{M.}~\bibnamefont{Ilyas}},
  \bibinfo{journal}{Int. J. Mod. Phys. D} \textbf{\bibinfo{volume}{27}},
  \bibinfo{pages}{1850044} (\bibinfo{year}{2017}).

\bibitem[{\citenamefont{Bowers and Liang}(1974)}]{Bowers:1974tgi}
\bibinfo{author}{\bibfnamefont{R.~L.} \bibnamefont{Bowers}} \bibnamefont{and}
  \bibinfo{author}{\bibfnamefont{E.~P.~T.} \bibnamefont{Liang}},
  \bibinfo{journal}{Astrophys. J.} \textbf{\bibinfo{volume}{188}},
  \bibinfo{pages}{657} (\bibinfo{year}{1974}).

\bibitem[{\citenamefont{Bhar}(2015)}]{Bhar:2014jta}
\bibinfo{author}{\bibfnamefont{P.}~\bibnamefont{Bhar}}, \bibinfo{journal}{Eur.
  Phys. J. C} \textbf{\bibinfo{volume}{75}}, \bibinfo{pages}{123}
  (\bibinfo{year}{2015}), \eprint{1408.6436}.

\bibitem[{\citenamefont{Antoniadis et~al.}(2013)}]{Antoniadis:2013pzd}
\bibinfo{author}{\bibfnamefont{J.}~\bibnamefont{Antoniadis}}
  \bibnamefont{et~al.}, \bibinfo{journal}{Science}
  \textbf{\bibinfo{volume}{340}}, \bibinfo{pages}{6131} (\bibinfo{year}{2013}),
  \eprint{1304.6875}.

\bibitem[{\citenamefont{Cromartie et~al.}(2019)}]{NANOGrav:2019jur}
\bibinfo{author}{\bibfnamefont{H.~T.} \bibnamefont{Cromartie}}
  \bibnamefont{et~al.} (\bibinfo{collaboration}{NANOGrav}),
  \bibinfo{journal}{Nature Astron.} \textbf{\bibinfo{volume}{4}},
  \bibinfo{pages}{72} (\bibinfo{year}{2019}), \eprint{1904.06759}.

\bibitem[{\citenamefont{Fonseca et~al.}(2021)}]{Fonseca:2021wxt}
\bibinfo{author}{\bibfnamefont{E.}~\bibnamefont{Fonseca}} \bibnamefont{et~al.},
  \bibinfo{journal}{Astrophys. J. Lett.} \textbf{\bibinfo{volume}{915}}
  (\bibinfo{year}{2021}), \eprint{2104.00880}.

\bibitem[{\citenamefont{Miller et~al.}(2021)\citenamefont{Miller, Lamb,
  Dittmann, Bogdanov, Arzoumanian, Gendreau, Guillot, Ho, Lattimer, Loewenstein
  et~al.}}]{Miller:2021qha}
\bibinfo{author}{\bibfnamefont{M.~C.} \bibnamefont{Miller}},
  \bibinfo{author}{\bibfnamefont{F.}~\bibnamefont{Lamb}},
  \bibinfo{author}{\bibfnamefont{A.}~\bibnamefont{Dittmann}},
  \bibinfo{author}{\bibfnamefont{S.}~\bibnamefont{Bogdanov}},
  \bibinfo{author}{\bibfnamefont{Z.}~\bibnamefont{Arzoumanian}},
  \bibinfo{author}{\bibfnamefont{K.}~\bibnamefont{Gendreau}},
  \bibinfo{author}{\bibfnamefont{S.}~\bibnamefont{Guillot}},
  \bibinfo{author}{\bibfnamefont{W.}~\bibnamefont{Ho}},
  \bibinfo{author}{\bibfnamefont{J.}~\bibnamefont{Lattimer}},
  \bibinfo{author}{\bibfnamefont{M.}~\bibnamefont{Loewenstein}},
  \bibnamefont{et~al.}, \bibinfo{journal}{The Astrophysical Journal Letters}
  \textbf{\bibinfo{volume}{918}}, \bibinfo{pages}{L28} (\bibinfo{year}{2021}).

\bibitem[{\citenamefont{Riley et~al.}(2021)\citenamefont{Riley, Watts, Ray,
  Bogdanov, Guillot, Morsink, Bilous, Arzoumanian, Choudhury, Deneva
  et~al.}}]{Riley:2021pdl}
\bibinfo{author}{\bibfnamefont{T.~E.} \bibnamefont{Riley}},
  \bibinfo{author}{\bibfnamefont{A.~L.} \bibnamefont{Watts}},
  \bibinfo{author}{\bibfnamefont{P.~S.} \bibnamefont{Ray}},
  \bibinfo{author}{\bibfnamefont{S.}~\bibnamefont{Bogdanov}},
  \bibinfo{author}{\bibfnamefont{S.}~\bibnamefont{Guillot}},
  \bibinfo{author}{\bibfnamefont{S.~M.} \bibnamefont{Morsink}},
  \bibinfo{author}{\bibfnamefont{A.~V.} \bibnamefont{Bilous}},
  \bibinfo{author}{\bibfnamefont{Z.}~\bibnamefont{Arzoumanian}},
  \bibinfo{author}{\bibfnamefont{D.}~\bibnamefont{Choudhury}},
  \bibinfo{author}{\bibfnamefont{J.~S.} \bibnamefont{Deneva}},
  \bibnamefont{et~al.}, \bibinfo{journal}{The Astrophysical Journal Letters}
  \textbf{\bibinfo{volume}{918}}, \bibinfo{pages}{L27} (\bibinfo{year}{2021}).

\bibitem[{\citenamefont{Demorest et~al.}(2010)\citenamefont{Demorest, Pennucci,
  Ransom, Roberts, and Hessels}}]{Demorest:2010bx}
\bibinfo{author}{\bibfnamefont{P.}~\bibnamefont{Demorest}},
  \bibinfo{author}{\bibfnamefont{T.}~\bibnamefont{Pennucci}},
  \bibinfo{author}{\bibfnamefont{S.}~\bibnamefont{Ransom}},
  \bibinfo{author}{\bibfnamefont{M.}~\bibnamefont{Roberts}}, \bibnamefont{and}
  \bibinfo{author}{\bibfnamefont{J.}~\bibnamefont{Hessels}},
  \bibinfo{journal}{Nature} \textbf{\bibinfo{volume}{467}},
  \bibinfo{pages}{1081} (\bibinfo{year}{2010}), \eprint{1010.5788}.

\bibitem[{\citenamefont{Fonseca et~al.}(2016)}]{Fonseca:2016tux}
\bibinfo{author}{\bibfnamefont{E.}~\bibnamefont{Fonseca}} \bibnamefont{et~al.},
  \bibinfo{journal}{Astrophys. J.} \textbf{\bibinfo{volume}{832}},
  \bibinfo{pages}{167} (\bibinfo{year}{2016}), \eprint{1603.00545}.

\bibitem[{\citenamefont{Arzoumanian et~al.}(2018)}]{NANOGRAV:2018hou}
\bibinfo{author}{\bibfnamefont{Z.}~\bibnamefont{Arzoumanian}}
  \bibnamefont{et~al.} (\bibinfo{collaboration}{NANOGRAV}),
  \bibinfo{journal}{Astrophys. J.} \textbf{\bibinfo{volume}{859}},
  \bibinfo{pages}{47} (\bibinfo{year}{2018}), \eprint{1801.02617}.

\bibitem[{\citenamefont{Reardon et~al.}(2016)\citenamefont{Reardon, Hobbs,
  Coles, Levin, Keith, Bailes, Bhat, Burke-Spolaor, Dai, Kerr
  et~al.}}]{Reardon:2015kba}
\bibinfo{author}{\bibfnamefont{D.}~\bibnamefont{Reardon}},
  \bibinfo{author}{\bibfnamefont{G.}~\bibnamefont{Hobbs}},
  \bibinfo{author}{\bibfnamefont{W.}~\bibnamefont{Coles}},
  \bibinfo{author}{\bibfnamefont{Y.}~\bibnamefont{Levin}},
  \bibinfo{author}{\bibfnamefont{M.}~\bibnamefont{Keith}},
  \bibinfo{author}{\bibfnamefont{M.}~\bibnamefont{Bailes}},
  \bibinfo{author}{\bibfnamefont{N.}~\bibnamefont{Bhat}},
  \bibinfo{author}{\bibfnamefont{S.}~\bibnamefont{Burke-Spolaor}},
  \bibinfo{author}{\bibfnamefont{S.}~\bibnamefont{Dai}},
  \bibinfo{author}{\bibfnamefont{M.}~\bibnamefont{Kerr}}, \bibnamefont{et~al.},
  \bibinfo{journal}{Monthly Notices of the Royal Astronomical Society}
  \textbf{\bibinfo{volume}{455}}, \bibinfo{pages}{1751} (\bibinfo{year}{2016}).

\bibitem[{\citenamefont{Gonzalez-Caniulef
  et~al.}(2019)\citenamefont{Gonzalez-Caniulef, Guillot, and
  Reisenegger}}]{Gonzalez-Caniulef:2019wzi}
\bibinfo{author}{\bibfnamefont{D.}~\bibnamefont{Gonzalez-Caniulef}},
  \bibinfo{author}{\bibfnamefont{S.}~\bibnamefont{Guillot}}, \bibnamefont{and}
  \bibinfo{author}{\bibfnamefont{A.}~\bibnamefont{Reisenegger}},
  \bibinfo{journal}{Mon. Not. Roy. Astron. Soc.}
  \textbf{\bibinfo{volume}{490}}, \bibinfo{pages}{5848} (\bibinfo{year}{2019}),
  \eprint{1904.12114}.

\bibitem[{\citenamefont{Miller et~al.}(2019)\citenamefont{Miller, Lamb,
  Dittmann, Bogdanov, Arzoumanian, Gendreau, Guillot, Harding, Ho, Lattimer
  et~al.}}]{Miller:2019cac}
\bibinfo{author}{\bibfnamefont{M.}~\bibnamefont{Miller}},
  \bibinfo{author}{\bibfnamefont{F.~K.} \bibnamefont{Lamb}},
  \bibinfo{author}{\bibfnamefont{A.}~\bibnamefont{Dittmann}},
  \bibinfo{author}{\bibfnamefont{S.}~\bibnamefont{Bogdanov}},
  \bibinfo{author}{\bibfnamefont{Z.}~\bibnamefont{Arzoumanian}},
  \bibinfo{author}{\bibfnamefont{K.~C.} \bibnamefont{Gendreau}},
  \bibinfo{author}{\bibfnamefont{S.}~\bibnamefont{Guillot}},
  \bibinfo{author}{\bibfnamefont{A.}~\bibnamefont{Harding}},
  \bibinfo{author}{\bibfnamefont{W.}~\bibnamefont{Ho}},
  \bibinfo{author}{\bibfnamefont{J.}~\bibnamefont{Lattimer}},
  \bibnamefont{et~al.}, \bibinfo{journal}{The Astrophysical Journal Letters}
  \textbf{\bibinfo{volume}{887}}, \bibinfo{pages}{L24} (\bibinfo{year}{2019}).

\bibitem[{\citenamefont{Raaijmakers et~al.}(2019)\citenamefont{Raaijmakers,
  Riley, Watts, Greif, Morsink, Hebeler, Schwenk, Hinderer, Nissanke, Guillot
  et~al.}}]{Raaijmakers:2019qny}
\bibinfo{author}{\bibfnamefont{G.}~\bibnamefont{Raaijmakers}},
  \bibinfo{author}{\bibfnamefont{T.~E.} \bibnamefont{Riley}},
  \bibinfo{author}{\bibfnamefont{A.~L.} \bibnamefont{Watts}},
  \bibinfo{author}{\bibfnamefont{S.}~\bibnamefont{Greif}},
  \bibinfo{author}{\bibfnamefont{S.}~\bibnamefont{Morsink}},
  \bibinfo{author}{\bibfnamefont{K.}~\bibnamefont{Hebeler}},
  \bibinfo{author}{\bibfnamefont{A.}~\bibnamefont{Schwenk}},
  \bibinfo{author}{\bibfnamefont{T.}~\bibnamefont{Hinderer}},
  \bibinfo{author}{\bibfnamefont{S.}~\bibnamefont{Nissanke}},
  \bibinfo{author}{\bibfnamefont{S.}~\bibnamefont{Guillot}},
  \bibnamefont{et~al.}, \bibinfo{journal}{The Astrophysical Journal Letters}
  \textbf{\bibinfo{volume}{887}}, \bibinfo{pages}{L22} (\bibinfo{year}{2019}).

\bibitem[{\citenamefont{Abbott et~al.}(2018)\citenamefont{Abbott, Abbott,
  Abbott, Acernese, Ackley, Adams, Adams, Addesso, Adhikari, Adya
  et~al.}}]{LIGOScientific:2018cki}
\bibinfo{author}{\bibfnamefont{B.~P.} \bibnamefont{Abbott}},
  \bibinfo{author}{\bibfnamefont{R.}~\bibnamefont{Abbott}},
  \bibinfo{author}{\bibfnamefont{T.}~\bibnamefont{Abbott}},
  \bibinfo{author}{\bibfnamefont{F.}~\bibnamefont{Acernese}},
  \bibinfo{author}{\bibfnamefont{K.}~\bibnamefont{Ackley}},
  \bibinfo{author}{\bibfnamefont{C.}~\bibnamefont{Adams}},
  \bibinfo{author}{\bibfnamefont{T.}~\bibnamefont{Adams}},
  \bibinfo{author}{\bibfnamefont{P.}~\bibnamefont{Addesso}},
  \bibinfo{author}{\bibfnamefont{R.~X.} \bibnamefont{Adhikari}},
  \bibinfo{author}{\bibfnamefont{V.~B.} \bibnamefont{Adya}},
  \bibnamefont{et~al.}, \bibinfo{journal}{Physical review letters}
  \textbf{\bibinfo{volume}{121}}, \bibinfo{pages}{161101}
  (\bibinfo{year}{2018}).

\bibitem[{\citenamefont{Abbott et~al.}(2020{\natexlab{a}})\citenamefont{Abbott,
  Abbott, Abraham, Acernese, Ackley, Adams, Adhikari, Adya, Affeldt, Agathos
  et~al.}}]{LIGOScientific:2020zkf}
\bibinfo{author}{\bibfnamefont{R.}~\bibnamefont{Abbott}},
  \bibinfo{author}{\bibfnamefont{T.}~\bibnamefont{Abbott}},
  \bibinfo{author}{\bibfnamefont{S.}~\bibnamefont{Abraham}},
  \bibinfo{author}{\bibfnamefont{F.}~\bibnamefont{Acernese}},
  \bibinfo{author}{\bibfnamefont{K.}~\bibnamefont{Ackley}},
  \bibinfo{author}{\bibfnamefont{C.}~\bibnamefont{Adams}},
  \bibinfo{author}{\bibfnamefont{R.~X.} \bibnamefont{Adhikari}},
  \bibinfo{author}{\bibfnamefont{V.}~\bibnamefont{Adya}},
  \bibinfo{author}{\bibfnamefont{C.}~\bibnamefont{Affeldt}},
  \bibinfo{author}{\bibfnamefont{M.}~\bibnamefont{Agathos}},
  \bibnamefont{et~al.}, \bibinfo{journal}{The Astrophysical Journal Letters}
  \textbf{\bibinfo{volume}{896}}, \bibinfo{pages}{L44}
  (\bibinfo{year}{2020}{\natexlab{a}}).

\bibitem[{\citenamefont{{Doroshenko} et~al.}(2022)\citenamefont{{Doroshenko},
  {Suleimanov}, {P{\"u}hlhofer}, and {Santangelo}}}]{2022NatAs...6.1444D}
\bibinfo{author}{\bibfnamefont{V.}~\bibnamefont{{Doroshenko}}},
  \bibinfo{author}{\bibfnamefont{V.}~\bibnamefont{{Suleimanov}}},
  \bibinfo{author}{\bibfnamefont{G.}~\bibnamefont{{P{\"u}hlhofer}}},
  \bibnamefont{and}
  \bibinfo{author}{\bibfnamefont{A.}~\bibnamefont{{Santangelo}}},
  \bibinfo{journal}{Nature Astronomy} \textbf{\bibinfo{volume}{6}},
  \bibinfo{pages}{1444} (\bibinfo{year}{2022}).

\bibitem[{\citenamefont{Romani et~al.}(2022)\citenamefont{Romani, Kandel,
  Filippenko, Brink, and Zheng}}]{Romani:2022jhd}
\bibinfo{author}{\bibfnamefont{R.~W.} \bibnamefont{Romani}},
  \bibinfo{author}{\bibfnamefont{D.}~\bibnamefont{Kandel}},
  \bibinfo{author}{\bibfnamefont{A.~V.} \bibnamefont{Filippenko}},
  \bibinfo{author}{\bibfnamefont{T.~G.} \bibnamefont{Brink}}, \bibnamefont{and}
  \bibinfo{author}{\bibfnamefont{W.}~\bibnamefont{Zheng}},
  \bibinfo{journal}{The Astrophysical Journal Letters}
  \textbf{\bibinfo{volume}{934}}, \bibinfo{pages}{L18} (\bibinfo{year}{2022}).

\bibitem[{\citenamefont{{El Hanafy} and {Awad}}(2023)}]{2023arXiv230514953E}
\bibinfo{author}{\bibfnamefont{W.}~\bibnamefont{{El Hanafy}}} \bibnamefont{and}
  \bibinfo{author}{\bibfnamefont{A.}~\bibnamefont{{Awad}}},
  \bibinfo{journal}{arXiv e-prints} \bibinfo{eid}{arXiv:2305.14953}
  (\bibinfo{year}{2023}), \eprint{2305.14953}.

\bibitem[{\citenamefont{Legred et~al.}(2021)\citenamefont{Legred,
  Chatziioannou, Essick, Han, and Landry}}]{Legred:2021hdx}
\bibinfo{author}{\bibfnamefont{I.}~\bibnamefont{Legred}},
  \bibinfo{author}{\bibfnamefont{K.}~\bibnamefont{Chatziioannou}},
  \bibinfo{author}{\bibfnamefont{R.}~\bibnamefont{Essick}},
  \bibinfo{author}{\bibfnamefont{S.}~\bibnamefont{Han}}, \bibnamefont{and}
  \bibinfo{author}{\bibfnamefont{P.}~\bibnamefont{Landry}},
  \bibinfo{journal}{Phys. Rev. D} \textbf{\bibinfo{volume}{104}},
  \bibinfo{pages}{063003} (\bibinfo{year}{2021}), \eprint{2106.05313}.

\bibitem[{\citenamefont{Abbott et~al.}(2019)\citenamefont{Abbott, Abbott,
  Abbott, Abraham, Acernese, Ackley, Adams, Adhikari, Adya, Affeldt
  et~al.}}]{LIGOScientific:2018jsj}
\bibinfo{author}{\bibfnamefont{B.}~\bibnamefont{Abbott}},
  \bibinfo{author}{\bibfnamefont{R.}~\bibnamefont{Abbott}},
  \bibinfo{author}{\bibfnamefont{T.}~\bibnamefont{Abbott}},
  \bibinfo{author}{\bibfnamefont{S.}~\bibnamefont{Abraham}},
  \bibinfo{author}{\bibfnamefont{F.}~\bibnamefont{Acernese}},
  \bibinfo{author}{\bibfnamefont{K.}~\bibnamefont{Ackley}},
  \bibinfo{author}{\bibfnamefont{C.}~\bibnamefont{Adams}},
  \bibinfo{author}{\bibfnamefont{R.~X.} \bibnamefont{Adhikari}},
  \bibinfo{author}{\bibfnamefont{V.}~\bibnamefont{Adya}},
  \bibinfo{author}{\bibfnamefont{C.}~\bibnamefont{Affeldt}},
  \bibnamefont{et~al.}, \bibinfo{journal}{The Astrophysical Journal Letters}
  \textbf{\bibinfo{volume}{882}}, \bibinfo{pages}{L24} (\bibinfo{year}{2019}).

\bibitem[{\citenamefont{Abbott
  et~al.}(2020{\natexlab{b}})}]{LIGOScientific:2020aai}
\bibinfo{author}{\bibfnamefont{B.~P.} \bibnamefont{Abbott}}
  \bibnamefont{et~al.} (\bibinfo{collaboration}{LIGO Scientific, Virgo}),
  \bibinfo{journal}{Astrophys. J. Lett.} \textbf{\bibinfo{volume}{892}},
  \bibinfo{pages}{L3} (\bibinfo{year}{2020}{\natexlab{b}}),
  \eprint{2001.01761}.

\bibitem[{\citenamefont{Yang et~al.}(2020)\citenamefont{Yang, Gayathri, Bartos,
  Haiman, Safarzadeh, and Tagawa}}]{Yang:2020xyi}
\bibinfo{author}{\bibfnamefont{Y.}~\bibnamefont{Yang}},
  \bibinfo{author}{\bibfnamefont{V.}~\bibnamefont{Gayathri}},
  \bibinfo{author}{\bibfnamefont{I.}~\bibnamefont{Bartos}},
  \bibinfo{author}{\bibfnamefont{Z.}~\bibnamefont{Haiman}},
  \bibinfo{author}{\bibfnamefont{M.}~\bibnamefont{Safarzadeh}},
  \bibnamefont{and} \bibinfo{author}{\bibfnamefont{H.}~\bibnamefont{Tagawa}},
  \bibinfo{journal}{Astrophys. J. Lett.} \textbf{\bibinfo{volume}{901}},
  \bibinfo{pages}{L34} (\bibinfo{year}{2020}), \eprint{2007.04781}.

\bibitem[{\citenamefont{Alho et~al.}(2022{\natexlab{a}})\citenamefont{Alho,
  Nat\'ario, Pani, and Raposo}}]{Alho:2022bki}
\bibinfo{author}{\bibfnamefont{A.}~\bibnamefont{Alho}},
  \bibinfo{author}{\bibfnamefont{J.}~\bibnamefont{Nat\'ario}},
  \bibinfo{author}{\bibfnamefont{P.}~\bibnamefont{Pani}}, \bibnamefont{and}
  \bibinfo{author}{\bibfnamefont{G.}~\bibnamefont{Raposo}},
  \bibinfo{journal}{Phys. Rev. D} \textbf{\bibinfo{volume}{106}},
  \bibinfo{pages}{L041502} (\bibinfo{year}{2022}{\natexlab{a}}),
  \eprint{2202.00043}.

\bibitem[{\citenamefont{Alho et~al.}(2022{\natexlab{b}})\citenamefont{Alho,
  Nat\'ario, Pani, and Raposo}}]{Alho:2021sli}
\bibinfo{author}{\bibfnamefont{A.}~\bibnamefont{Alho}},
  \bibinfo{author}{\bibfnamefont{J.}~\bibnamefont{Nat\'ario}},
  \bibinfo{author}{\bibfnamefont{P.}~\bibnamefont{Pani}}, \bibnamefont{and}
  \bibinfo{author}{\bibfnamefont{G.}~\bibnamefont{Raposo}},
  \bibinfo{journal}{Phys. Rev. D} \textbf{\bibinfo{volume}{105}},
  \bibinfo{pages}{044025} (\bibinfo{year}{2022}{\natexlab{b}}),
  \bibinfo{note}{[Erratum: Phys.Rev.D 105, 129903 (2022)]},
  \eprint{2107.12272}.

\bibitem[{\citenamefont{Roupas and Nashed}(2020)}]{Roupas:2020mvs}
\bibinfo{author}{\bibfnamefont{Z.}~\bibnamefont{Roupas}} \bibnamefont{and}
  \bibinfo{author}{\bibfnamefont{G.~G.} \bibnamefont{Nashed}},
  \bibinfo{journal}{The European Physical Journal C}
  \textbf{\bibinfo{volume}{80}}, \bibinfo{pages}{1} (\bibinfo{year}{2020}).

\bibitem[{\citenamefont{Raposo et~al.}(2019)\citenamefont{Raposo, Pani,
  Bezares, Palenzuela, and Cardoso}}]{Raposo:2018rjn}
\bibinfo{author}{\bibfnamefont{G.}~\bibnamefont{Raposo}},
  \bibinfo{author}{\bibfnamefont{P.}~\bibnamefont{Pani}},
  \bibinfo{author}{\bibfnamefont{M.}~\bibnamefont{Bezares}},
  \bibinfo{author}{\bibfnamefont{C.}~\bibnamefont{Palenzuela}},
  \bibnamefont{and} \bibinfo{author}{\bibfnamefont{V.}~\bibnamefont{Cardoso}},
  \bibinfo{journal}{Phys. Rev. D} \textbf{\bibinfo{volume}{99}},
  \bibinfo{pages}{104072} (\bibinfo{year}{2019}), \eprint{1811.07917}.

\bibitem[{\citenamefont{Cardoso and Pani}(2019)}]{Cardoso:2019rvt}
\bibinfo{author}{\bibfnamefont{V.}~\bibnamefont{Cardoso}} \bibnamefont{and}
  \bibinfo{author}{\bibfnamefont{P.}~\bibnamefont{Pani}},
  \bibinfo{journal}{Living Rev. Rel.} \textbf{\bibinfo{volume}{22}},
  \bibinfo{pages}{4} (\bibinfo{year}{2019}), \eprint{1904.05363}.

\bibitem[{\citenamefont{Nashed and El~Hanafy}(2022)}]{Nashed:2022zyi}
\bibinfo{author}{\bibfnamefont{G.}~\bibnamefont{Nashed}} \bibnamefont{and}
  \bibinfo{author}{\bibfnamefont{W.}~\bibnamefont{El~Hanafy}},
  \bibinfo{journal}{The European Physical Journal C}
  \textbf{\bibinfo{volume}{82}}, \bibinfo{pages}{679} (\bibinfo{year}{2022}).

\bibitem[{\citenamefont{El~Hanafy}(2022)}]{ElHanafy:2022kjl}
\bibinfo{author}{\bibfnamefont{W.}~\bibnamefont{El~Hanafy}},
  \bibinfo{journal}{Astrophys. J.} \textbf{\bibinfo{volume}{940}},
  \bibinfo{pages}{51} (\bibinfo{year}{2022}), \eprint{2209.10287}.

\bibitem[{\citenamefont{Kobayashi and Maeda}(2008)}]{Kobayashi:2008tq}
\bibinfo{author}{\bibfnamefont{T.}~\bibnamefont{Kobayashi}} \bibnamefont{and}
  \bibinfo{author}{\bibfnamefont{K.-i.} \bibnamefont{Maeda}},
  \bibinfo{journal}{Phys. Rev. D} \textbf{\bibinfo{volume}{78}},
  \bibinfo{pages}{064019} (\bibinfo{year}{2008}), \eprint{0807.2503}.

\bibitem[{\citenamefont{Upadhye and Hu}(2009)}]{Upadhye:2009kt}
\bibinfo{author}{\bibfnamefont{A.}~\bibnamefont{Upadhye}} \bibnamefont{and}
  \bibinfo{author}{\bibfnamefont{W.}~\bibnamefont{Hu}}, \bibinfo{journal}{Phys.
  Rev. D} \textbf{\bibinfo{volume}{80}}, \bibinfo{pages}{064002}
  (\bibinfo{year}{2009}), \eprint{0905.4055}.

\bibitem[{\citenamefont{Feng et~al.}(2017)\citenamefont{Feng, Geng, Kao, and
  Luo}}]{Feng:2017hje}
\bibinfo{author}{\bibfnamefont{W.-X.} \bibnamefont{Feng}},
  \bibinfo{author}{\bibfnamefont{C.-Q.} \bibnamefont{Geng}},
  \bibinfo{author}{\bibfnamefont{W.~F.} \bibnamefont{Kao}}, \bibnamefont{and}
  \bibinfo{author}{\bibfnamefont{L.-W.} \bibnamefont{Luo}},
  \bibinfo{journal}{Int. J. Mod. Phys. D} \textbf{\bibinfo{volume}{27}},
  \bibinfo{pages}{1750186} (\bibinfo{year}{2017}), \eprint{1702.05936}.

\bibitem[{\citenamefont{Teppa~Pannia et~al.}(2017)\citenamefont{Teppa~Pannia,
  Garc\'\i{}a, Perez~Bergliaffa, Orellana, and Romero}}]{TeppaPannia:2016vsb}
\bibinfo{author}{\bibfnamefont{F.~A.} \bibnamefont{Teppa~Pannia}},
  \bibinfo{author}{\bibfnamefont{F.}~\bibnamefont{Garc\'\i{}a}},
  \bibinfo{author}{\bibfnamefont{S.~E.} \bibnamefont{Perez~Bergliaffa}},
  \bibinfo{author}{\bibfnamefont{M.}~\bibnamefont{Orellana}}, \bibnamefont{and}
  \bibinfo{author}{\bibfnamefont{G.~E.} \bibnamefont{Romero}},
  \bibinfo{journal}{Gen. Rel. Grav.} \textbf{\bibinfo{volume}{49}},
  \bibinfo{pages}{25} (\bibinfo{year}{2017}), \eprint{1607.03508}.

\bibitem[{\citenamefont{Wojnar and Velten}(2016)}]{Wojnar:2016bzk}
\bibinfo{author}{\bibfnamefont{A.}~\bibnamefont{Wojnar}} \bibnamefont{and}
  \bibinfo{author}{\bibfnamefont{H.}~\bibnamefont{Velten}},
  \bibinfo{journal}{Eur. Phys. J. C} \textbf{\bibinfo{volume}{76}},
  \bibinfo{pages}{697} (\bibinfo{year}{2016}), \eprint{1604.04257}.

\bibitem[{\citenamefont{Arapo\u{g}lu et~al.}(2017)\citenamefont{Arapo\u{g}lu,
  \c{C}\i{}k\i{}nto\u{g}lu, and Ek\c{s}i}}]{Arapoglu:2016ozr}
\bibinfo{author}{\bibfnamefont{S.}~\bibnamefont{Arapo\u{g}lu}},
  \bibinfo{author}{\bibfnamefont{S.}~\bibnamefont{\c{C}\i{}k\i{}nto\u{g}lu}},
  \bibnamefont{and} \bibinfo{author}{\bibfnamefont{K.~Y.}
  \bibnamefont{Ek\c{s}i}}, \bibinfo{journal}{Phys. Rev. D}
  \textbf{\bibinfo{volume}{96}}, \bibinfo{pages}{084040}
  (\bibinfo{year}{2017}), \eprint{1604.02328}.

\bibitem[{\citenamefont{Katsuragawa et~al.}(2016)\citenamefont{Katsuragawa,
  Nojiri, Odintsov, and Yamazaki}}]{Katsuragawa:2015lbl}
\bibinfo{author}{\bibfnamefont{T.}~\bibnamefont{Katsuragawa}},
  \bibinfo{author}{\bibfnamefont{S.}~\bibnamefont{Nojiri}},
  \bibinfo{author}{\bibfnamefont{S.~D.} \bibnamefont{Odintsov}},
  \bibnamefont{and} \bibinfo{author}{\bibfnamefont{M.}~\bibnamefont{Yamazaki}},
  \bibinfo{journal}{Phys. Rev. D} \textbf{\bibinfo{volume}{93}},
  \bibinfo{pages}{124013} (\bibinfo{year}{2016}), \eprint{1512.00660}.

\bibitem[{\citenamefont{Fiziev}(2015)}]{Fiziev:2015xpa}
\bibinfo{author}{\bibfnamefont{P.~P.} \bibnamefont{Fiziev}}
  (\bibinfo{year}{2015}), \eprint{1506.08585}.

\bibitem[{\citenamefont{Hendi et~al.}(2015)\citenamefont{Hendi, Bordbar,
  Eslam~Panah, and Najafi}}]{Hendi:2015pua}
\bibinfo{author}{\bibfnamefont{S.~H.} \bibnamefont{Hendi}},
  \bibinfo{author}{\bibfnamefont{G.~H.} \bibnamefont{Bordbar}},
  \bibinfo{author}{\bibfnamefont{B.}~\bibnamefont{Eslam~Panah}},
  \bibnamefont{and} \bibinfo{author}{\bibfnamefont{M.}~\bibnamefont{Najafi}},
  \bibinfo{journal}{Astrophys. Space Sci.} \textbf{\bibinfo{volume}{358}},
  \bibinfo{pages}{30} (\bibinfo{year}{2015}), \eprint{1503.01011}.

\bibitem[{\citenamefont{Momeni et~al.}(2015)\citenamefont{Momeni, Gholizade,
  Raza, and Myrzakulov}}]{Momeni:2015vwa}
\bibinfo{author}{\bibfnamefont{D.}~\bibnamefont{Momeni}},
  \bibinfo{author}{\bibfnamefont{H.}~\bibnamefont{Gholizade}},
  \bibinfo{author}{\bibfnamefont{M.}~\bibnamefont{Raza}}, \bibnamefont{and}
  \bibinfo{author}{\bibfnamefont{R.}~\bibnamefont{Myrzakulov}},
  \bibinfo{journal}{Int. J. Mod. Phys. A} \textbf{\bibinfo{volume}{30}},
  \bibinfo{pages}{1550093} (\bibinfo{year}{2015}), \eprint{1502.05000}.

\bibitem[{\citenamefont{Zubair and Abbas}(2016)}]{Zubair:2016kov}
\bibinfo{author}{\bibfnamefont{M.}~\bibnamefont{Zubair}} \bibnamefont{and}
  \bibinfo{author}{\bibfnamefont{G.}~\bibnamefont{Abbas}},
  \bibinfo{journal}{Astrophys. Space Sci.} \textbf{\bibinfo{volume}{361}},
  \bibinfo{pages}{342} (\bibinfo{year}{2016}).

\bibitem[{\citenamefont{Bakirova and Folomeev}(2016)}]{Bakirova:2016ffk}
\bibinfo{author}{\bibfnamefont{E.}~\bibnamefont{Bakirova}} \bibnamefont{and}
  \bibinfo{author}{\bibfnamefont{V.}~\bibnamefont{Folomeev}},
  \bibinfo{journal}{Gen. Rel. Grav.} \textbf{\bibinfo{volume}{48}},
  \bibinfo{pages}{135} (\bibinfo{year}{2016}), \bibinfo{note}{[Erratum:
  Gen.Rel.Grav. 48, 164 (2016)]}, \eprint{1603.01936}.

\bibitem[{\citenamefont{Aparicio~Resco
  et~al.}(2016)\citenamefont{Aparicio~Resco, de~la Cruz-Dombriz,
  Llanes~Estrada, and Zapatero~Castrillo}}]{AparicioResco:2016xcm}
\bibinfo{author}{\bibfnamefont{M.}~\bibnamefont{Aparicio~Resco}},
  \bibinfo{author}{\bibfnamefont{A.}~\bibnamefont{de~la Cruz-Dombriz}},
  \bibinfo{author}{\bibfnamefont{F.~J.} \bibnamefont{Llanes~Estrada}},
  \bibnamefont{and}
  \bibinfo{author}{\bibfnamefont{V.}~\bibnamefont{Zapatero~Castrillo}},
  \bibinfo{journal}{Phys. Dark Univ.} \textbf{\bibinfo{volume}{13}},
  \bibinfo{pages}{147} (\bibinfo{year}{2016}), \eprint{1602.03880}.

\bibitem[{\citenamefont{Moraes et~al.}(2016)\citenamefont{Moraes, Arba\~nil,
  and Malheiro}}]{Moraes:2015uxq}
\bibinfo{author}{\bibfnamefont{P.~H. R.~S.} \bibnamefont{Moraes}},
  \bibinfo{author}{\bibfnamefont{J.~D.~V.} \bibnamefont{Arba\~nil}},
  \bibnamefont{and} \bibinfo{author}{\bibfnamefont{M.}~\bibnamefont{Malheiro}},
  \bibinfo{journal}{JCAP} \textbf{\bibinfo{volume}{06}}, \bibinfo{pages}{005}
  (\bibinfo{year}{2016}), \eprint{1511.06282}.

\bibitem[{\citenamefont{Sharif and Yousaf}(2015)}]{Sharif:2015jaa}
\bibinfo{author}{\bibfnamefont{M.}~\bibnamefont{Sharif}} \bibnamefont{and}
  \bibinfo{author}{\bibfnamefont{Z.}~\bibnamefont{Yousaf}},
  \bibinfo{journal}{Can. J. Phys.} \textbf{\bibinfo{volume}{93}},
  \bibinfo{pages}{905} (\bibinfo{year}{2015}).

\bibitem[{\citenamefont{Sotani and Kokkotas}(2017)}]{Sotani:2017pfj}
\bibinfo{author}{\bibfnamefont{H.}~\bibnamefont{Sotani}} \bibnamefont{and}
  \bibinfo{author}{\bibfnamefont{K.~D.} \bibnamefont{Kokkotas}},
  \bibinfo{journal}{Phys. Rev. D} \textbf{\bibinfo{volume}{95}},
  \bibinfo{pages}{044032} (\bibinfo{year}{2017}), \eprint{1702.00874}.

\bibitem[{\citenamefont{Capozziello et~al.}(2011)\citenamefont{Capozziello,
  De~Laurentis, Odintsov, and Stabile}}]{Capozziello:2011nr}
\bibinfo{author}{\bibfnamefont{S.}~\bibnamefont{Capozziello}},
  \bibinfo{author}{\bibfnamefont{M.}~\bibnamefont{De~Laurentis}},
  \bibinfo{author}{\bibfnamefont{S.~D.} \bibnamefont{Odintsov}},
  \bibnamefont{and} \bibinfo{author}{\bibfnamefont{A.}~\bibnamefont{Stabile}},
  \bibinfo{journal}{Phys. Rev. D} \textbf{\bibinfo{volume}{83}},
  \bibinfo{pages}{064004} (\bibinfo{year}{2011}), \eprint{1101.0219}.

\bibitem[{\citenamefont{Arapoglu et~al.}(2011)\citenamefont{Arapoglu,
  Deliduman, and Eksi}}]{Arapoglu:2010rz}
\bibinfo{author}{\bibfnamefont{A.~S.} \bibnamefont{Arapoglu}},
  \bibinfo{author}{\bibfnamefont{C.}~\bibnamefont{Deliduman}},
  \bibnamefont{and} \bibinfo{author}{\bibfnamefont{K.~Y.} \bibnamefont{Eksi}},
  \bibinfo{journal}{JCAP} \textbf{\bibinfo{volume}{07}}, \bibinfo{pages}{020}
  (\bibinfo{year}{2011}), \eprint{1003.3179}.

\bibitem[{\citenamefont{Astashenok et~al.}(2013)\citenamefont{Astashenok,
  Capozziello, and Odintsov}}]{Astashenok:2013vza}
\bibinfo{author}{\bibfnamefont{A.~V.} \bibnamefont{Astashenok}},
  \bibinfo{author}{\bibfnamefont{S.}~\bibnamefont{Capozziello}},
  \bibnamefont{and} \bibinfo{author}{\bibfnamefont{S.~D.}
  \bibnamefont{Odintsov}}, \bibinfo{journal}{JCAP}
  \textbf{\bibinfo{volume}{12}}, \bibinfo{pages}{040} (\bibinfo{year}{2013}),
  \eprint{1309.1978}.

\bibitem[{\citenamefont{Astashenok et~al.}(2014)\citenamefont{Astashenok,
  Capozziello, and Odintsov}}]{Astashenok:2014pua}
\bibinfo{author}{\bibfnamefont{A.~V.} \bibnamefont{Astashenok}},
  \bibinfo{author}{\bibfnamefont{S.}~\bibnamefont{Capozziello}},
  \bibnamefont{and} \bibinfo{author}{\bibfnamefont{S.~D.}
  \bibnamefont{Odintsov}}, \bibinfo{journal}{Phys. Rev. D}
  \textbf{\bibinfo{volume}{89}}, \bibinfo{pages}{103509}
  (\bibinfo{year}{2014}), \eprint{1401.4546}.

\bibitem[{\citenamefont{Astashenok
  et~al.}(2015{\natexlab{a}})\citenamefont{Astashenok, Capozziello, and
  Odintsov}}]{Astashenok:2014gda}
\bibinfo{author}{\bibfnamefont{A.~V.} \bibnamefont{Astashenok}},
  \bibinfo{author}{\bibfnamefont{S.}~\bibnamefont{Capozziello}},
  \bibnamefont{and} \bibinfo{author}{\bibfnamefont{S.~D.}
  \bibnamefont{Odintsov}}, \bibinfo{journal}{Astrophys. Space Sci.}
  \textbf{\bibinfo{volume}{355}}, \bibinfo{pages}{333}
  (\bibinfo{year}{2015}{\natexlab{a}}), \eprint{1405.6663}.

\bibitem[{\citenamefont{Astashenok
  et~al.}(2015{\natexlab{b}})\citenamefont{Astashenok, Capozziello, and
  Odintsov}}]{Astashenok:2014nua}
\bibinfo{author}{\bibfnamefont{A.~V.} \bibnamefont{Astashenok}},
  \bibinfo{author}{\bibfnamefont{S.}~\bibnamefont{Capozziello}},
  \bibnamefont{and} \bibinfo{author}{\bibfnamefont{S.~D.}
  \bibnamefont{Odintsov}}, \bibinfo{journal}{JCAP}
  \textbf{\bibinfo{volume}{01}}, \bibinfo{pages}{001}
  (\bibinfo{year}{2015}{\natexlab{b}}), \eprint{1408.3856}.

\bibitem[{\citenamefont{Harko et~al.}(2011)\citenamefont{Harko, Lobo, Nojiri,
  and Odintsov}}]{Harko:2011kv}
\bibinfo{author}{\bibfnamefont{T.}~\bibnamefont{Harko}},
  \bibinfo{author}{\bibfnamefont{F.~S.~N.} \bibnamefont{Lobo}},
  \bibinfo{author}{\bibfnamefont{S.}~\bibnamefont{Nojiri}}, \bibnamefont{and}
  \bibinfo{author}{\bibfnamefont{S.~D.} \bibnamefont{Odintsov}},
  \bibinfo{journal}{Phys. Rev. D} \textbf{\bibinfo{volume}{84}},
  \bibinfo{pages}{024020} (\bibinfo{year}{2011}), \eprint{1104.2669}.

\bibitem[{\citenamefont{Das et~al.}(2016)\citenamefont{Das, Rahaman, Guha, and
  Ray}}]{Das:2016mxq}
\bibinfo{author}{\bibfnamefont{A.}~\bibnamefont{Das}},
  \bibinfo{author}{\bibfnamefont{F.}~\bibnamefont{Rahaman}},
  \bibinfo{author}{\bibfnamefont{B.~K.} \bibnamefont{Guha}}, \bibnamefont{and}
  \bibinfo{author}{\bibfnamefont{S.}~\bibnamefont{Ray}}, \bibinfo{journal}{Eur.
  Phys. J. C} \textbf{\bibinfo{volume}{76}}, \bibinfo{pages}{654}
  (\bibinfo{year}{2016}), \eprint{1608.00566}.

\bibitem[{\citenamefont{Deb et~al.}(2018)\citenamefont{Deb, Rahaman, Ray, and
  Guha}}]{Deb:2017rhd}
\bibinfo{author}{\bibfnamefont{D.}~\bibnamefont{Deb}},
  \bibinfo{author}{\bibfnamefont{F.}~\bibnamefont{Rahaman}},
  \bibinfo{author}{\bibfnamefont{S.}~\bibnamefont{Ray}}, \bibnamefont{and}
  \bibinfo{author}{\bibfnamefont{B.}~\bibnamefont{Guha}},
  \bibinfo{journal}{Journal of Cosmology and Astroparticle Physics}
  \textbf{\bibinfo{volume}{2018}}, \bibinfo{pages}{044} (\bibinfo{year}{2018}).

\bibitem[{\citenamefont{Lobato et~al.}(2020)\citenamefont{Lobato,
  Louren{\c{c}}o, Moraes, Lenzi, De~Avellar, De~Paula, Dutra, and
  Malheiro}}]{Lobato:2020fxt}
\bibinfo{author}{\bibfnamefont{R.}~\bibnamefont{Lobato}},
  \bibinfo{author}{\bibfnamefont{O.}~\bibnamefont{Louren{\c{c}}o}},
  \bibinfo{author}{\bibfnamefont{P.}~\bibnamefont{Moraes}},
  \bibinfo{author}{\bibfnamefont{C.}~\bibnamefont{Lenzi}},
  \bibinfo{author}{\bibfnamefont{M.}~\bibnamefont{De~Avellar}},
  \bibinfo{author}{\bibfnamefont{W.}~\bibnamefont{De~Paula}},
  \bibinfo{author}{\bibfnamefont{M.}~\bibnamefont{Dutra}}, \bibnamefont{and}
  \bibinfo{author}{\bibfnamefont{M.}~\bibnamefont{Malheiro}},
  \bibinfo{journal}{Journal of Cosmology and Astroparticle Physics}
  \textbf{\bibinfo{volume}{2020}}, \bibinfo{pages}{039} (\bibinfo{year}{2020}).

\bibitem[{\citenamefont{Deb et~al.}(2019)\citenamefont{Deb, Ketov, Maurya,
  Khlopov, Moraes, and Ray}}]{Deb:2018sgt}
\bibinfo{author}{\bibfnamefont{D.}~\bibnamefont{Deb}},
  \bibinfo{author}{\bibfnamefont{S.~V.} \bibnamefont{Ketov}},
  \bibinfo{author}{\bibfnamefont{S.}~\bibnamefont{Maurya}},
  \bibinfo{author}{\bibfnamefont{M.}~\bibnamefont{Khlopov}},
  \bibinfo{author}{\bibfnamefont{P.}~\bibnamefont{Moraes}}, \bibnamefont{and}
  \bibinfo{author}{\bibfnamefont{S.}~\bibnamefont{Ray}},
  \bibinfo{journal}{Monthly Notices of the Royal Astronomical Society}
  \textbf{\bibinfo{volume}{485}}, \bibinfo{pages}{5652} (\bibinfo{year}{2019}).

\bibitem[{\citenamefont{Maurya et~al.}(2019)\citenamefont{Maurya, Errehymy,
  Deb, Tello-Ortiz, and Daoud}}]{Maurya:2019sfm}
\bibinfo{author}{\bibfnamefont{S.}~\bibnamefont{Maurya}},
  \bibinfo{author}{\bibfnamefont{A.}~\bibnamefont{Errehymy}},
  \bibinfo{author}{\bibfnamefont{D.}~\bibnamefont{Deb}},
  \bibinfo{author}{\bibfnamefont{F.}~\bibnamefont{Tello-Ortiz}},
  \bibnamefont{and} \bibinfo{author}{\bibfnamefont{M.}~\bibnamefont{Daoud}},
  \bibinfo{journal}{Physical Review D} \textbf{\bibinfo{volume}{100}},
  \bibinfo{pages}{044014} (\bibinfo{year}{2019}).

\bibitem[{\citenamefont{Maurya and Tello-Ortiz}(2020)}]{Maurya:2019iup}
\bibinfo{author}{\bibfnamefont{S.}~\bibnamefont{Maurya}} \bibnamefont{and}
  \bibinfo{author}{\bibfnamefont{F.}~\bibnamefont{Tello-Ortiz}},
  \bibinfo{journal}{Annals of Physics} \textbf{\bibinfo{volume}{414}},
  \bibinfo{pages}{168070} (\bibinfo{year}{2020}).

\bibitem[{\citenamefont{Nashed}(2023{\natexlab{a}})}]{Nashed:2023uvk}
\bibinfo{author}{\bibfnamefont{G.~G.~L.} \bibnamefont{Nashed}},
  \bibinfo{journal}{Eur. Phys. J. C} \textbf{\bibinfo{volume}{83}},
  \bibinfo{pages}{698} (\bibinfo{year}{2023}{\natexlab{a}}).

\bibitem[{\citenamefont{Chandrasekhar}(1964)}]{Chandrasekhar:1964zz}
\bibinfo{author}{\bibfnamefont{S.}~\bibnamefont{Chandrasekhar}},
  \bibinfo{journal}{Astrophys. J.} \textbf{\bibinfo{volume}{140}},
  \bibinfo{pages}{417} (\bibinfo{year}{1964}), \bibinfo{note}{[Erratum:
  Astrophys.J. 140, 1342 (1964)]}.

\bibitem[{\citenamefont{Das et~al.}(2024)\citenamefont{Das, Debnath, Ashraf,
  and Khurana}}]{Das:2023bff}
\bibinfo{author}{\bibfnamefont{K.~P.} \bibnamefont{Das}},
  \bibinfo{author}{\bibfnamefont{U.}~\bibnamefont{Debnath}},
  \bibinfo{author}{\bibfnamefont{A.}~\bibnamefont{Ashraf}}, \bibnamefont{and}
  \bibinfo{author}{\bibfnamefont{M.}~\bibnamefont{Khurana}},
  \bibinfo{journal}{Phys. Dark Univ.} \textbf{\bibinfo{volume}{43}},
  \bibinfo{pages}{101398} (\bibinfo{year}{2024}).

\bibitem[{\citenamefont{Shamir and Komal}(2017)}]{Shamir:2017ndy}
\bibinfo{author}{\bibfnamefont{M.~F.} \bibnamefont{Shamir}} \bibnamefont{and}
  \bibinfo{author}{\bibfnamefont{A.}~\bibnamefont{Komal}},
  \bibinfo{journal}{Int. J. Geom. Meth. Mod. Phys.}
  \textbf{\bibinfo{volume}{14}}, \bibinfo{pages}{1750169}
  (\bibinfo{year}{2017}), \eprint{1708.06602}.

\bibitem[{\citenamefont{Shamir and Ahmad}(2017)}]{Shamir:2017rjz}
\bibinfo{author}{\bibfnamefont{M.~F.} \bibnamefont{Shamir}} \bibnamefont{and}
  \bibinfo{author}{\bibfnamefont{M.}~\bibnamefont{Ahmad}},
  \bibinfo{journal}{Eur. Phys. J. C} \textbf{\bibinfo{volume}{77}},
  \bibinfo{pages}{674} (\bibinfo{year}{2017}), \eprint{1705.06910}.

\bibitem[{\citenamefont{Landau}(2013)}]{landau2013classical}
\bibinfo{author}{\bibfnamefont{L.~D.} \bibnamefont{Landau}},
  \emph{\bibinfo{title}{The classical theory of fields}},
  vol.~\bibinfo{volume}{2} (\bibinfo{publisher}{Elsevier},
  \bibinfo{year}{2013}).

\bibitem[{\citenamefont{Einstein}(1916)}]{Einstein:1916vd}
\bibinfo{author}{\bibfnamefont{A.}~\bibnamefont{Einstein}},
  \bibinfo{journal}{Annalen Phys.} \textbf{\bibinfo{volume}{49}},
  \bibinfo{pages}{769} (\bibinfo{year}{1916}).

\bibitem[{\citenamefont{Starobinsky}(2007)}]{Starobinsky:2007hu}
\bibinfo{author}{\bibfnamefont{A.~A.} \bibnamefont{Starobinsky}},
  \bibinfo{journal}{JETP Lett.} \textbf{\bibinfo{volume}{86}},
  \bibinfo{pages}{157} (\bibinfo{year}{2007}), \eprint{0706.2041}.

\bibitem[{\citenamefont{Sotiriou and Faraoni}(2010)}]{Sotiriou:2008rp}
\bibinfo{author}{\bibfnamefont{T.~P.} \bibnamefont{Sotiriou}} \bibnamefont{and}
  \bibinfo{author}{\bibfnamefont{V.}~\bibnamefont{Faraoni}},
  \bibinfo{journal}{Reviews of Modern Physics} \textbf{\bibinfo{volume}{82}},
  \bibinfo{pages}{451} (\bibinfo{year}{2010}).

\bibitem[{\citenamefont{Nashed and Capozziello}(2019)}]{Nashed:2019tuk}
\bibinfo{author}{\bibfnamefont{G.~G.~L.} \bibnamefont{Nashed}}
  \bibnamefont{and}
  \bibinfo{author}{\bibfnamefont{S.}~\bibnamefont{Capozziello}},
  \bibinfo{journal}{Phys. Rev. D} \textbf{\bibinfo{volume}{99}},
  \bibinfo{pages}{104018} (\bibinfo{year}{2019}), \eprint{1902.06783}.

\bibitem[{\citenamefont{Nashed and
  Nojiri}(2023{\natexlab{a}})}]{Nashed:2022xmv}
\bibinfo{author}{\bibfnamefont{G.~G.~L.} \bibnamefont{Nashed}}
  \bibnamefont{and} \bibinfo{author}{\bibfnamefont{S.}~\bibnamefont{Nojiri}},
  \bibinfo{journal}{Fortsch. Phys.} \textbf{\bibinfo{volume}{71}},
  \bibinfo{pages}{2200091} (\bibinfo{year}{2023}{\natexlab{a}}),
  \eprint{2206.04836}.

\bibitem[{\citenamefont{Nashed and Nojiri}(2022)}]{Nashed:2022mij}
\bibinfo{author}{\bibfnamefont{G.~G.~L.} \bibnamefont{Nashed}}
  \bibnamefont{and} \bibinfo{author}{\bibfnamefont{S.}~\bibnamefont{Nojiri}},
  \bibinfo{journal}{Phys. Rev. D} \textbf{\bibinfo{volume}{106}},
  \bibinfo{pages}{044024} (\bibinfo{year}{2022}), \eprint{2207.13915}.

\bibitem[{\citenamefont{Nashed and
  Nojiri}(2023{\natexlab{b}})}]{Nashed:2021cfs}
\bibinfo{author}{\bibfnamefont{G.~G.~L.} \bibnamefont{Nashed}}
  \bibnamefont{and} \bibinfo{author}{\bibfnamefont{S.}~\bibnamefont{Nojiri}},
  \bibinfo{journal}{Eur. Phys. J. C} \textbf{\bibinfo{volume}{83}},
  \bibinfo{pages}{68} (\bibinfo{year}{2023}{\natexlab{b}}),
  \eprint{2112.13391}.

\bibitem[{\citenamefont{Bamba et~al.}(2010)\citenamefont{Bamba, Odintsov,
  Sebastiani, and Zerbini}}]{Bamba:2010wfw}
\bibinfo{author}{\bibfnamefont{K.}~\bibnamefont{Bamba}},
  \bibinfo{author}{\bibfnamefont{S.~D.} \bibnamefont{Odintsov}},
  \bibinfo{author}{\bibfnamefont{L.}~\bibnamefont{Sebastiani}},
  \bibnamefont{and} \bibinfo{author}{\bibfnamefont{S.}~\bibnamefont{Zerbini}},
  \bibinfo{journal}{Eur. Phys. J. C} \textbf{\bibinfo{volume}{67}},
  \bibinfo{pages}{295} (\bibinfo{year}{2010}), \eprint{0911.4390}.

\bibitem[{\citenamefont{Nashed and Capozziello}(2020)}]{Nashed:2020kjh}
\bibinfo{author}{\bibfnamefont{G.~G.} \bibnamefont{Nashed}} \bibnamefont{and}
  \bibinfo{author}{\bibfnamefont{S.}~\bibnamefont{Capozziello}},
  \bibinfo{journal}{The European Physical Journal C}
  \textbf{\bibinfo{volume}{80}}, \bibinfo{pages}{1} (\bibinfo{year}{2020}).

\bibitem[{\citenamefont{Ganguly et~al.}(2014)\citenamefont{Ganguly, Gannouji,
  Goswami, and Ray}}]{Ganguly:2013taa}
\bibinfo{author}{\bibfnamefont{A.}~\bibnamefont{Ganguly}},
  \bibinfo{author}{\bibfnamefont{R.}~\bibnamefont{Gannouji}},
  \bibinfo{author}{\bibfnamefont{R.}~\bibnamefont{Goswami}}, \bibnamefont{and}
  \bibinfo{author}{\bibfnamefont{S.}~\bibnamefont{Ray}},
  \bibinfo{journal}{Physical Review D} \textbf{\bibinfo{volume}{89}},
  \bibinfo{pages}{064019} (\bibinfo{year}{2014}).

\bibitem[{\citenamefont{Zeldovich and Novikov}(1971)}]{1971reas.book.....Z}
\bibinfo{author}{\bibfnamefont{Y.~B.} \bibnamefont{Zeldovich}}
  \bibnamefont{and} \bibinfo{author}{\bibfnamefont{I.~D.}
  \bibnamefont{Novikov}}, \bibinfo{journal}{Chicago: University of Chicago
  Press}  (\bibinfo{year}{1971}).

\bibitem[{\citenamefont{Herrera}(1992)}]{Herrera:1992lwz}
\bibinfo{author}{\bibfnamefont{L.}~\bibnamefont{Herrera}},
  \bibinfo{journal}{Phys. Lett. A} \textbf{\bibinfo{volume}{165}},
  \bibinfo{pages}{206} (\bibinfo{year}{1992}).

\bibitem[{\citenamefont{Buchdahl}(1959{\natexlab{a}})}]{Buchdahl:1959zz}
\bibinfo{author}{\bibfnamefont{H.~A.} \bibnamefont{Buchdahl}},
  \bibinfo{journal}{Physical Review} \textbf{\bibinfo{volume}{116}},
  \bibinfo{pages}{1027} (\bibinfo{year}{1959}{\natexlab{a}}).

\bibitem[{\citenamefont{Ivanov}(2002)}]{Ivanov:2002xf}
\bibinfo{author}{\bibfnamefont{B.~V.} \bibnamefont{Ivanov}},
  \bibinfo{journal}{Physical Review D} \textbf{\bibinfo{volume}{65}},
  \bibinfo{pages}{104011} (\bibinfo{year}{2002}).

\bibitem[{\citenamefont{Barraco et~al.}(2003)\citenamefont{Barraco, Hamity, and
  Gleiser}}]{Barraco:2003jq}
\bibinfo{author}{\bibfnamefont{D.~E.} \bibnamefont{Barraco}},
  \bibinfo{author}{\bibfnamefont{V.~H.} \bibnamefont{Hamity}},
  \bibnamefont{and} \bibinfo{author}{\bibfnamefont{R.~J.}
  \bibnamefont{Gleiser}}, \bibinfo{journal}{Physical Review D}
  \textbf{\bibinfo{volume}{67}}, \bibinfo{pages}{064003}
  (\bibinfo{year}{2003}).

\bibitem[{\citenamefont{B{\"o}hmer and Harko}(2006)}]{Boehmer:2006ye}
\bibinfo{author}{\bibfnamefont{C.}~\bibnamefont{B{\"o}hmer}} \bibnamefont{and}
  \bibinfo{author}{\bibfnamefont{T.}~\bibnamefont{Harko}},
  \bibinfo{journal}{Classical and Quantum Gravity}
  \textbf{\bibinfo{volume}{23}}, \bibinfo{pages}{6479} (\bibinfo{year}{2006}).

\bibitem[{\citenamefont{Chan et~al.}(1993)\citenamefont{Chan, Herrera, and
  Santos}}]{chan1993dynamical}
\bibinfo{author}{\bibfnamefont{R.}~\bibnamefont{Chan}},
  \bibinfo{author}{\bibfnamefont{L.}~\bibnamefont{Herrera}}, \bibnamefont{and}
  \bibinfo{author}{\bibfnamefont{N.}~\bibnamefont{Santos}},
  \bibinfo{journal}{Monthly Notices of the Royal Astronomical Society}
  \textbf{\bibinfo{volume}{265}}, \bibinfo{pages}{533} (\bibinfo{year}{1993}).

\bibitem[{\citenamefont{Heintzmann and
  Hillebrandt}(1975)}]{1975A&A....38...51H}
\bibinfo{author}{\bibfnamefont{H.}~\bibnamefont{Heintzmann}} \bibnamefont{and}
  \bibinfo{author}{\bibfnamefont{W.}~\bibnamefont{Hillebrandt}},
  \bibinfo{journal}{Astronomy and Astrophysics} \textbf{\bibinfo{volume}{38}},
  \bibinfo{pages}{51} (\bibinfo{year}{1975}).

\bibitem[{\citenamefont{Tolman}(1930)}]{1930PhRv...35..896T}
\bibinfo{author}{\bibfnamefont{R.~C.} \bibnamefont{Tolman}},
  \bibinfo{journal}{Physical Review} \textbf{\bibinfo{volume}{35}},
  \bibinfo{pages}{896} (\bibinfo{year}{1930}).

\bibitem[{\citenamefont{Buchdahl}(1959{\natexlab{b}})}]{PhysRev.116.1027}
\bibinfo{author}{\bibfnamefont{H.~A.} \bibnamefont{Buchdahl}},
  \bibinfo{journal}{Phys. Rev.} \textbf{\bibinfo{volume}{116}},
  \bibinfo{pages}{1027} (\bibinfo{year}{1959}{\natexlab{b}}),
  \urlprefix\url{https://link.aps.org/doi/10.1103/PhysRev.116.1027}.

\bibitem[{\citenamefont{Abubekerov et~al.}(2008)\citenamefont{Abubekerov,
  Antokhina, Cherepashchuk, and Shimanskii}}]{Abubekerov:2008inw}
\bibinfo{author}{\bibfnamefont{M.}~\bibnamefont{Abubekerov}},
  \bibinfo{author}{\bibfnamefont{E.}~\bibnamefont{Antokhina}},
  \bibinfo{author}{\bibfnamefont{A.}~\bibnamefont{Cherepashchuk}},
  \bibnamefont{and}
  \bibinfo{author}{\bibfnamefont{V.}~\bibnamefont{Shimanskii}},
  \bibinfo{journal}{Astronomy Reports} \textbf{\bibinfo{volume}{52}},
  \bibinfo{pages}{379} (\bibinfo{year}{2008}).

\bibitem[{\citenamefont{Pons et~al.}(2002)\citenamefont{Pons, Walter, Lattimer,
  Prakash, Neuh{\"a}user, and An}}]{Pons:2001px}
\bibinfo{author}{\bibfnamefont{J.~A.} \bibnamefont{Pons}},
  \bibinfo{author}{\bibfnamefont{F.~M.} \bibnamefont{Walter}},
  \bibinfo{author}{\bibfnamefont{J.~M.} \bibnamefont{Lattimer}},
  \bibinfo{author}{\bibfnamefont{M.}~\bibnamefont{Prakash}},
  \bibinfo{author}{\bibfnamefont{R.}~\bibnamefont{Neuh{\"a}user}},
  \bibnamefont{and} \bibinfo{author}{\bibfnamefont{P.}~\bibnamefont{An}},
  \bibinfo{journal}{The Astrophysical Journal} \textbf{\bibinfo{volume}{564}},
  \bibinfo{pages}{981} (\bibinfo{year}{2002}).

\bibitem[{\citenamefont{Rawls et~al.}(2011)\citenamefont{Rawls, Orosz,
  McClintock, Torres, Bailyn, and Buxton}}]{Rawls:2011jw}
\bibinfo{author}{\bibfnamefont{M.~L.} \bibnamefont{Rawls}},
  \bibinfo{author}{\bibfnamefont{J.~A.} \bibnamefont{Orosz}},
  \bibinfo{author}{\bibfnamefont{J.~E.} \bibnamefont{McClintock}},
  \bibinfo{author}{\bibfnamefont{M.~A.} \bibnamefont{Torres}},
  \bibinfo{author}{\bibfnamefont{C.~D.} \bibnamefont{Bailyn}},
  \bibnamefont{and} \bibinfo{author}{\bibfnamefont{M.~M.}
  \bibnamefont{Buxton}}, \bibinfo{journal}{The Astrophysical Journal}
  \textbf{\bibinfo{volume}{730}}, \bibinfo{pages}{25} (\bibinfo{year}{2011}).

\bibitem[{\citenamefont{Ozel et~al.}(2009)\citenamefont{Ozel, Guver, and
  Psaltis}}]{Ozel:2008kb}
\bibinfo{author}{\bibfnamefont{F.}~\bibnamefont{Ozel}},
  \bibinfo{author}{\bibfnamefont{T.}~\bibnamefont{Guver}}, \bibnamefont{and}
  \bibinfo{author}{\bibfnamefont{D.}~\bibnamefont{Psaltis}},
  \bibinfo{journal}{Astrophys. J.} \textbf{\bibinfo{volume}{693}},
  \bibinfo{pages}{1775} (\bibinfo{year}{2009}), \eprint{0810.1521}.

\bibitem[{\citenamefont{Webb and Barret}(2007)}]{Webb:2007tc}
\bibinfo{author}{\bibfnamefont{N.~A.} \bibnamefont{Webb}} \bibnamefont{and}
  \bibinfo{author}{\bibfnamefont{D.}~\bibnamefont{Barret}},
  \bibinfo{journal}{The Astrophysical Journal} \textbf{\bibinfo{volume}{671}},
  \bibinfo{pages}{727} (\bibinfo{year}{2007}).

\bibitem[{\citenamefont{Heinke et~al.}(2006)\citenamefont{Heinke, Rybicki,
  Narayan, and Grindlay}}]{Rybicki:2005id}
\bibinfo{author}{\bibfnamefont{C.~O.} \bibnamefont{Heinke}},
  \bibinfo{author}{\bibfnamefont{G.~B.} \bibnamefont{Rybicki}},
  \bibinfo{author}{\bibfnamefont{R.}~\bibnamefont{Narayan}}, \bibnamefont{and}
  \bibinfo{author}{\bibfnamefont{J.~E.} \bibnamefont{Grindlay}},
  \bibinfo{journal}{The Astrophysical Journal} \textbf{\bibinfo{volume}{644}},
  \bibinfo{pages}{1090} (\bibinfo{year}{2006}).

\bibitem[{\citenamefont{G{\"u}ver et~al.}(2010)\citenamefont{G{\"u}ver,
  Wroblewski, Camarota, and {\"O}zel}}]{Guver:2010td}
\bibinfo{author}{\bibfnamefont{T.}~\bibnamefont{G{\"u}ver}},
  \bibinfo{author}{\bibfnamefont{P.}~\bibnamefont{Wroblewski}},
  \bibinfo{author}{\bibfnamefont{L.}~\bibnamefont{Camarota}}, \bibnamefont{and}
  \bibinfo{author}{\bibfnamefont{F.}~\bibnamefont{{\"O}zel}},
  \bibinfo{journal}{The Astrophysical Journal} \textbf{\bibinfo{volume}{719}},
  \bibinfo{pages}{1807} (\bibinfo{year}{2010}).

\bibitem[{\citenamefont{Naik et~al.}(2011)\citenamefont{Naik, Paul, and
  Ali}}]{Naik:2011qc}
\bibinfo{author}{\bibfnamefont{S.}~\bibnamefont{Naik}},
  \bibinfo{author}{\bibfnamefont{B.}~\bibnamefont{Paul}}, \bibnamefont{and}
  \bibinfo{author}{\bibfnamefont{Z.}~\bibnamefont{Ali}}, \bibinfo{journal}{The
  Astrophysical Journal} \textbf{\bibinfo{volume}{737}}, \bibinfo{pages}{79}
  (\bibinfo{year}{2011}).

\bibitem[{\citenamefont{Marshall and Angelini}(1996)}]{1996IAUC.6331....1M}
\bibinfo{author}{\bibfnamefont{F.}~\bibnamefont{Marshall}} \bibnamefont{and}
  \bibinfo{author}{\bibfnamefont{L.}~\bibnamefont{Angelini}},
  \bibinfo{journal}{International Astronomical Union Circular}
  \textbf{\bibinfo{volume}{6331}}, \bibinfo{pages}{1} (\bibinfo{year}{1996}).

\bibitem[{\citenamefont{Nashed}(2023{\natexlab{b}})}]{Nashed:2023pxd}
\bibinfo{author}{\bibfnamefont{G.~G.~L.} \bibnamefont{Nashed}},
  \bibinfo{journal}{Astrophys. J.} \textbf{\bibinfo{volume}{950}},
  \bibinfo{pages}{129} (\bibinfo{year}{2023}{\natexlab{b}}),
  \eprint{2306.10273}.

\end{thebibliography}

\end{document}